\documentclass[3p,twocolumn,letter]{elsarticle}

\usepackage{mdwlist}
\usepackage{subfig}
\usepackage{wrapfig}
\usepackage{amssymb}
\usepackage{psfrag}
\usepackage{graphicx}
\usepackage{alltt}
\usepackage[fleqn]{amsmath}
\usepackage{color}
\usepackage{colortbl}
\usepackage[colorlinks,breaklinks]{hyperref}

\journal{Nuclear Engineering and Design; Accepted for publication on Jun.\ 12, 2013}

\newcommand{\be}{\begin{equation}}
\newcommand{\ee}{\end{equation}}
\newcommand{\beqn}{\begin{eqnarray}}
\newcommand{\eeqn}{\end{eqnarray}}

\newcommand{\bdm}{\begin{displaymath}}
\newcommand{\edm}{\end{displaymath}}
\newcommand{\beqa}{\begin{eqnarray}}
\newcommand{\eeqa}{\end{eqnarray}}

\newcommand{\vv}{{\bf v}}

\usepackage{color}

\newcommand{\bv}[1]{{\mbox{\boldmath$\mathbf{#1}$}}}    
\newcommand{\irmean}[1]{{\langle{#1}\rangle}}           

\newcommand{\ti}{Large-Eddy Simulations of Turbulent Flow for
                 Grid-to-Rod Fretting in Nuclear Reactors\\[0.2cm]
                 \normalsize\texttt{\laur}\\[-0.2cm]}

\newcommand{\laur}{\textbf{LA-UR 12-26572}}

\begin{document}

\begin{frontmatter}

\title{\textbf{\ti}}

\author[LANL]{J.\ Bakosi}
\author[LANL]{M.A.\ Christon}
\author[LANL]{R.B.\ Lowrie}
\author[LANL]{L.A.\ Pritchett-Sheats}
\author[INL]{R.R.\ Nourgaliev}

\address[LANL]{\{jbakosi,christon,lowrie,lpritch\}@lanl.gov \\
  Corresponding author: J. Bakosi
  (jbakosi@lanl.gov, tel: +1-505-663-5607, fax: +1-505-663-5504) \\
  Computational Physics Group (CCS-2) \\
  Computer, Computational and Statistical Sciences Division \\
  Los Alamos National Laboratory \\
  Los Alamos, NM 87544}

\address[INL]{robert.nourgaliev@inl.gov \\
  Reactor Safety Simulation Group \\
  Thermal Science and Safety Analysis Department \\
  Idaho National Laboratory \\
  Idaho Falls, ID 83415}

\hypersetup{citecolor=blue,
            linkcolor=blue,
            urlcolor=blue,
            pdftitle=\ti,
            pdfauthor={J. Bakosi,
                       M.A. Christon, 
                       R.B. Lowrie,
                       L.A. Pritchett-Sheats,
                       R.R. Nourgaliev}}

\begin{abstract}
The grid-to-rod fretting (GTRF) problem in pressurized water reactors
is a flow-induced vibration problem that results in wear and failure
of the fuel rods in nuclear assemblies. In order to understand the
fluid dynamics of GTRF and to build an archival database of turbulence
statistics for various configurations, implicit large-eddy simulations
of time-dependent single-phase turbulent flow have been performed in
$3\times3$ and $5\times5$ rod bundles with a single grid spacer. To
assess the computational mesh and resolution requirements, a method
for quantitative assessment of unstructured meshes with no-slip walls
is described. The calculations have been carried out using Hydra-TH, a
thermal-hydraulics code developed at Los Alamos for the Consortium for
Advanced Simulation of Light water reactors, a United States
Department of Energy Innovation Hub. Hydra-TH uses a second-order
implicit incremental projection method to solve the single-phase
incompressible Navier-Stokes equations. The simulations explicitly
resolve the large scale motions of the turbulent flow field using
first principles and rely on a monotonicity-preserving numerical
technique to represent the unresolved scales.  Each series of
simulations for the $3\times3$ and $5\times5$ rod-bundle geometries is
an analysis of the flow field statistics combined with a
mesh-refinement study and validation with available experimental data.
Our primary focus is the time history and statistics of the forces
loading the fuel rods. These hydrodynamic forces are believed to be
the key player resulting in rod vibration and GTRF wear, one of the
leading causes for leaking nuclear fuel which costs power utilities
millions of dollars in preventive measures. We demonstrate that
implicit large-eddy simulation of rod-bundle flows is a viable way to
calculate the excitation forces for the GTRF problem.

\end{abstract}


\end{frontmatter}

\section{Introduction\label{sec:intro}}
Within the core of a pressurized-water nuclear reactor (PWR), water flow is used
to cool the hot irradiated fuel rods. The grid-to-rod fretting (GTRF) problem in
such reactors is a flow-induced vibration problem that results in wear and
failure of the rods. GTRF wear is one of the leading causes for leaking nuclear
fuel and costs power utilities millions of dollars in preventive measures. In
order to understand the root causes of such fuel leaks, we investigate the
complex turbulent coolant flow around fuel-rod bundles. Our ultimate goal is to
accurately predict the turbulent excitation forces on the fuel rods, along with
the coupled structural response of the rods and their supports. To date, it has
not been possible to completely characterize the flow-induced fluid-structure
interaction (FSI) problem for GTRF. Indeed, given the incompressible nature of
the coolant, the relatively high Reynolds number, and the flexible character of
the fuel rods and spacers, the FSI problem at the reactor core scale is
daunting.

As pointed out by Pa\"{i}doussis \cite{paidoussis:1982}, there are
a number of flow-induced vibration problems in a nuclear power plant that
involve the reactor, associated piping, heat exchangers, steam generators, and
ancillary diagnostic equipment. Pettigrew, et al.\ \cite{pettigrew:1991} also
consider a broad array of flow-induced vibration problems albeit specialized to
the CANDU reactor configuration. A complete review of the work associated with
all possible flow-induced reactor vibration problems is far beyond the scope of
this work. However, a brief overview of some of the work related to the
application of computational fluid dynamics (CFD) for fretting problems is
presented here.

The work by Ikeno and Kajishima \cite{ikeno:2006} relied on large-eddy
simulation (LES) of the flow downstream of mixing vanes in a rod bundle.  They
used an immersed-boundary technique to treat the complex geometry and a dynamic
subgrid-scale model to examine the mixing grid wake and downstream swirl.
Benhamadouce, et al.\ \cite{benhamadouche:2009} performed an LES of the flow in
the subchannels surrounding a single rod, and subsequently used the turbulent
forces to compute the elastic vibration of the fuel rod. Here, a relatively
coarse mesh with 8 million cells was used for the $Re=30,000$ flow. Related work
by Kim \cite{kim:2009,kim:2010a,kim:2010b} has considered grid-to-rod fretting
wear models for PWRs as well as the effects of the rod
support conditions on fuel rod vibration.

Conner, et al.\ \cite{conner:2010} present a validation study using a $5 \times
5$ rod bundle, representative of a fuel assembly and compare the
mixing-vane-induced swirl with particle image velocimetry (PIV) data. Here, the
RNG $k$-$\epsilon$ model was used with $40 \leq y^+ \leq 100$ to compute a
steady-state solution. Yan, et al.\ \cite{yan:2011} performed time-accurate CFD
computations and compared the effect of the so-called ``protective grid'' at the
fuel-inlet region of a reactor.  For this study, meshes with 7, 16 and 60
million cells were used.  Here, it was shown that a time-accurate CFD
calculation can be used to determine transient fuel rod forces for subsequent
dynamic analyses. This work also demonstrated that the protective grid
significantly reduces flow-induced vibration at the reactor core inlet. Zhang
and Yu \cite{zhang:2011}, and Bhattacharya, et al.\ \cite{bhattacharya:2012}
have performed large-eddy simulations using the CANDU fuel bundles and the
vortex shedding phenomena associated with the endplates.  The work by
Delafontaine and Ricciardi \cite{delafontaine:2012} used LES to determine the
time-dependent rod forces downstream of a $3 \times 3$ rod bundle.  Here
detailed information about the angular variation of pressure forces on the fuel
rod are presented.  The work by Liu, et al.\ \cite{liu:2013} considered
fluid-structure interaction in simplified fuel assemblies where a rod buckling
instability was demonstrated to occur with large axial flow velocities.  Related
work was performed by Mohany and Hassan \cite{mohany:2013} to represent the
flow-induced vibration and associated fretting wear in a CANDU fuel bundle.

\begin{figure*}[t]
\begin{center}
\setlength{\belowcaptionskip}{-0.6cm}
\subfloat[Surface mesh for the center rod and spacer in the 47M-cell $3\times3$
          rod bundle.]{
\includegraphics[width=1.0\columnwidth]{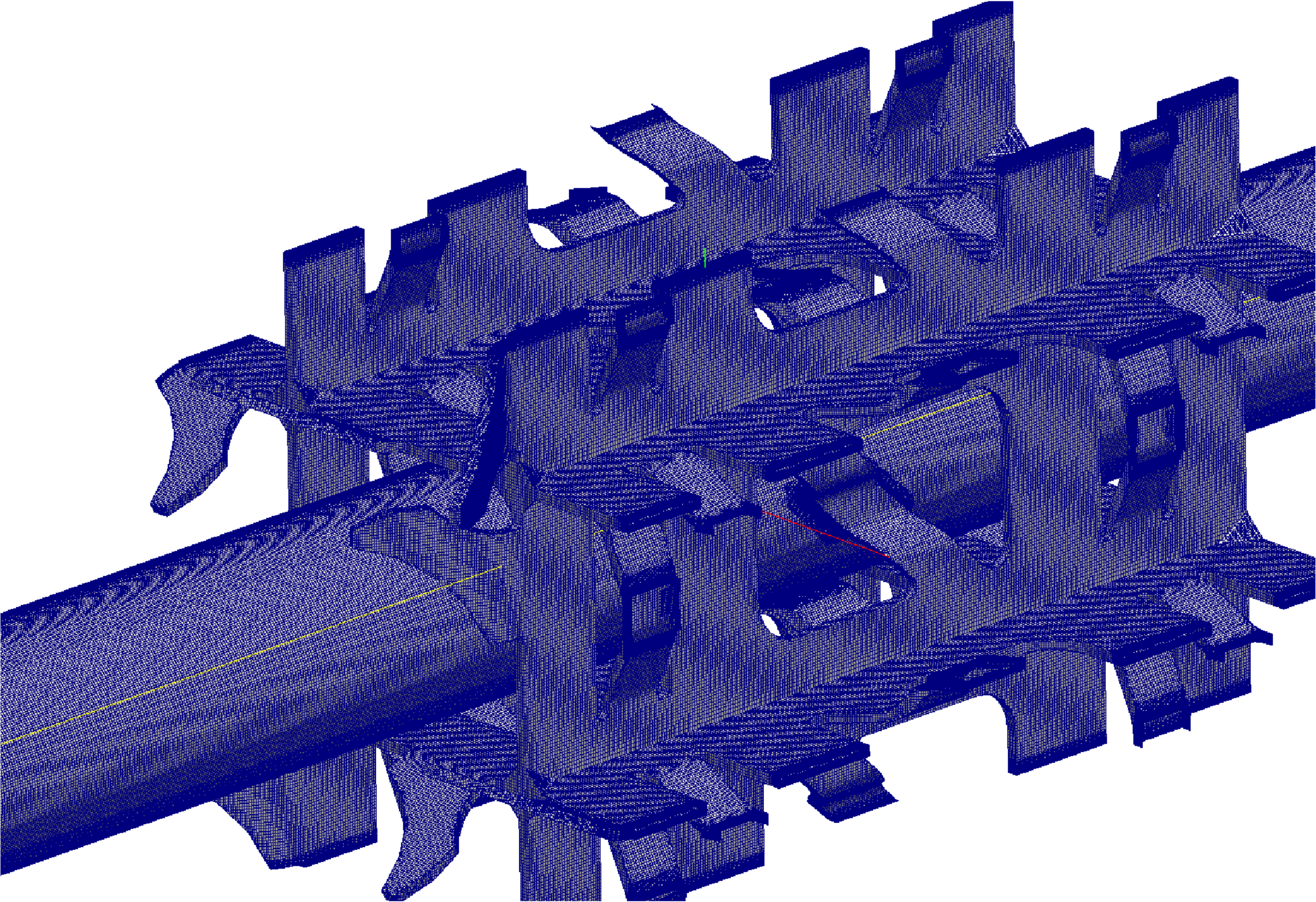}
\includegraphics[width=1.0\columnwidth]{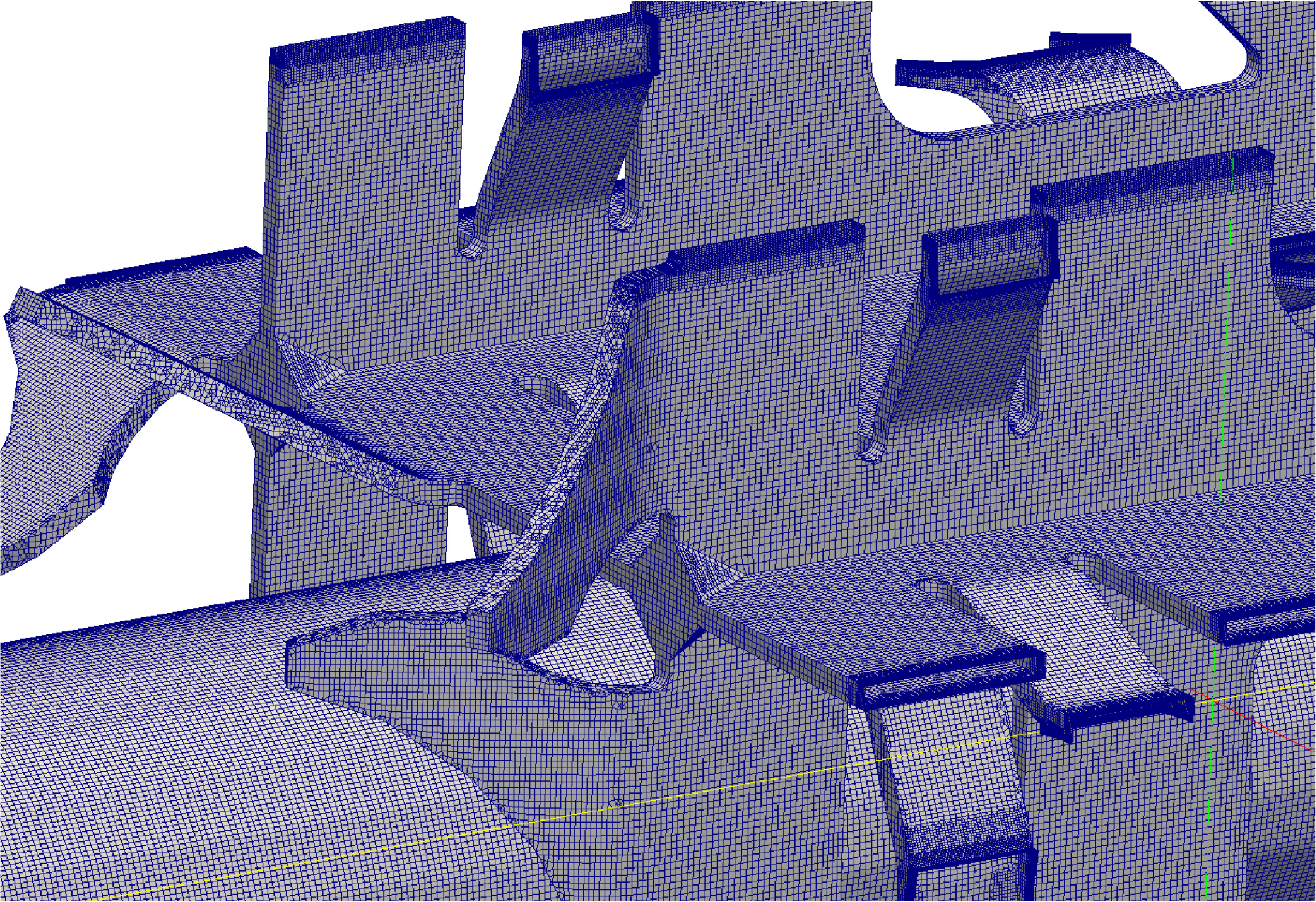}}\\
\subfloat[Surface mesh of the spacer with mixing vanes (left) and of the rods
          and spacer (right) int the 14M-cell $5\times5$ rod bundle.]{
\includegraphics[width=1.0\columnwidth]{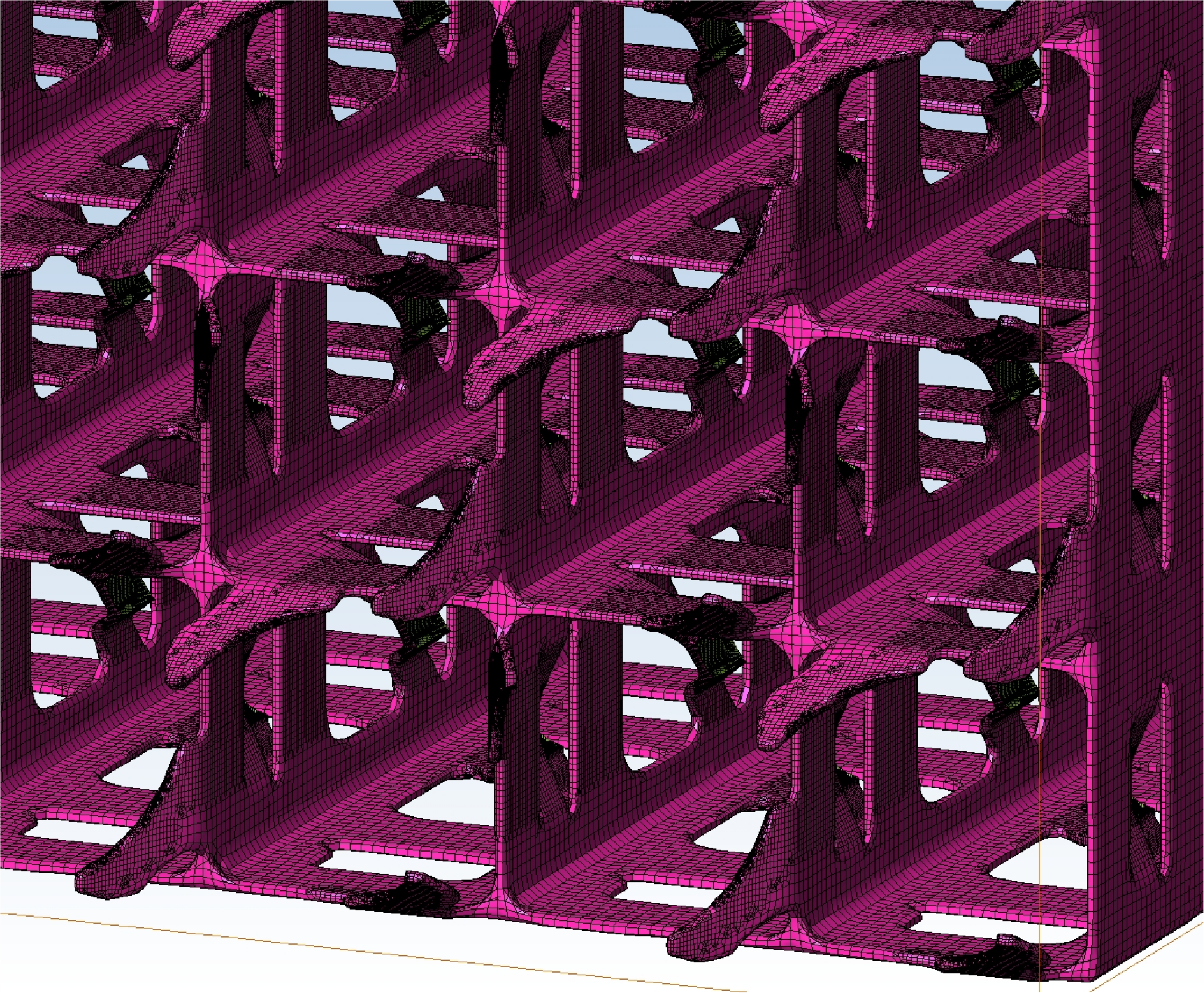}
\includegraphics[width=1.0\columnwidth]{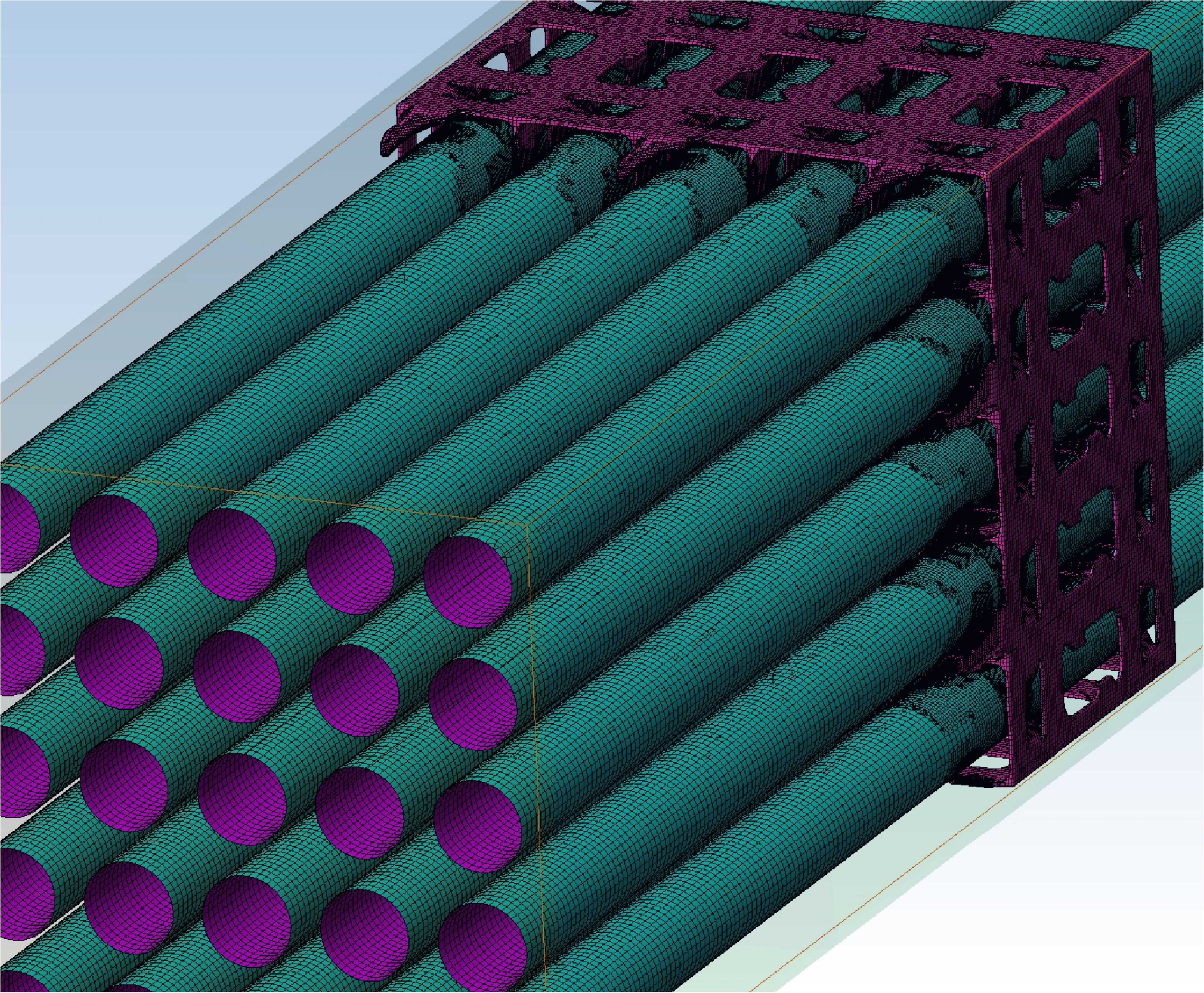}}
\caption{Surface meshes of the $5\times5$ and $3\times3$ rod bundle geometries.}
\label{fig:mesh}
\end{center}
\end{figure*}

Our investigation of LES for GTRF centers on CFD
simulations using Hydra-TH, a thermal-hydraulics code developed at Los
Alamos National Laboratory. Hydra-TH can compute high-Reynolds
number turbulent flows over arbitrarily complex geometries. There are
tens of thousands of fuel rods in a PWR, and computing the flow over
all of them simultaneously, with the fidelity required for GTRF, is
not feasible on today's computers. Instead, this study concentrates on
representative $3\times3$ and $5\times5$ rod-bundle configurations,
for which sample computational meshs are shown in Figure
\ref{fig:mesh}. The $3\times3$ and $5\times5$ geometries were
extracted from a $17\times17$ fuel assembly found in a typical PWR.
There is a large degree of symmetry in the fuel assembly which makes
this geometrical simplification a reasonable approximation. The
coolant flow generally moves axially on the outside of the fuel rods.
A ``grid spacer'' (rod support structure) is also shown, which
contains mixing vanes that stir the flow in order to enhance heat
transfer from the rod to the coolant flow. This stirring, along with 
the high Reynolds number of the flow, results in a complex turbulent
flow that is believed to be the primary driver of GTRF-induced failure
downstream of the grid spacer and mixing vanes.

The current engineering practice for the GTRF problem is to compute turbulent
flow solving the Reynolds-averaged Navier-Stokes (RANS) equations augmented by a
turbulence model. RANS models, used in either steady (RANS) or unsteady (URANS)
mode, directly compute the flow statistics by solving for only the mean field
values, and depending on the model, the second moments. For example, the
$k$-$\epsilon$ model, see e.g.\ \cite{Pope_00}, is a popular way to approximate
the \emph{effects of fluctuations on the mean} velocity. However,
it is important to appreciate that the main goal of the $k$-$\epsilon$ model is
to provide closure for the mean velocity via the time scale $\sim k/\epsilon$.
Consequently, one can expect a statistically meaningful description of the mean
but less of the fluctuations, e.g., $k$ or $\epsilon$, themselves, as only their
effect on the mean is represented.  If the second moments, e.g.\
the fluctuations about the mean, are also important, a model with a higher level
of description is required. For example, second-order RANS models directly
compute not only the means but the root-mean-square (RMS) fields as well.

RANS models, including the $k$-$\epsilon$ model, widely adopt the Boussinesq (or
turbulent viscosity) hypothesis whose limitations and their failure to
adequately predict even simple turbulent flows are well-known \cite{Pope_00}.
Therefore, it is best practice to compute a problem by a method that provides a
higher level of description before a RANS model can be operationally used with
confidence. One such technique is LES, which provides a direct representation of
the energy-containing motions of a turbulent flow.  Compared to RANS, LES has
the advantage of describing unsteady, large-scale turbulent structures, and
hence can be used to study phenomena such as the unsteady loads of the GTRF
problem. In LES, the dynamically small scales are modeled, while the large
unsteady motions are computed without approximation. If the small scales are
universal, i.e.\ Kolmogorov's hypothesis holds, as in many engineering
applications, LES is known to provide excellent results, as the effect of the
small (modeled) scales on the large scales are negligible.

We believe a problem such as GTRF must be studied with LES before operational
use of unsteady RANS can be attempted. Consistent with this idea, this study is
a step towards (1) understanding the fluid dynamics of GTRF, (2) assessing the
computational resolution requirements, and (3) building a database of turbulence
statistics for different configurations based on which rational decisions for
future computations can be made and the development of a GTRF-specific RANS
model can be attempted.

The rest of this paper is organized as follows. Hydra-TH is briefly reviewed in
\S\ref{sec:hydra}, then \S\ref{sec:mesh} discusses mesh generation and presents
a method for quantitative mesh assessment for complex geometries. The flow
computations are discussed in \S\ref{sec:calculations}. Finally,
\S\ref{sec:summary} gives a summary, while \S\ref{sec:future} points to future
directions.

\section{Overview of Hydra-TH\label{sec:hydra}}
Hydra-TH is a thermal hydraulics code developed at Los Alamos National
Laboratory for the Consortium for Advanced Simulation of Light water reactors
(CASL).\footnote{\href{http://www.casl.gov}{www.casl.gov}} The code is being
developed to address a number of single and multiphase problems ranging from
GTRF to departure from nucleate boiling.

Hydra-TH is a massively parallel code built on the Hydra Toolkit. The Hydra
Toolkit is written in C++ and provides a rich suite of lightweight high
performance components that permit rapid application development, supports
multiple discretization techniques, provides I/O interfaces to permit reading
and writing multiple file formats for meshes, plot data, time-history, surface
and restart output. The Toolkit also provides run-time parallel domain
decomposition with data-migration for both static and dynamic load-balancing.
Linear algebra is handled through an abstract virtual interface that enables the
use of both native and external libraries such as
PETSc\footnote{\href{http://www.mcs.anl.gov/petsc}{www.mcs.anl.gov/petsc}} and
Trilinos\footnote{\href{http://trilinos.sandia.gov}{trilinos.sandia.gov}}.

Hydra-TH also contains a rich suite of turbulence models that range from LES to
detached-eddy and various RANS models. The code relies on a hybrid
finite-volume/finite-element discretization for incompressible flow that
provides a stable and accurate discretization while preserving local
conservation properties important in many thermal hydraulics applications.
Hydra-TH also supports the use of hybrid meshes that permit the resolution of
boundary layers on very complex geometries.

For the incompressible Navier-Stokes formulation, all transport
variables are cell-centered and treated with a locally conservative
discretization that includes a high-resolution
monotonicity-preser\-ving advection algorithm. The spatial
discretization is formally derived using a discontinuous-Galerkin
framework that, in the limit, reduces to a locally-conservative
finite-volume method with reconstruction and second-order spatial
accuracy. The advection algorithm is designed to permit both implicit
and explicit advection with the explicit advection targeted primarily
at volume-tracking with interface reconstruction. The available
time-integration methods include backward-Euler and the
neutrally-dissipative trapezoidal method. The implicit advective
treatment delivers unconditional stability for scalar transport
equations and conditional stability for the momentum equation.
As opposed to the transport variables, which are
cell-centered, the pressure is node-centered. A Poisson-equation is
solved for a Lagrange multiplier from which the pressure is computed.
The divergence-free constraint on the velocity field is enforced via a
projection algorithm, similar to \cite{gresho90a, gresho90b}.
The code has been run on up to ten thousand compute
cores using 100-million-cell meshes, and is being developed to exploit the
multiple levels of hybrid parallelism of future compute architectures. More
details on the Hydra Toolkit and its incompressible solver are given
in the Hydra-TH Theory manual \citep{MAC2011a}.

\section{Mesh Generation and Quality Assessment\label{sec:mesh}}
This section discusses the computational meshes and
associated quality metrics for GTRF LES computations. \S
\ref{sec:meshgen} presents two different meshing technologies used in
this work. \S \ref{sec:quality} presents a method for quantitative
assessment of unstructured meshes with no-slip walls. In \S
\ref{sec:evaluation} an evaluation of the two series of meshes is given.

\subsection{Mesh generation for GTRF}
\label{sec:meshgen}
In one of our earlier studies \cite{GTRF_2011}, several desirable
characteristics of the computational meshes for LES were discussed,
and include the following.
\begin{enumerate*}
\item Sufficient overall mesh resolution for capturing the important
energy-containing features of the flow.
\item Smooth transitions in regions downstream of the spacer to avoid unphysical
aliasing (i.e., numerical back-scatter) of the kinetic energy from smaller to
larger scales.
\item High quality boundary layers to adequately but economically resolve the
complex turbulent flow in the vicinity of walls.
\end{enumerate*}
In calculations with heat transfer, point 3 above may be particularly important
due to (1) the highly inhomogeneous and anisotropic nature of the flow near
walls, and (2) the potentially first-order influence of the mesh quality on the
simulation.

As an alternative to the Cubit mesh
generator,\footnote{\href{http://cubit.sandia.gov} {http://cubit.sandia.gov}}
used to generate the meshes for the calculations in \cite{GTRF_2011}, we
explored Numeca's Hexpress/Hybrid, a.k.a.\
Spider,\footnote{\href{http://www.numeca.be/index.php?id=hexhyb}
{http://www.numeca.be/index.php?id=hexhyb}} meshing tool. Spider's shrink-wrap
meshing technology is quite different from that of Cubit and allows for fully
automatic generation of body fitted meshes on arbitrarily complex geometries.
Spider meshes are unstructured, hex-dominant, and conformal, containing
hexahedra, tetrahedra, wedges, and pyramids. As a consequence, extremely complex
geometries can be discretized with good quality elements. Furthermore, Spider is
capable of generating high-quality viscous layers; its configuration is based on
a simple text file, though a graphical interface is also available; and it is
easy to use in batch mode, yielding fast throughput for generating a series
of meshes for convergence studies and uncertainty quantification.

We have generated a series of Spider meshes for the rod-bundle geometry
developed at Westinghouse Electric Company for both $3\times3$ and $5\times5$
rod configurations. The dimensions of the geometry, used for mesh
generation and computations, are the same as in \cite{elmahdi:2011}.
 The approximate cell-count for the $3\times3$ meshes are 2 million
(M), 7M, 14M, 27M, 30M, 47M, 80M, and 185M, and for the $5\times5$ meshes are
14M and 96M. Example snapshots of the surface mesh for the rod and spacer
geometries in the $3\times3$ and $5\times5$ rod-bundle geometries are displayed
in Figure \ref{fig:mesh}. Visual inspection reveals uniform cell sizes inside of
the domain with targeted refinement in corners and edges in the vicinity of the
spacer and symmetry planes (not shown). Compared to the Cubit meshes, discussed
in \cite{GTRF_2011}, which had abrupt $\sim4\Delta x$ jumps downstream of the
spacer, the Spider meshes have no visible abrupt transitions in cell sizes and
there are smooth transitions from refined corners inside the domain. These
features are desirable from the viewpoint of obtaining quality LES results.
The meshes, resulting in different number of total cells, have been
generated by changing the single parameter, \texttt{BASEH}, in the Spider input
deck. \texttt{BASEH} defines the minimum resolution, characteristic of the
largest cell size. This yields different-resolution and similar meshes.
The meshes discussed in this study contain no targeted boundary
layer refinement close to walls. Generating meshes with power-law-graded
boundary layers using Spider is a subject of future
work.

In summary, we are satisfied with Spider as a tool for mesh generation: it is
relatively easy to use, fast, and automatically generates high-quality meshes
for extremely complex geometries, required for the GTRF problem. As an example,
the 96M $5\times5$ mesh is generated in only 80 minutes on an 8-core workstation
with 48GM RAM. Spider is a shared-memory parallel code and its approximate
memory requirement is 0.5GB RAM per million cells generated; it can export the
mesh in the latest Exodus-II file format with an HDF5 container, required for
mesh sizes beyond $\sim$60M hex cells.

\subsection{Mesh quality assessment}
\label{sec:quality}
In CFD simulations of wall-bounded turbulent flows, the mesh quality along
no-slip walls has a first-order influence on the accuracy of the numerical
solution. Boundary elements need to be small and uniform in the wall-normal
direction for adequately resolved boundary layers.  However, for very complex
flow geometries the size and uniformity of the cells along walls are difficult
to maintain during mesh generation. Clearly, there is a need for a quantitative
assessment of the quality of complex meshes preferably \emph{a priori} of
large-scale LES simulations.

In CFD engineering practice, the quantity $y^+$ is commonly used to assess the
mesh at no-slip walls. The distance from the wall, measured in viscous lengths,
or wall-units, is defined by
\begin{equation}
y^+\equiv\frac{y}{\nu}\sqrt{\frac{\tau_\mathrm{w}}{\rho}},
\label{eq:yplus}
\end{equation}
where $y$, $\nu$, $\tau_\mathrm{w}$, and $\rho$ denote the (dimensional)
distance from the wall (in the wall-normal direction), kinematic viscosity,
wall-shear stress, and fluid density, respectively. As $y^+$ is similar to a
local Reynolds-number, its magnitude can be expected to determine the relative
importance of viscous and turbulent processes. Values of $y^+\approx1$ are
recommended for LES with full wall-resolution, and $y^+\gtrsim20$ for RANS
models with wall-functions. The value of $y^+$ can be computed in each
computational cell along no-slip walls for a given Reynolds number and mesh,
in general, yielding a different value in each cell. For unstructured
meshes, the $y^+$ field is generally non-uniform
along a surface.

To assess a mesh \emph{a priori} of computation, a physically realistic $y^+$
field, defined by Eq.\ \ref{eq:yplus}, must be obtained with minimal
computational effort. In a numerical solution of the Navier-Stokes
equations, the $y^+$ field obtained after the first
time step can be used to assess the mesh if (and only if) the initial and
boundary conditions are consistent with the dynamical level of approximation of
the computed flow. In constant-density single-phase flows, this consistency
amounts to a divergence-free velocity field and a consistent pressure, both
satisfying the prescribed initial and boundary conditions.

The incompressible flow solver in Hydra-TH is based on a second-order implicit
projection algorithm that uses a pressure-Poisson equation to
continuously project the velocity field to a divergence-free velocity
space at each time-step. Given a set of user-prescribed initial
and boundary conditions, the initial startup procedure (before $t\!=\!0$)
computes the solution of a Poisson equation for a Lagrange multiplier. Then a
subsequent projection of the prescribed velocity field to a divergence-free
subspace ensures that (1) the velocity field is divergence-free, (2) the
velocity is consistent with the pressure, and (3) both fields satisfy the
prescribed initial and boundary conditions for an incompressible flow. This
procedure guarantees that basic solvability conditions \cite{MAC2011a} are
satisfied at $t\!=\!0$ and that a mathematically and numerically well-posed
Navier-Stokes problem is integrated for $t\!>\!0$. For details on the startup
procedure in Hydra-TH, see \cite{MAC2011a}.

The $y^+$ fields, discussed below, have been obtained after the above startup
procedure, solving the constant-density Navier-Stokes equations, for
a single time step, and serve as the basis of the mesh quality
assessment \emph{a priori} of computations.  We emphasize that the
$y^+$ field computed after the first time step is approximate: it depends on the
mesh, the Reynolds number, and the turbulence model employed to compute
$\tau_\mathrm{w}$, and is only constant when the flow is statistically
stationary.  Ideally, a more accurate $y+$ could be obtained after the flow has
reached a statistically stationary state. However, for large meshes, obtaining a
statistically stationary state may require thousands of time steps and thus it
is not economical as a quick \emph{a priori} mesh assessment.  Instead, we rely
on the approximate but physically and mathematically consistent $y^+$ field
after the first time step. As $y^+$ depends on the turbulence
model, its values (in an absolute sense) are of limited value: the main goal of
the \emph{a priori} mesh assessment is the evaluation of the mesh quality with
respect to the mesh generators and the generated meshes.
\begin{figure*}[t]
\begin{center}
\setlength{\belowcaptionskip}{-0.7cm}
\subfloat[$y^+$ on a 7 million-cell Spider mesh.]{
\includegraphics[width=1.0\columnwidth]{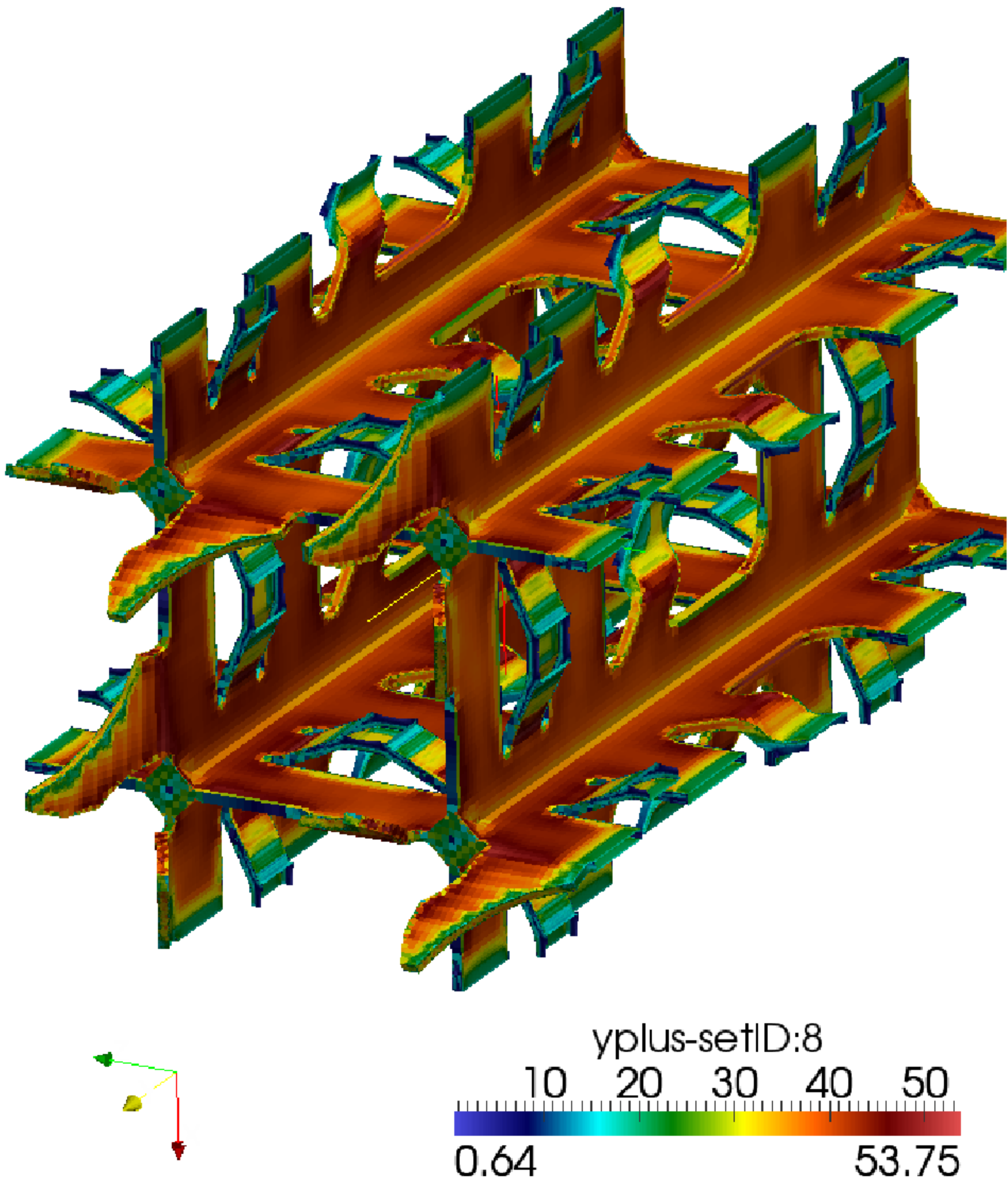}}
\subfloat[$y^+$ on a 8.3 million-cell Cubit mesh.]{
\includegraphics[width=1.0\columnwidth]{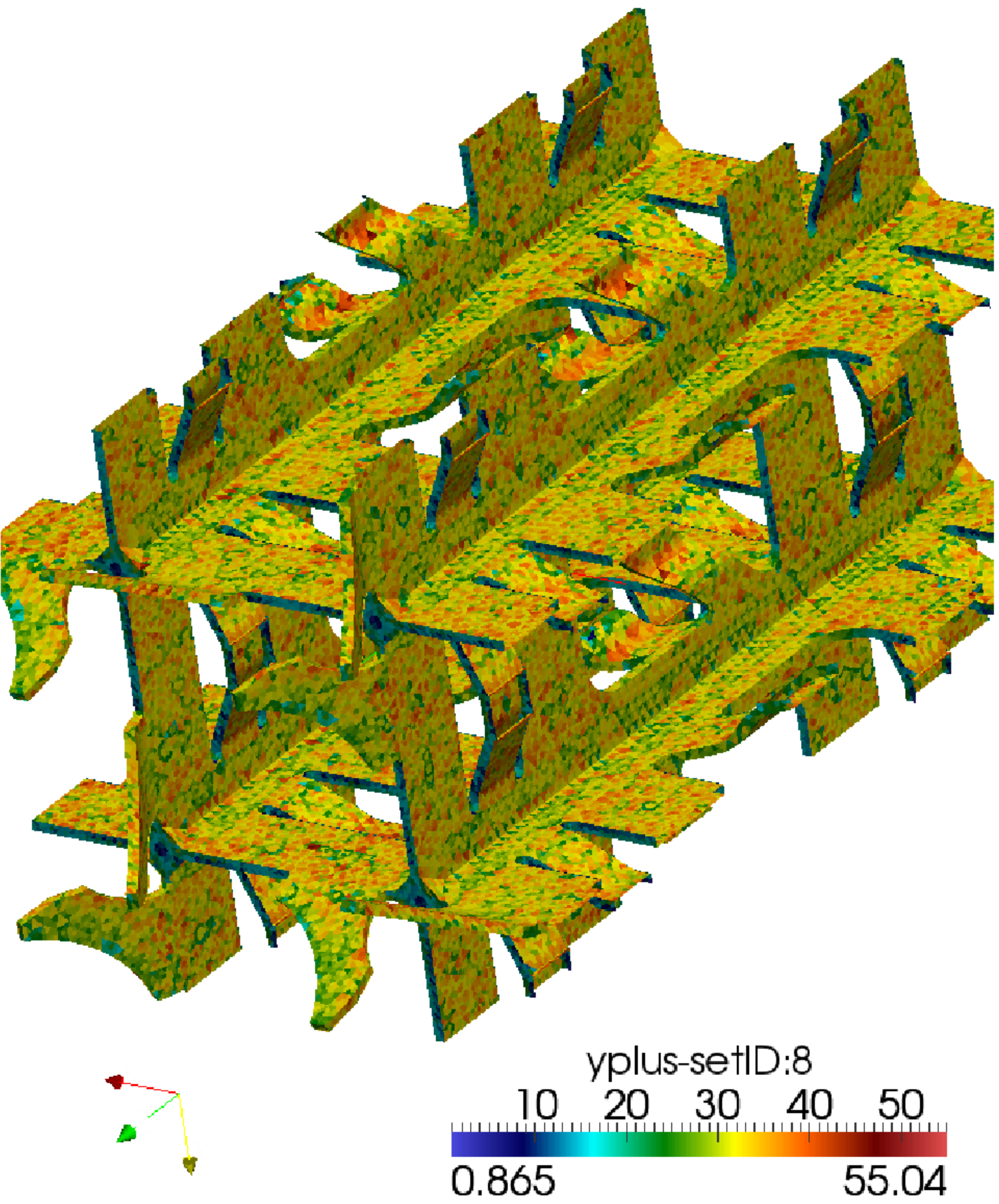}}
\caption{Spatial distribution of $y^+$ for two different meshes for the same
geometry and Reynolds number.}
\label{fig:yplus}
\end{center}
\end{figure*}

\subsection{Evaluation of the mesh quality}
\label{sec:evaluation}
Figure \ref{fig:yplus} shows the spatial $y^+$ distribution for the same
Reynolds number and geometry for two different meshes, generated by different
mesh generation technologies. The mesh in Figure \ref{fig:yplus}(a) is generated
by Spider, while the mesh in Figure \ref{fig:yplus}(b) is produced by Cubit.

While the smallest and largest $y^+$ values on these two meshes are comparable,
the fields are very different. The $y^+$ on the Spider mesh appears to be much
smoother and low $y^+$ values indicate highly refined edges and corners.
Compared to the Spider mesh, the Cubit mesh exhibits a much more
checkerboard-like pattern, indicating a larger spatial variation of $y^+$ on the
surface. A uniform $y^+$ distribution is desirable for a predictable simulation
quality. For example, sudden changes in $y^+$ (e.g., due to abrupt changes in
the cell sizes along walls) may perturb an otherwise smooth boundary layer,
resulting in artificial adverse pressure gradients and unphysical boundary-layer
separation. This is particularly important in wall-resolving LES, as the
simulations must adequately represent the highly inhomogeneous and anisotropic
nature of the turbulent flow in the vicinity of walls. In RANS simulations, more
uniform $y^+$ fields are reassuring in that the assumptions, such as zero mean
pressure gradient, used in the development of wall-functions, are satisfied.

\begin{figure}
\begin{center}
\setlength{\abovecaptionskip}{0pt}
\setlength{\belowcaptionskip}{-0.7cm}
\includegraphics[width=1.0\columnwidth]{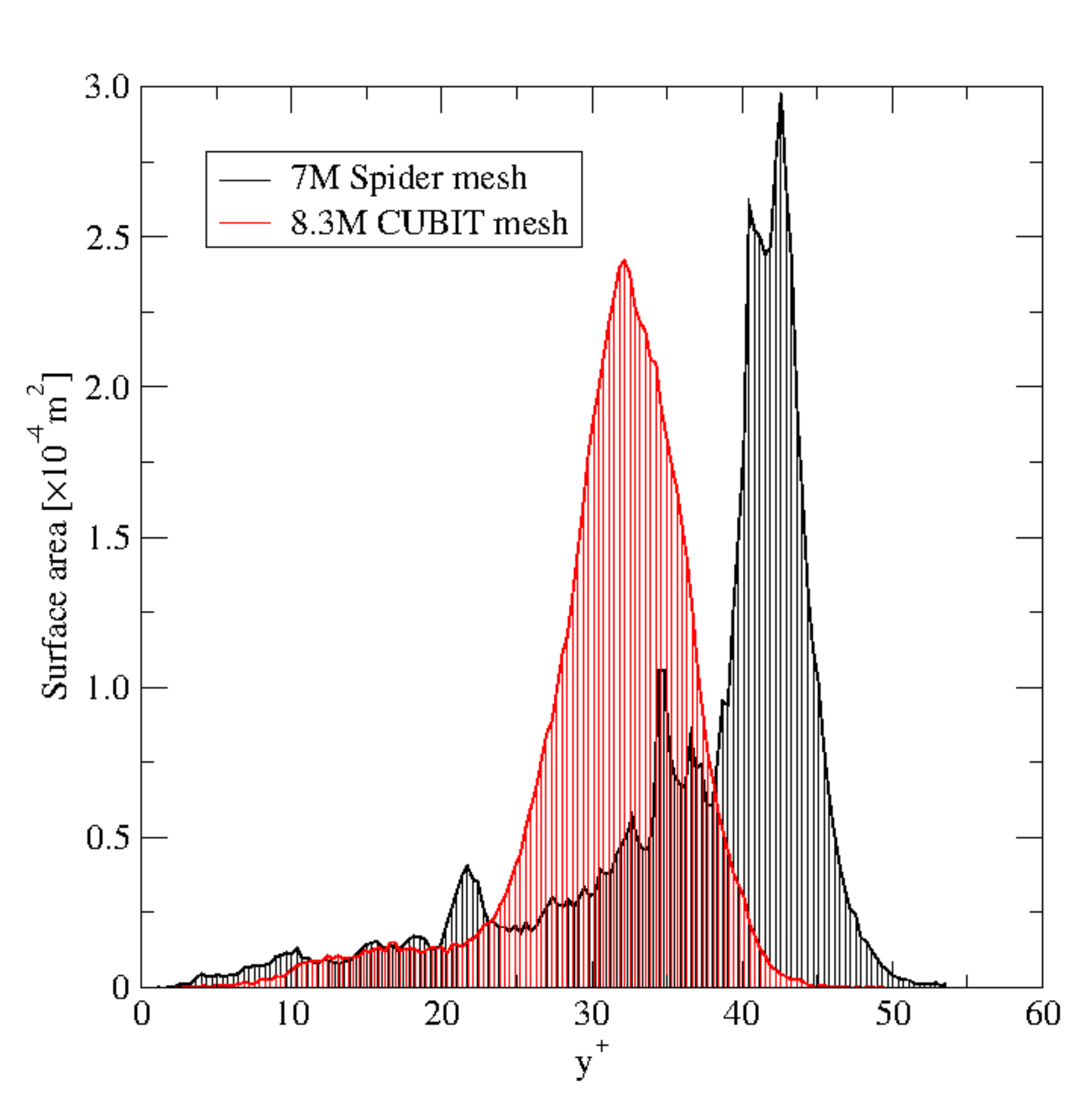}
\caption{Cell-area-weighted histograms of $y^+$ for the 7M Spider and 8.3M Cubit
meshes for the same geometry and Reynolds number. The histograms correspond to the meshes and spatial $y^+$ distributions in Figure \ref{fig:yplus}.}
\label{fig:yplus_distribution}
\end{center}
\end{figure}

\begin{figure}
\begin{center}
\setlength{\abovecaptionskip}{0pt}
\setlength{\belowcaptionskip}{-0.8cm}
\includegraphics[width=1.0\columnwidth]{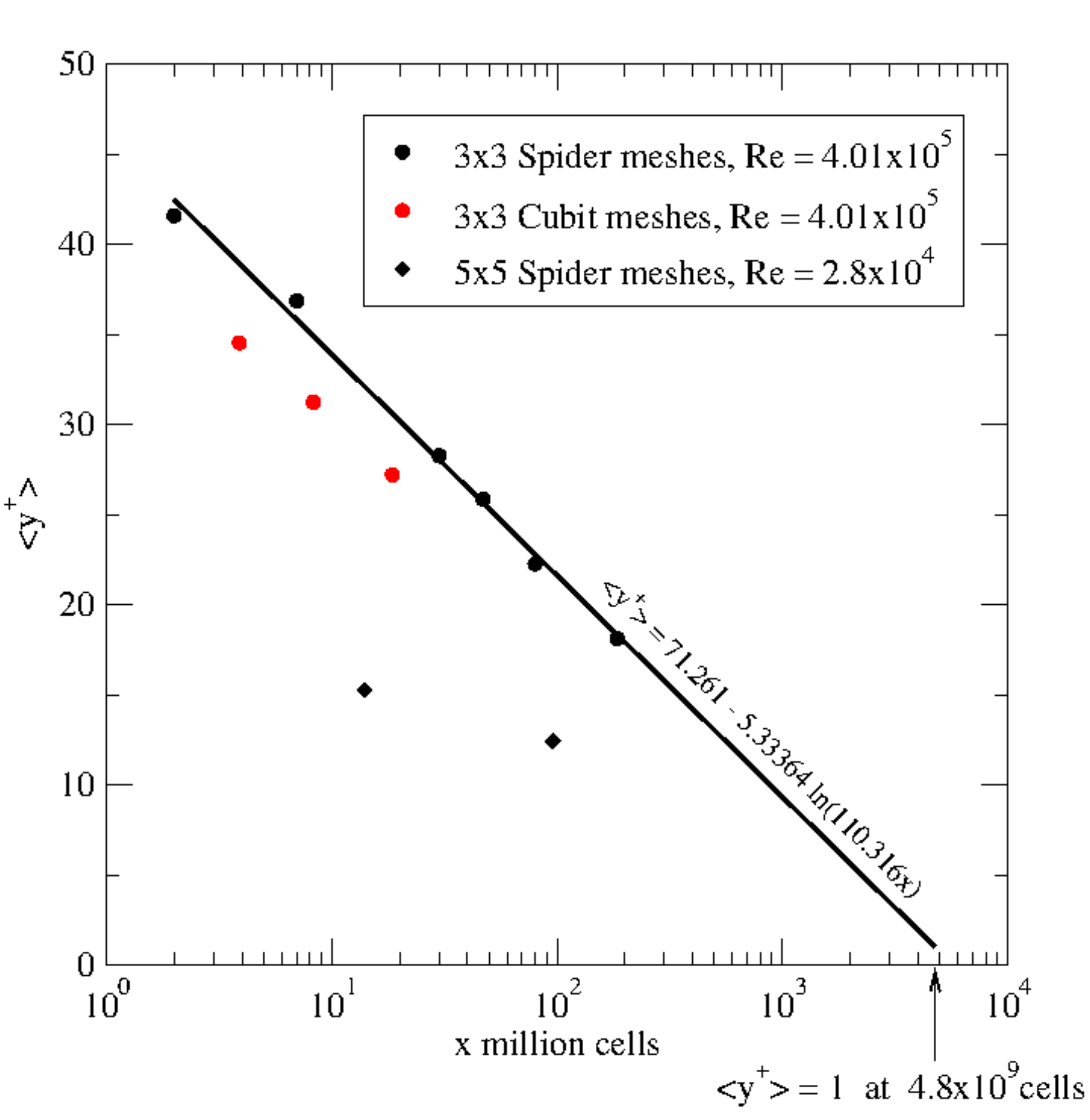}
\caption{$\irmean{y^+}$ vs.\ number of cells for all meshes generated for the
$3\times3$ and $5\times5$ configurations.}
\label{fig:mean_yplus}
\end{center}
\end{figure}

\begin{figure*}
\begin{center}
\setlength{\abovecaptionskip}{0.6cm}
\setlength{\belowcaptionskip}{-0.4cm}
\subfloat[$\mathrm{d}y^+$ on the 7M Spider mesh.]{
\includegraphics[width=1.0\columnwidth]{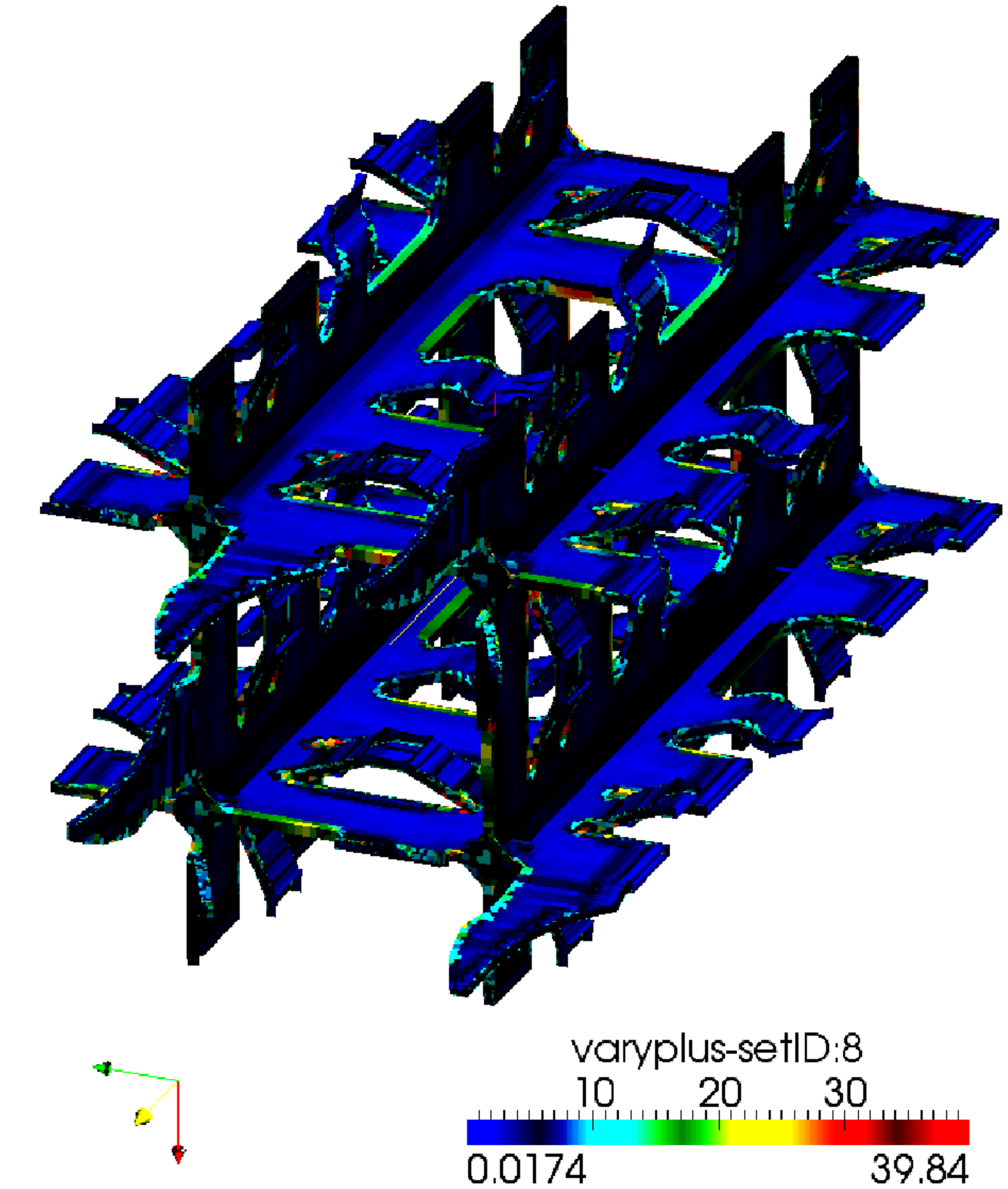}}
\subfloat[$\mathrm{d}y^+$ on the 8.3M Cubit mesh.]{
\includegraphics[width=1.0\columnwidth]{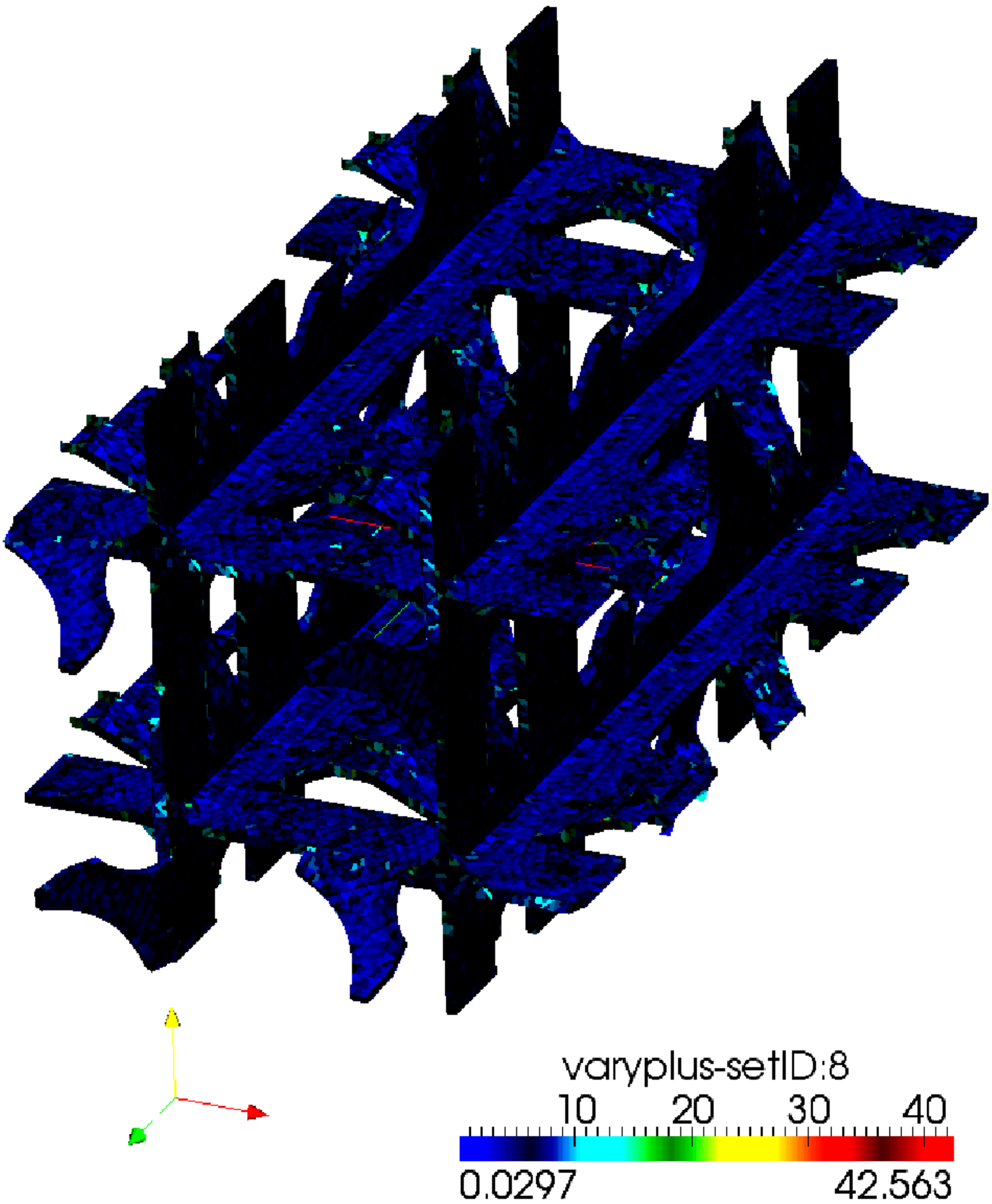}}
\caption{Spatial distribution of the variation of $y^+$, $\mathrm{d}y^+$, for
Spider and Cubit meshes for the same geometry and Reynolds number.}
\label{fig:dyplus}
\end{center}
\end{figure*}

\begin{figure*}
\begin{center}
\setlength{\abovecaptionskip}{0.5cm}
\setlength{\belowcaptionskip}{-0.5cm}
\subfloat[$\mathrm{d}y^+$ histograms for the 7M Spider and 8.3M Cubit meshes,
corresponding to the meshes and spatial $\mathrm{d}y^+$ distributions in Figure
\ref{fig:dyplus}.]{
\includegraphics[width=1.0\columnwidth]{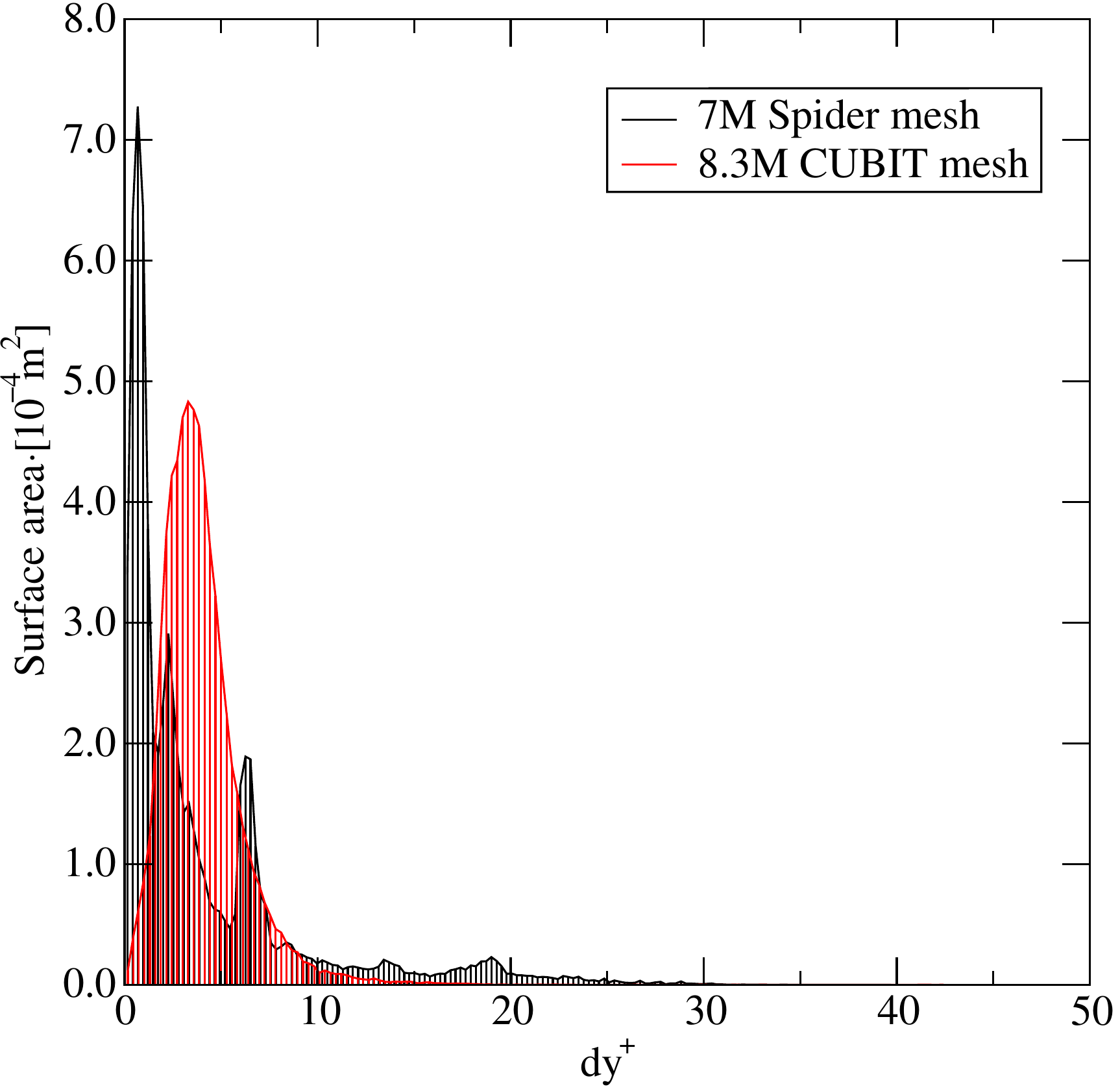}}
\subfloat[$\mathrm{d}y^+$ histogram for the 47M Spider mesh.]{
\includegraphics[width=1.0\columnwidth]{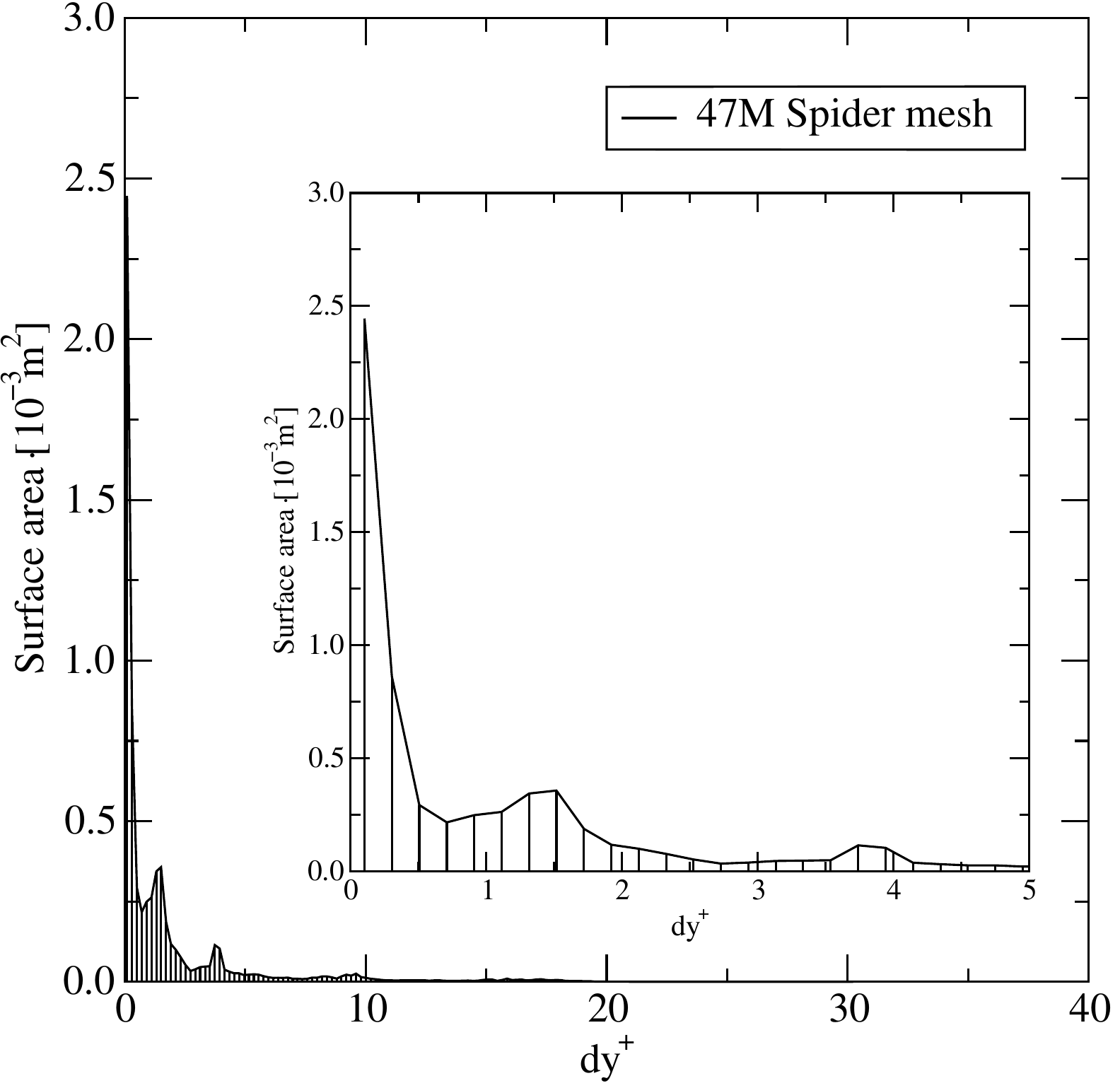}}
\caption{Cell-area-weighted histograms of the variation of $y^+$, for the spacer
meshes for the same geometry and Reynolds number. The closer the histogram to a
delta-peak at $\mathrm{d}y^+\!=\!0$, the more uniform the $y^+$ field is
distributed along the surface.}
\label{fig:dyplus_distribution}
\end{center}
\end{figure*}

\begin{table*}
\setlength{\abovecaptionskip}{0pt}
\setlength{\belowcaptionskip}{-0.1cm}
\begin{center}
\begin{tabular}{r|r|r|r||c|c|c|c}
Rod bundle & Mesh & Generator & No.\ of elements & $y^+_{min}$ & $y^+_{max}$ & $\irmean{y^+}$ & $\mathrm{TV}(y^+)/A$ \\
\hline
$3\times3$ &   2M &    Spider &   2.6M &        1.55 &       73.41 &          41.58 &                 0.38 \\
           &   7M &           &   7.8M &        0.82 &       53.57 &          36.85 &                 0.27 \\
           &  30M &           &   30.0M &        0.72 &       43.18 &          28.26 &                 0.21 \\
           &  47M &           &   46.8M &        0.70 &       40.92 &          25.85 &                 0.20 \\
           &  80M &           &   83.2M &        0.34 &       34.25 &          22.25 &                 0.15 \\
           & 185M &           &   185.4M &        0.35 &       29.87 &          18.10 &                 0.16 \\
\hline
$5\times5$ &  14M &    Spider &   14.2M &        0.29 &       22.04 &          15.24 &                 0.11 \\
           &  96M &           &   96.3M &        0.31 &       16.42 &          12.41 &                 0.09 \\
\hline
$3\times3$ & 3.9M &     Cubit &  3.9M  &        1.87 &       61.38 &          34.52 &                 0.40 \\
           & 8.3M &           &  8.3M  &        1.99 &       53.70 &          31.23 &                 0.35 \\
           &18.6M &           &  18.7M &        0.85 &       45.41 &          27.20 &                 0.30 \\
\hline
\end{tabular}
\end{center}
\caption{Statistics of computational meshes generated for the GTRF problem
on $3\times3$ and $5\times5$ rod-bundle geometries. The Spider meshes are all
hex-dominated hybrid meshes, containing hexahedra, pyramids, tetrahedra, and
wedges; the Cubit meshes contain pure hexahedra. The $y^+$ fields are computed
along the spacer surfaces, the most complex part of the geometry where
no-slip/no-penetration boundary conditions are imposed. The Reynolds number for
the $3\times3$ rod-bundle is $\textit{Re}_{D_h}=4.01\times10^5$, corresponding
to \cite{GTRF_2011} and \cite{elmahdi:2011}. The Reynolds number for the
$5\times5$ geometry is $\textit{Re}_{D_h}=28.0\times10^3$, corresponding to
\cite{yan:2012}. The extremes of the $y^+$ fields, $y^+_{min}$ and $y^+_{max}$,
denote the extents of the $y^+$ histograms, while $\irmean{y^+}$ and the (total
variation) $\mathrm{TV}(y^+)$ are the mean of the $y^+$ histograms and the
integral of the $\mathrm{d}y^+$ histograms, respectively. The total variation is
normalized by the total surface of the spacer, $A$.
\label{tbl:mesh_stats}}
\end{table*}

Quantitative \emph{a priori} metrics that could be used to assess non-uniform
$y^+$ fields on a complex surface, such as around the spacer and mixing vanes,
depicted in Figure \ref{fig:yplus}, may be defined based on the statistical
distributions of the $y^+$ field. Such a distribution may be generated by
counting up the $y^+$ values of the cells adjacent to a surface and grouping
them into equally-sized bins between their extremes. Instead of
counting each $y^+$ value as 1, we generate an area-weighted $y^+$ histogram by
counting values of the wall-attached cell area times the $y^+$ along the
surface. Such an area-weighted histogram is displayed in Figure
\ref{fig:yplus_distribution}, corresponding to the $y^+$ distributions of the
two meshes in Figure \ref{fig:yplus}.

The mean of the histograms in Figure \ref{fig:yplus_distribution} may be used
as a quantitative metric to assess the \emph{average} $y^+$ on a complex
surface with no-slip boundary conditions. The mean can be computed by
numerically estimating the integral
\begin{equation}
\irmean{y^+} \equiv  \frac{\int y^+ f(y^+)\mathrm{d}y^+}
                          {\int f(y^+)\mathrm{d}y^+}
             \approx \frac{\sum y^+ f(y^+) \Delta y^+}
                          {\sum f(y^+) \Delta y^+},
\end{equation}
where $f(y^+)$ denotes the function values of the area-weighted
histogram, such as in Figure \ref{fig:yplus_distribution}. The mean
$y^+$, $\irmean{y^+}$, computed for all the meshes generated is
shown in Table \ref{tbl:mesh_stats} and displayed in Figure
\ref{fig:mean_yplus}.

The spatial uniformity of the $y^+$ field is also of interest. To define a
useful metric that characterizes the uniformity, the spatial variation of $y^+$
is extracted based on the $y^+$ field. The variation of $y^+$ is computed by
visiting each surface cell and finding the maximum difference among the $y^+$ of
the given cell and that of its immediate neighbors. This yields a new scalar
field, which we call $\mathrm{d}y^+$, whose area-weighted histogram is computed
using the same method as that of $y^+$, discussed previously. The spatial
distribution of the $y^+$ variation fields for the 7M Spider and 8.3M Cubit
meshes are shown in Figure \ref{fig:dyplus}.  This confirms the earlier
observation that the Spider mesh is smoother, while the $y^+$ varies more
significantly in the Cubit mesh.  This is quantified in the histograms shown in
Figure \ref{fig:dyplus_distribution}. Compared to the $y^+$ histograms, the
$\mathrm{d}y^+$ histograms are only defined for $\mathrm{d}y^+\!\ge\!0$. The
theoretical ideal $\mathrm{d}y^+$ histogram is a Dirac delta function at
$\mathrm{d}y^+=0$. Larger $\mathrm{d}y^+$ values correspond to a larger surface
area covered by given $y^+$ variation. While the $\mathrm{d}y^+$ histogram for
the Spider mesh has a large peak close to 0, the peak for the Cubit mesh is
displaced to the right, indicating that there is significant variation in $y^+$
for most of the surface area. The differences in the uniformity of $y^+$ can be
explained by the differences in the mesh generation strategies. The Spider
meshes are generated by first inserting an initial layer of similar-size cells
closely following the complex boundary and using different cell-types without
truncating cells or geometry.  In Cubit, the mesh is first generated using
tetrahedra which then are dissected to yield an all-hex mesh, resulting in
non-uniform cell sizes along walls.  Figure \ref{fig:dyplus_distribution} shows
the $\mathrm{d}y^+$ histogram for a finer, 47M-cell, Spider mesh: the variations
in $y^+$ is significantly reduced compared to the coarser 7M mesh. To quantify
the uniformity with a single scalar, one can estimate the total variation of
$y^+$ by computing the total surface area under the $\mathrm{d}y^+$ histograms:
\begin{equation}
\mathrm{TV}(y^+) \equiv \int g(\mathrm{d}y^+)\mathrm{d}(\mathrm{d}y^+)
                 \approx\sum g(\Delta y^+)\Delta(\Delta y^+),
\end{equation}
where $g(\mathrm{d}y^+)$ denotes the function values of the area-weighted
$\mathrm{d}y^+$ histogram, such as in Figure \ref{fig:dyplus_distribution}.  The
total variation (TV) has also been computed for all meshes generated and
displayed in Table \ref{tbl:mesh_stats} for comparison. The following
observations can be made based on the data in Table
\ref{tbl:mesh_stats}:
\begin{itemize*}
\item The mean $y^+$ monotonically decreases with increasing cell count,
      signaling an overall uniform increase of refinement with larger meshes;
      a prerequisite for meaningful mesh convergence studies and uncertainty
      quantification.
\item The total variation of $y^+$ also monotonically decreases with increasing
      cell-count, i.e., relatively larger surface area is covered by more
      uniform-size wall-cells, an assurance of increasing mesh quality at walls,
      which minimizes unphysical perturbations in the wall-treatment, important
      for both LES and RANS simulations with wall-functions.
\item Based on the $\mathrm{d}y^+$ histograms and the total variation of $y^+$,
      the quality of Spider meshes are clearly superior to those produced using
      Cubit.
\end{itemize*}
Figure \ref{fig:mean_yplus} shows a graphical representation of the
$\irmean{y^+}$ column in Table \ref{tbl:mesh_stats}. The
$\irmean{y^+}$ is plotted for both series of Spider meshes for the
$3\times3$ and $5\times5$ geometries, as well as for the series of
Cubit meshes. While the Reynolds number is the same for the 
Cubit and Spider series of meshes for the $3\times3$ geometry, the
$\irmean{y^+}$ for the $5\times5$ meshes are computed for a lower
Reynolds number. This is the main reason for a significantly lower
$\irmean{y^+}$ for the $5\times5$ meshes for similar total cell counts
when compared to the $3\times3$\ meshes. Both the Spider and the Cubit meshes
exhibit a monotonic decrease in the mean $y^+$ with increasing cell
count. The trends are all logarithmic. This is expected as none of
these meshes have power-law-graded boundary-layer refinement at walls.
The logarithmic fit extrapolates the trend for the $3\times3$ Spider
meshes and predicts that to achieve $\irmean{y^+}\sim1$ with this
meshing strategy would require approximately 5 billion cells. This is
clearly not practical, therefore the next step in mesh generation for
wall-resolving LES is to add power-law-graded boundary layer
refinement.

In summary, a method for quantitative assessment of unstructured meshes with
no-slip walls has been described. The two metrics, $\irmean{y^+}$ and
$\mathrm{TV}(y^+)$, have been used to assess mesh quality generated by two mesh
generators for two rod-bundle geometries.

\section{LES Calculations and Analysis\label{sec:calculations}}
This section discusses the LES simulations using the Spider meshes for the
$3\times3$ (\S \ref{sec:3x3_spider}) and $5\times5$ (\S \ref{sec:5x5_spider})
rod-bundle geometries, respectively. Our earlier study, \cite{GTRF_2011},
provides details on calculations using LES, detached-eddy, and the
Spalart-Allmaras (URANS) turbulence models using the Cubit meshes, where it
was determined that the most accurate GTRF forces are obtained with
LES.

The flow geometry and the computational setup closely resembles
that of Elmahdi, et al.\ \cite{elmahdi:2011}, in which LES is documented using
Star-CCM+ with the wall-adapted large-eddy unresolved-scale model of
\cite{nicoud:1999}. In \cite{elmahdi:2011} a central difference scheme,
available in Star-CCM+, has been used with a ``wall-blending factor''.  In
comparison, the Hydra-TH calculations, discussed below, do not use any
particular treatment at no-slip, no-penetration boundaries. While Elmahdi, et
al.\ use a maximum $CFL$ of 1.0, the Hydra-TH simulations employ $CFL=4.0$.

\subsection{LES on the $3\times3$ Spider meshes}
\label{sec:3x3_spider}
\begin{figure*}
\begin{center}
\subfloat[Instantaneous velocity streamlines colored by helicity for the 2M
          Spider mesh.]{
\includegraphics[width=0.9\textwidth]{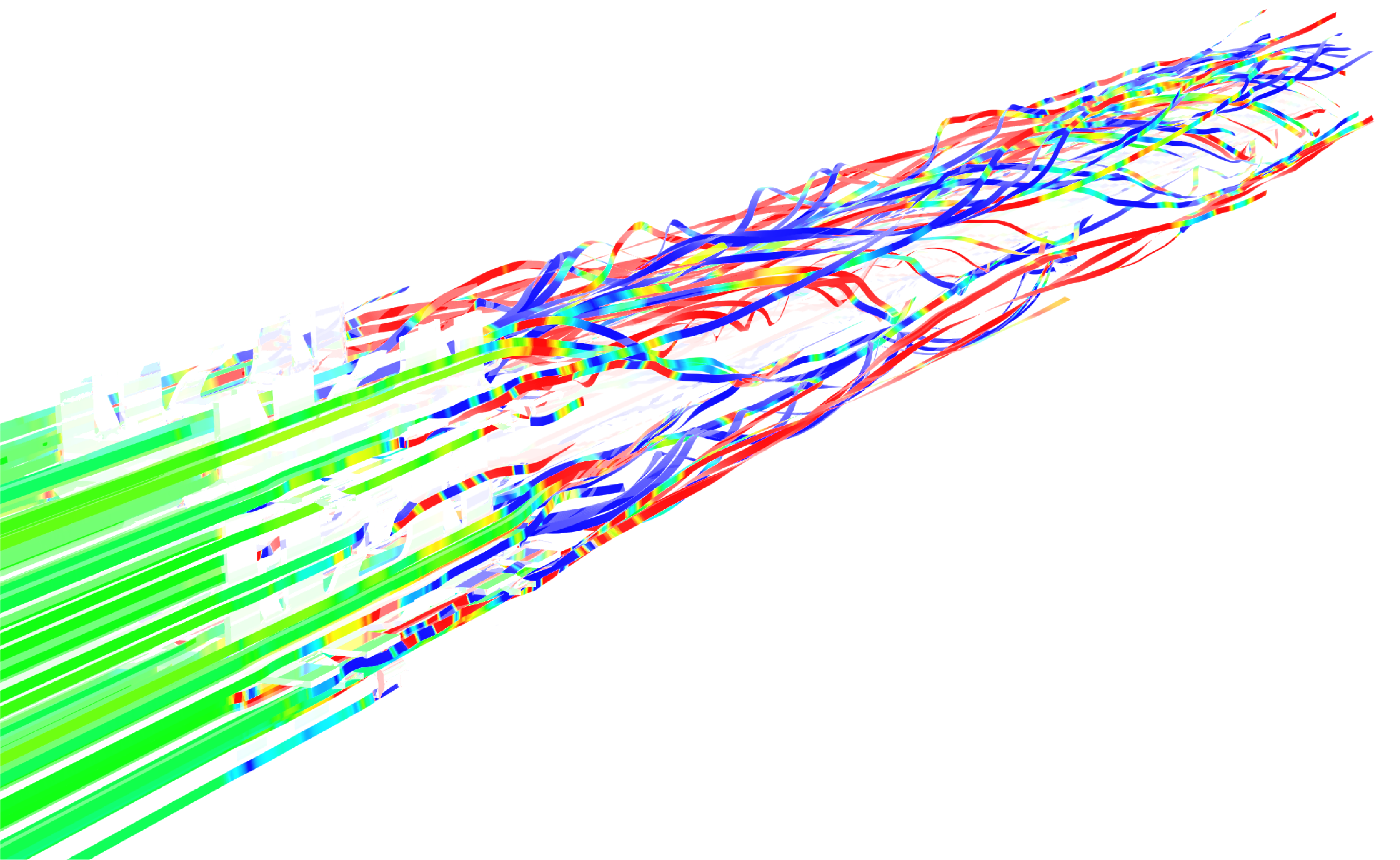}} \\
\subfloat[Instantaneous helicity isosurfaces for the 47M Spider mesh.]{
\includegraphics[width=0.9\textwidth]{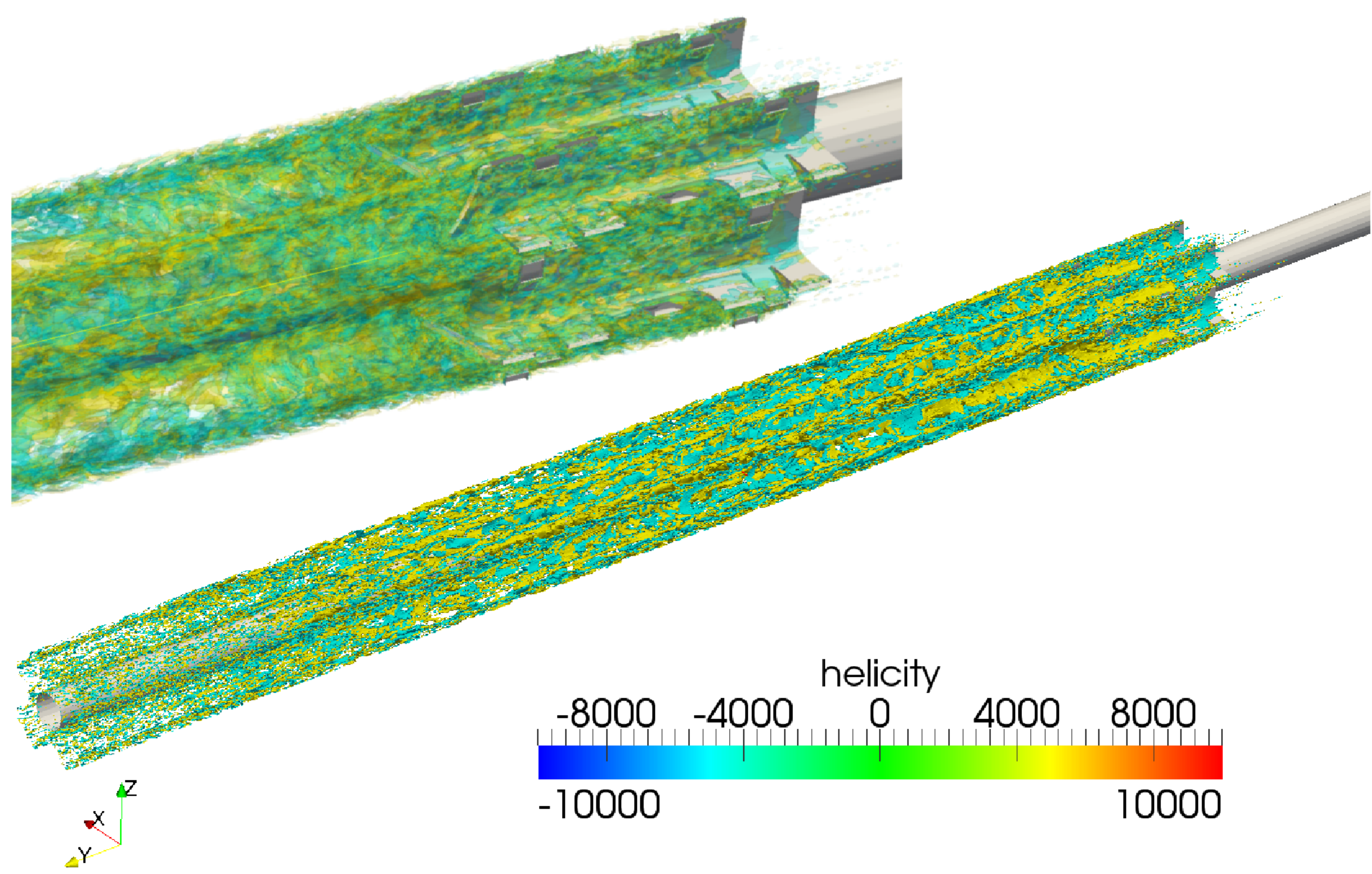}}
\caption{Instantaneous snapshots of velocity ($\bv{v}$) and helicity
($\bv{v}\cdot\omega$) isosurfaces from using the 2M and 47M Spider meshes.}
\label{fig:spider_3x3_helicity}
\end{center}
\end{figure*}

\begin{figure}
\begin{center}
\setlength{\abovecaptionskip}{0pt}
\setlength{\belowcaptionskip}{-0.7cm}
\includegraphics[width=1.0\columnwidth]{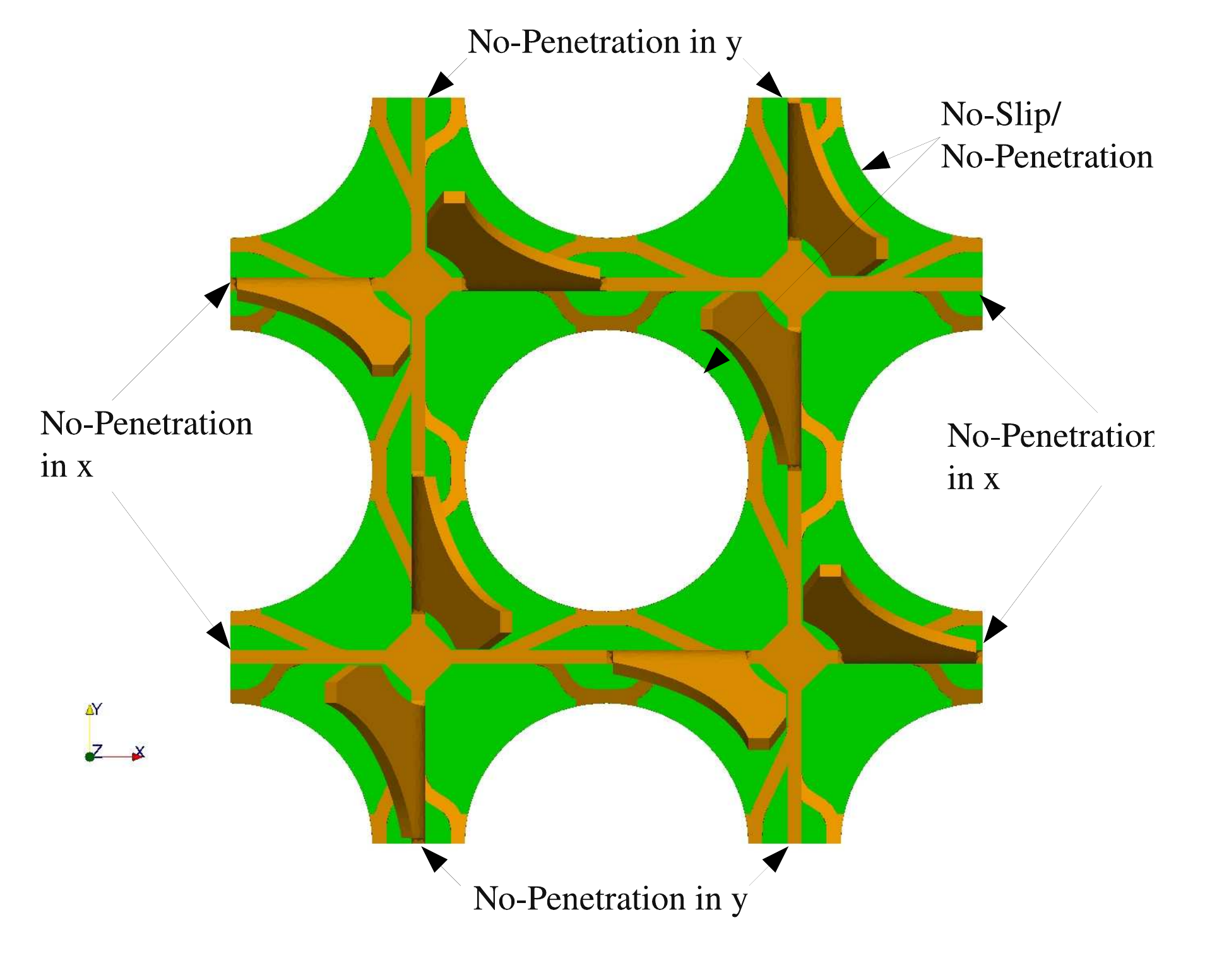}
\end{center}
\caption{Boundary conditions on rod and spacer surfaces, and
  subchannel boundaries.\label{fig:bcs}}
\end{figure}

For the $3\times3$ calculations, the working fluid is water at a
temperature of $394.2 \ K$, a density of $942.0 \ kg/m^3$, and a dynamic
viscosity of $2.32 \times 10^{-4} \ kg/m/s$. The inlet velocity is prescribed as
$\vv = (0, 0, 5) \ m/s$. This corresponds to a Reynolds number, based on the rod
diameter, of $Re_D = 1.93 \times 10^5$, while the Reynolds number based on the
hydraulic diameter is $Re_{D_h} = 4.01 \times 10^5$. The hydraulic diameter is
defined as $D_h =4 A_{flow}/P_{wet}$. The inlet velocity is constant. While this
is certainly not a good approximation for modeling several grid-spans, a main
goal of the current calculations is to facilitate a direct comparison to the data
of Elmahdi, et al.\ \cite{elmahdi:2011}, which prescribes the same inlet
conditions. No-slip, no-penetration conditions are prescribed at the rod and
spacer surfaces. At the outlet, the hydrostatic pressure is specified to be $p_h
= 0.0$ in conjunction with a zero shear stress condition. No-penetration
conditions with in-plane slip were applied at the subchannel boundaries as shown
in Figure \ref{fig:bcs}. As only one grid-span of a $3\times3$ rod-bundle of the
full reactor core is modeled, no attempt is made here to prescribe realistic PWR
operating conditions. Our goal is to reproduce the simulation conditions in
\cite{elmahdi:2011}.

A qualitative picture of the instantaneous flow behind the mixing vanes is
obtained by depicting the velocity streamlines with ribbons and isosurfaces of
the instantaneous helicity field in Figure \ref{fig:spider_3x3_helicity}. The
stream ribbons are for the coarsest 2M simulations, while the isosurfaces are
for the 47M case. The vortices generated by the spacer and the mixing vanes are
advected downstream. Figure \ref{fig:spider_3x3_helicity}(b) shows that the
neutrally dissipative advection algorithm in Hydra-TH does an excellent job in
maintaining the complex vortex structures far downstream.

Similar to our earlier LES calculations on the $3\times3$ rod-bundle
\cite{GTRF_2011}, a series of preliminary coarse-mesh simulations were conducted
using the Spider meshes to determine when a statistically stationary flow is
achieved. The time-evolution of the domain-integrated kinetic energy (not shown)
was used as an indicator. Based on the kinetic energy the time of approximately
$0.1\mathrm{s}$, which corresponds to approximately 1.25 flow
transits by which the domain-integrated kinetic energy has reached a statistically
stationary state, was chosen as the starting point for collecting
time-averaged flow statistics until the end of the simulation at
$t=1.0\mathrm{s}$.

\begin{figure*}[t]
\begin{center}
\setlength{\abovecaptionskip}{0pt}
\setlength{\belowcaptionskip}{0pt}
\includegraphics[width=0.49\textwidth]{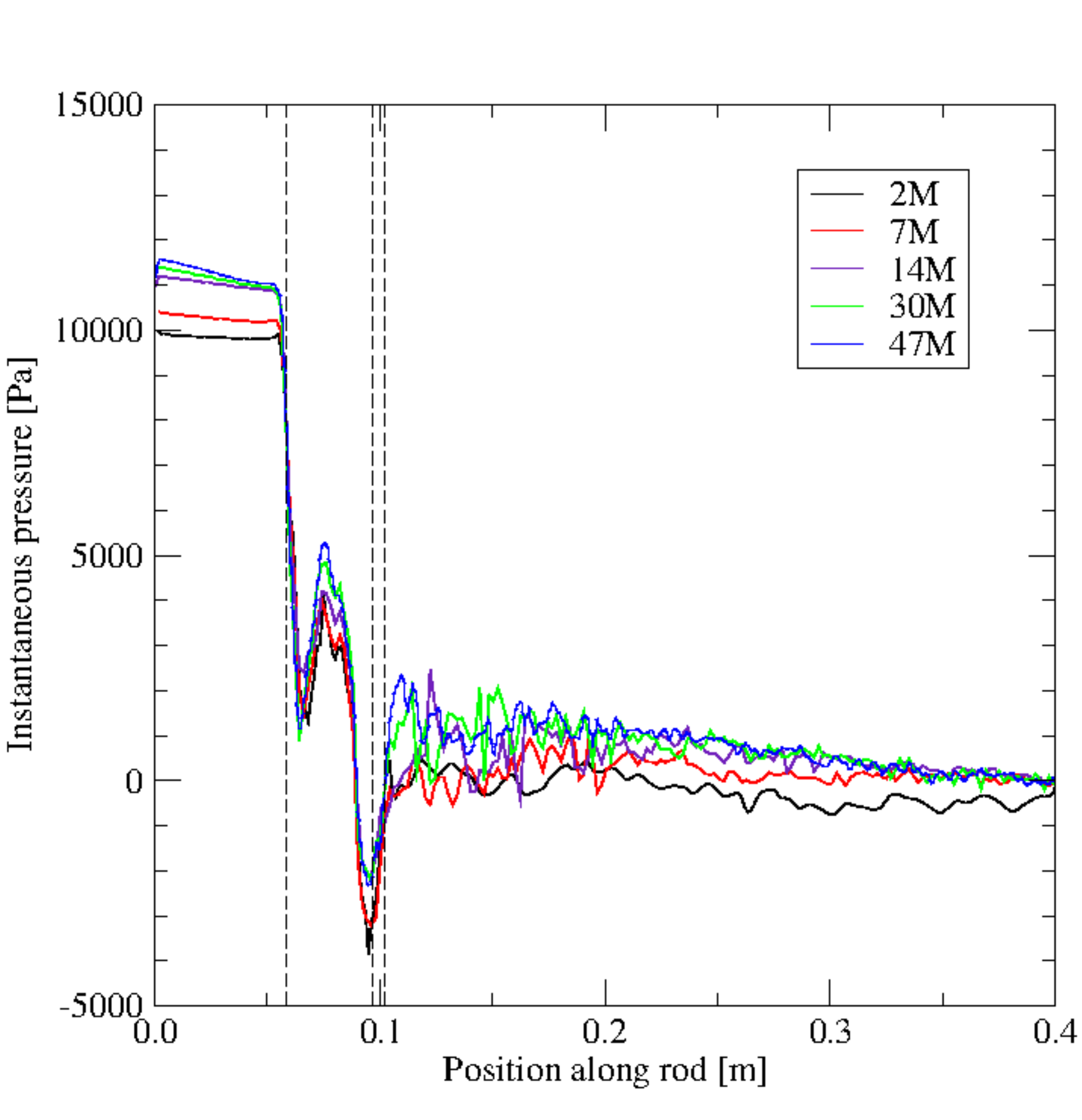}
\includegraphics[width=0.49\textwidth]{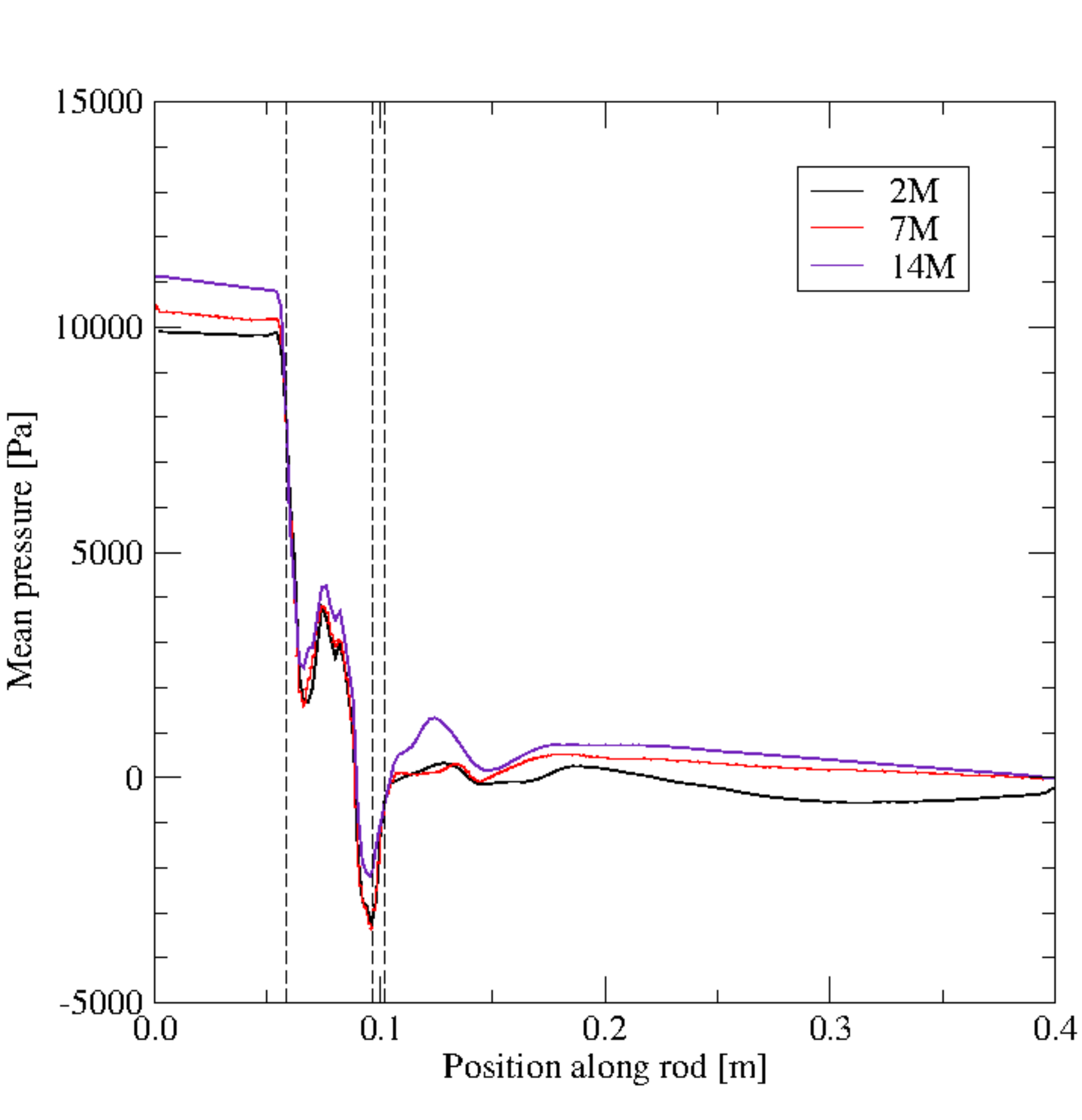}
\vspace{-0.18cm}
\end{center}
\caption{Instantaneous (left) and mean (right) pressure line plots for
         different meshes.
\label{fig:spider_pressure}}
\end{figure*}

LES calculations, using the 2M, 7M, 30M, 47M Spider
meshes, discussed earlier, and additional ones with 14M and 27M
meshes, have been carried out. The instantaneous
pressure is plotted in Figure \ref{fig:spider_pressure} for five
different meshes. The pressure line plots have been
extracted using a line along the rod, offset from the surface by
$6.3698 \times 10^{-6} \  D$ where $D$ is the rod diameter, and 
extending for the full length of the rod.
The vertical lines in Figure \ref{fig:spider_pressure}
delineate the bounds of the spacer and the mixing vanes. It is
reassuring that the pressure lines are qualitatively very similar for
all mesh resolutions.  Since the hydrostatic pressure at the outflow
is fixed at $p=0$, the value of the calculated inlet pressure
determines the pressure drop over the whole domain. The pressure drop
for the 47M Spider mesh is 11.425 kPa, however, this value does not
seem to have converged for this series of calculations.  In contrast
to the pressure observed with the Cubit meshes \cite{GTRF_2011}, the
pressure drop is monotonically increasing with mesh resolution --
largely a consequence of the better boundary layer resolution compared to the
Cubit meshes.

The mean pressure along the rod is also plotted in Figure
\ref{fig:spider_pressure} and indicates that the bulk of the 
of the pressure loss is due to the spacer. In spite of the
turbulent flow induced by the spacer, the characteristic peaks and troughs in
the profile of the mean pressure is very much reproducible throughout the spacer
using the 2M, 7M, and 14M meshes. Downstream of the mixing vanes a slight wave
in the mean pressure is apparent in the coarsest 2M-mesh simulation. The mean
pressure using the 7M mesh appears as what one would intuitively expect for a
turbulent pipe flow: from approximately $y=0.175m$, the 
mean pressure decreases linearly.

\begin{figure*}[t]
\begin{center}
\setlength{\abovecaptionskip}{0pt}
\setlength{\belowcaptionskip}{0pt}
\subfloat[RMS pressure integrated over the full length of the central rod for
three different meshes.]{
\includegraphics[width=0.48\textwidth]{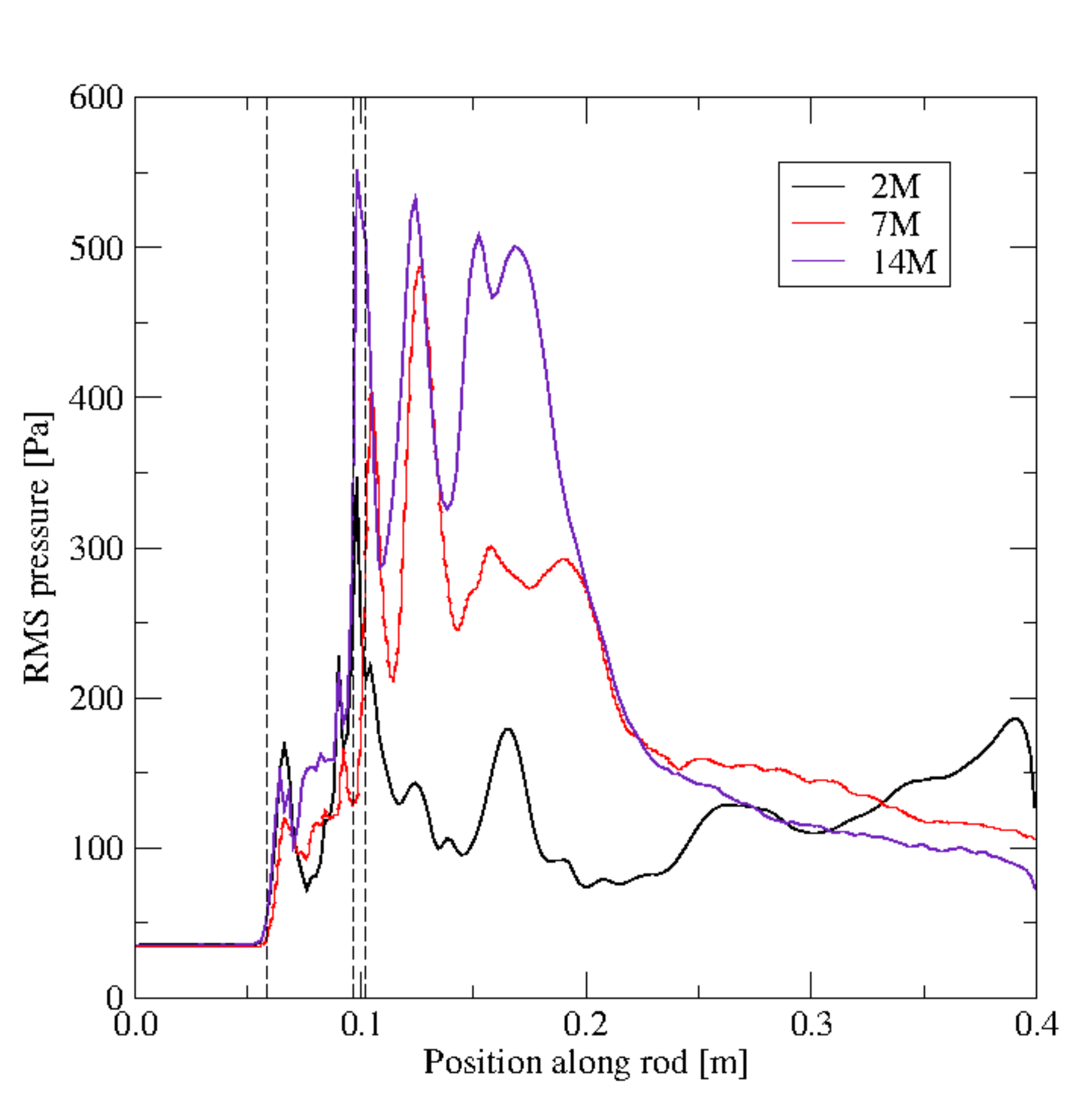}}
\hspace{0.15cm}
\subfloat[RMS total force on the central rod integrated in 1-inch segments
downstream of the mixing vanes. The Star-CCM+ results are from the LES
calculations in \cite{elmahdi:2011}.]{
\includegraphics[width=0.48\textwidth]{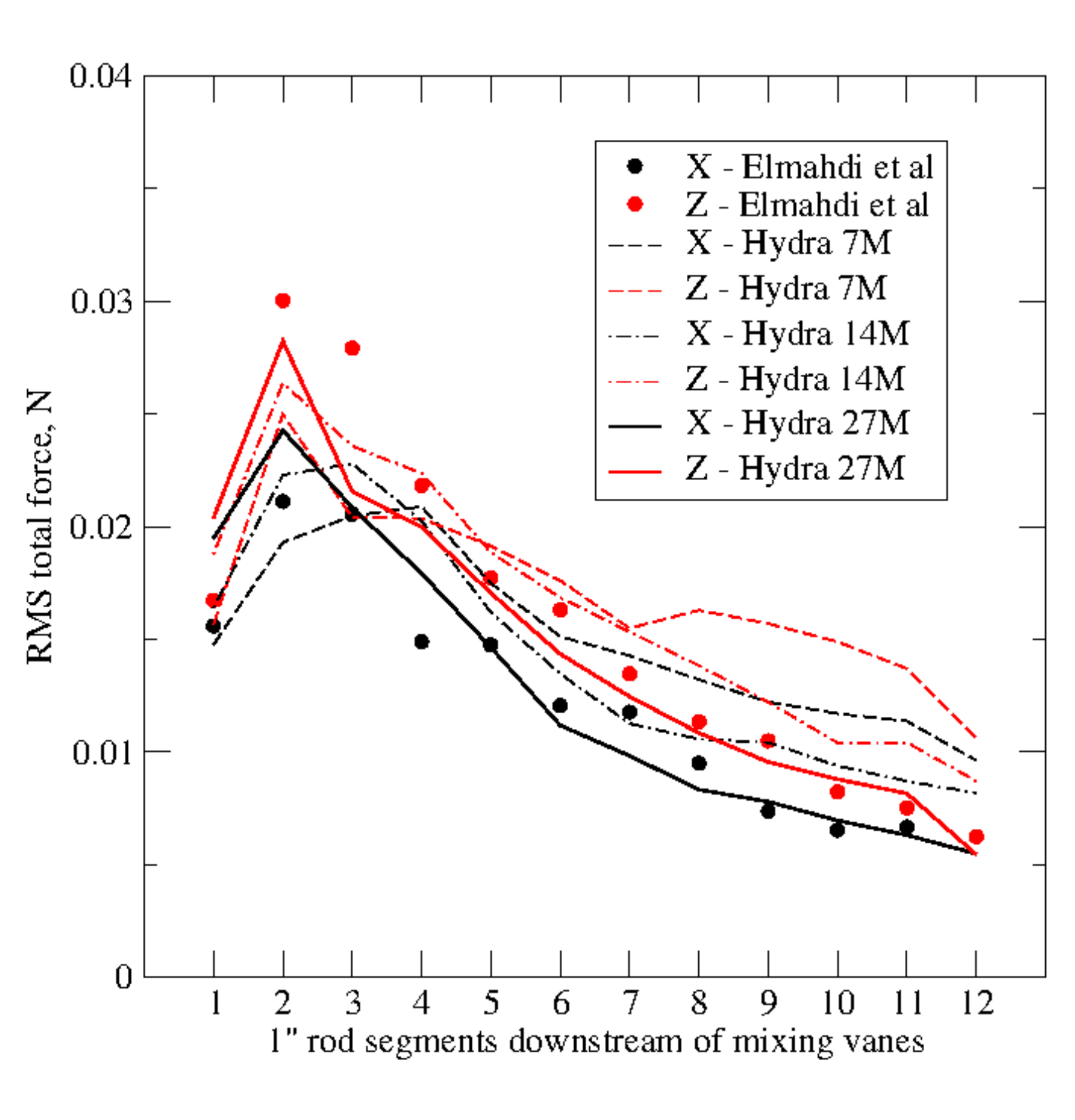}}
\end{center}
\caption{Second moments of the pressure (integrated for the full length) and
         the total force (dominated by the pressure force) in segments.}
\label{fig:spider_rmspress}
\end{figure*}

\begin{figure*}[t]
\begin{center}
\setlength{\abovecaptionskip}{0pt}
\setlength{\belowcaptionskip}{0pt}
\subfloat[Turbulent kinetic energy along the rod for three different meshes.]{
\includegraphics[width=0.48\textwidth]{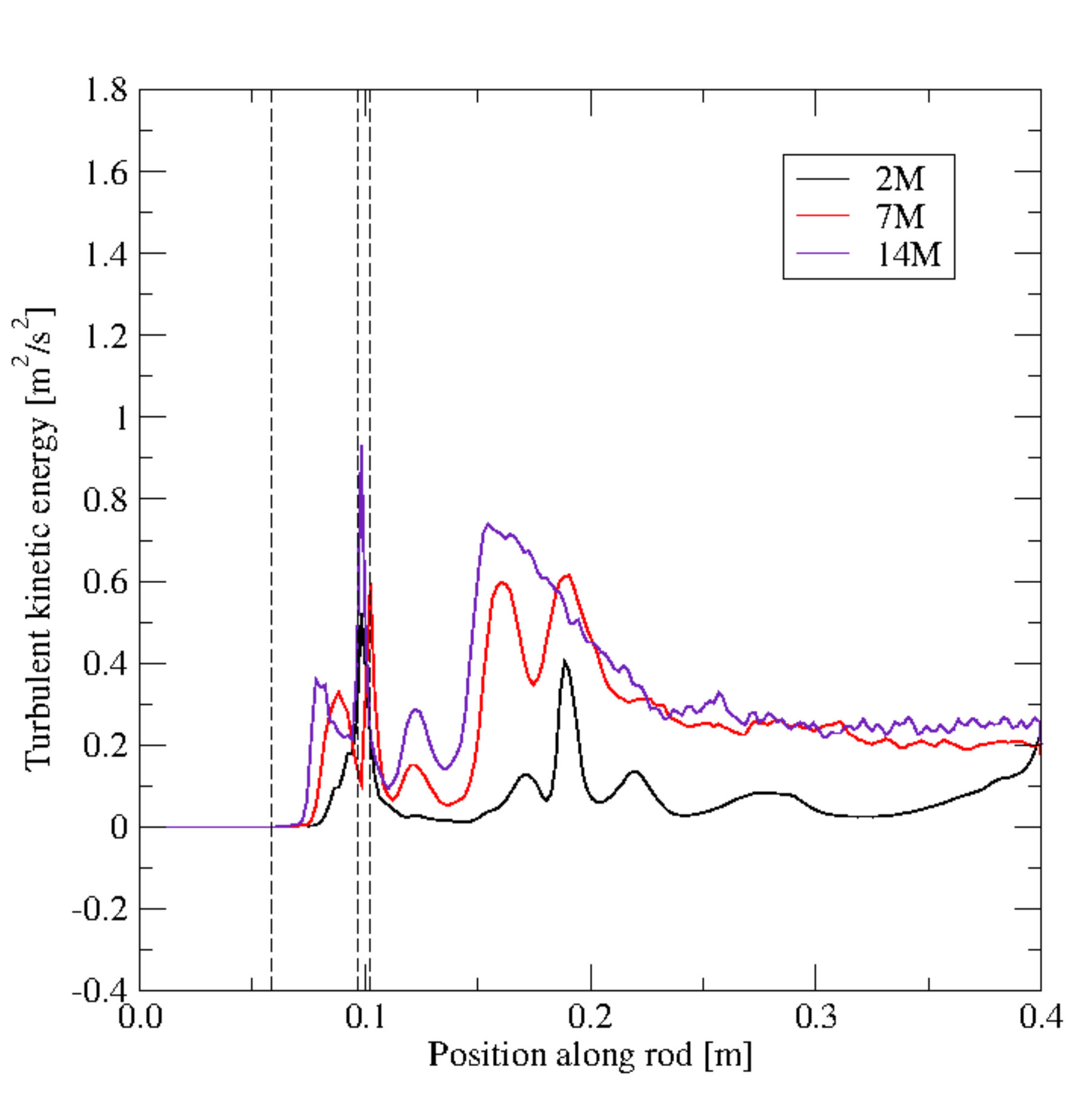}}
\hspace{0.15cm}
\subfloat[Reynolds stress along the rod for the 14M mesh.]{
\includegraphics[width=0.48\textwidth]{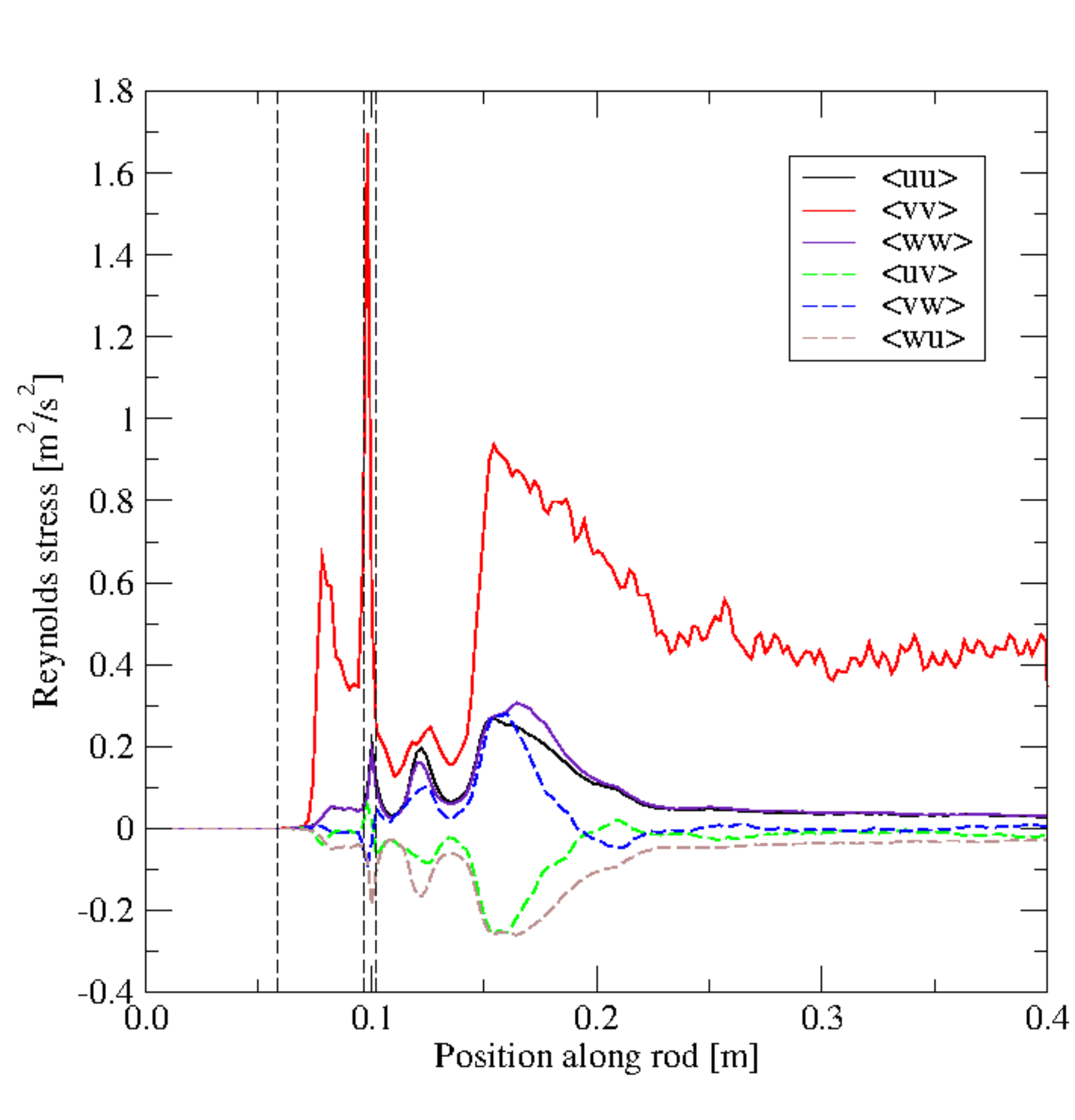}}
\end{center}
\caption{Second moments of the fluctuating velocity field for three different meshes.}
\label{fig:spider_reynoldsstress}
\end{figure*}

\begin{figure*}
\begin{center}
\setlength{\abovecaptionskip}{0pt}
\setlength{\belowcaptionskip}{0pt}
\subfloat[Total force on the central rod.]{
\includegraphics[height=2.4in]{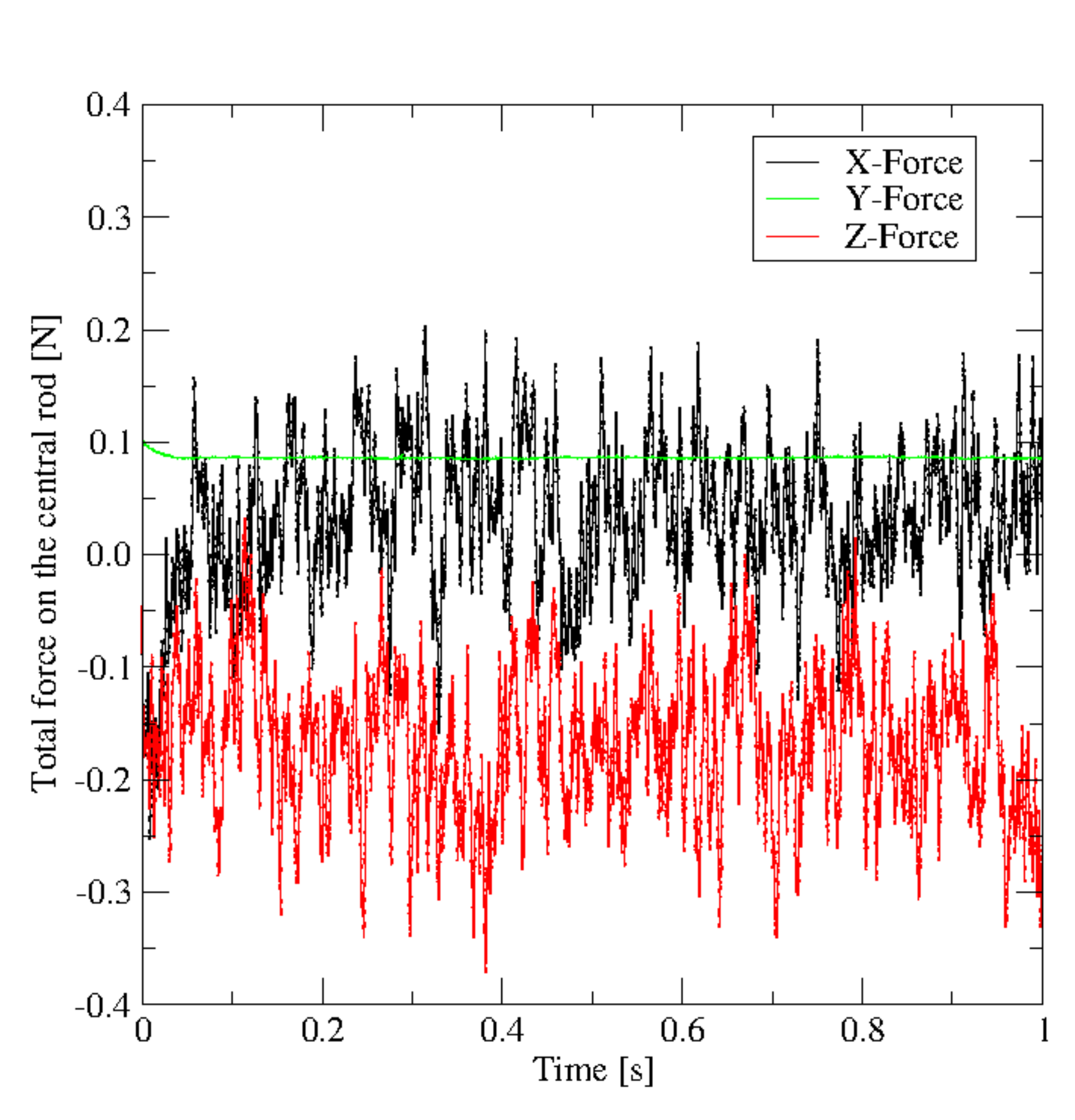}} 
\subfloat[Total force on the spacer.]{ 
\includegraphics[height=2.4in]{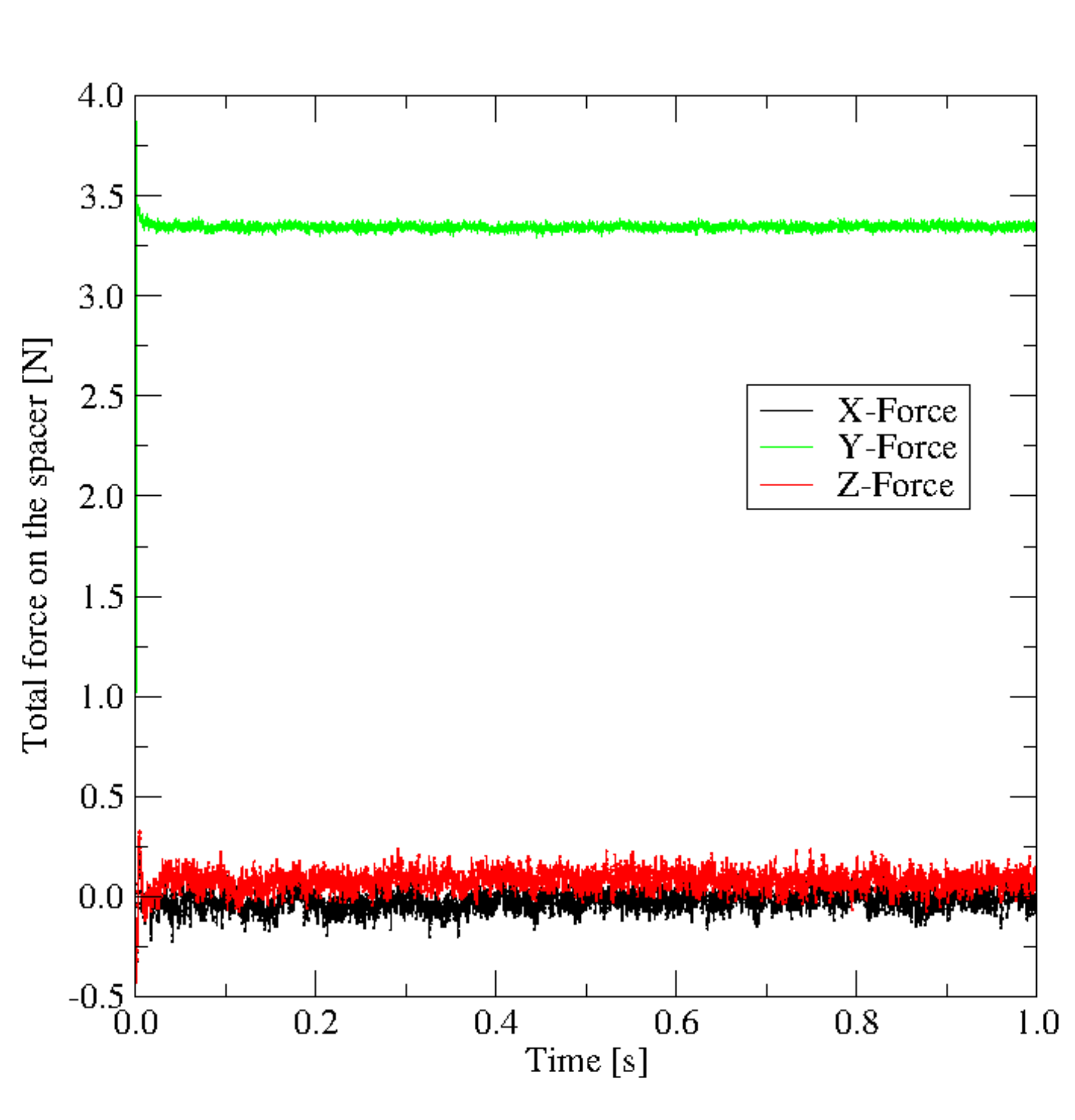}} \\
\subfloat[Pressure force on the central rod.]{
\includegraphics[height=2.4in]{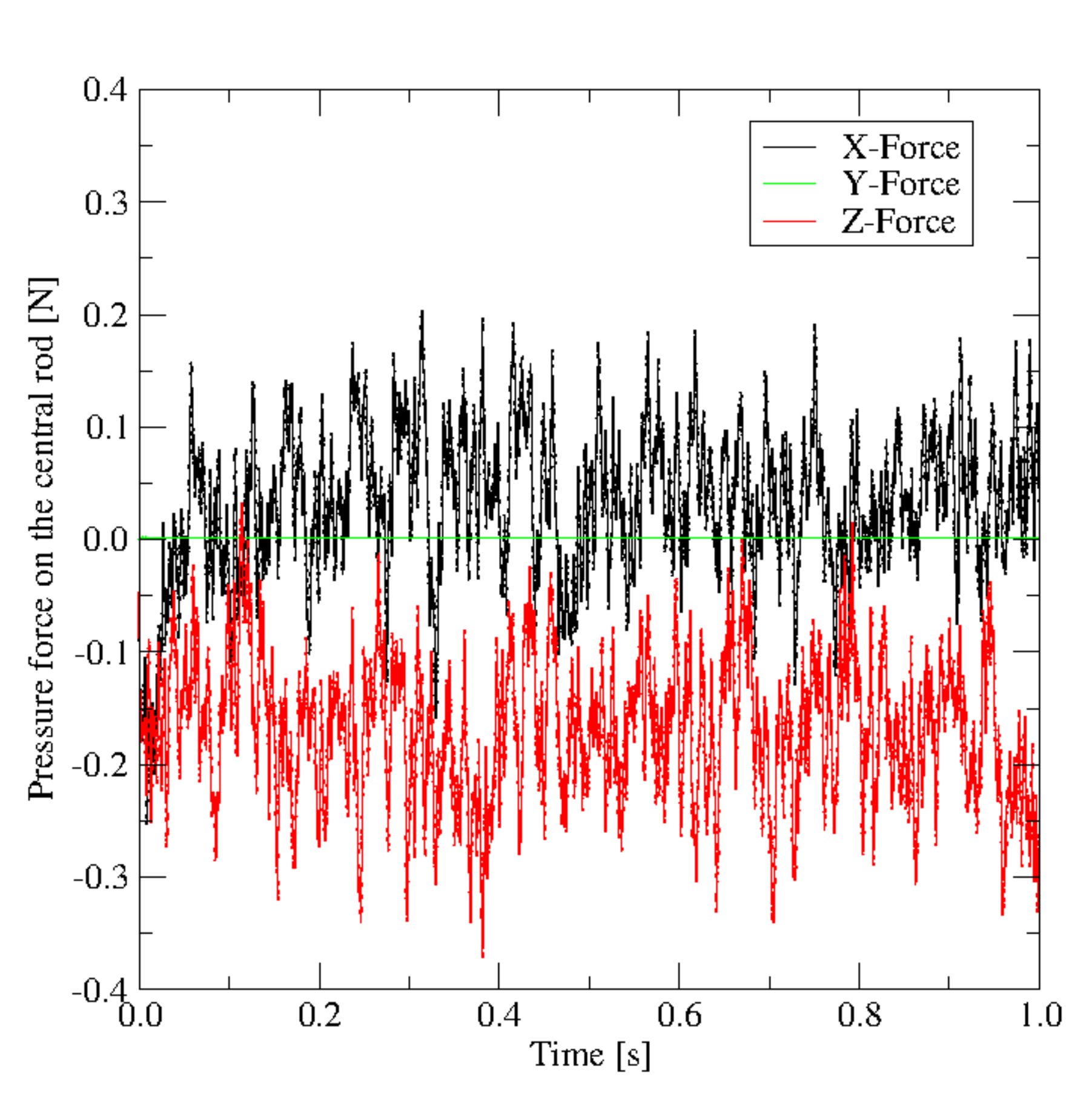}} 
\subfloat[Pressure force on the spacer.]{
\includegraphics[height=2.4in]{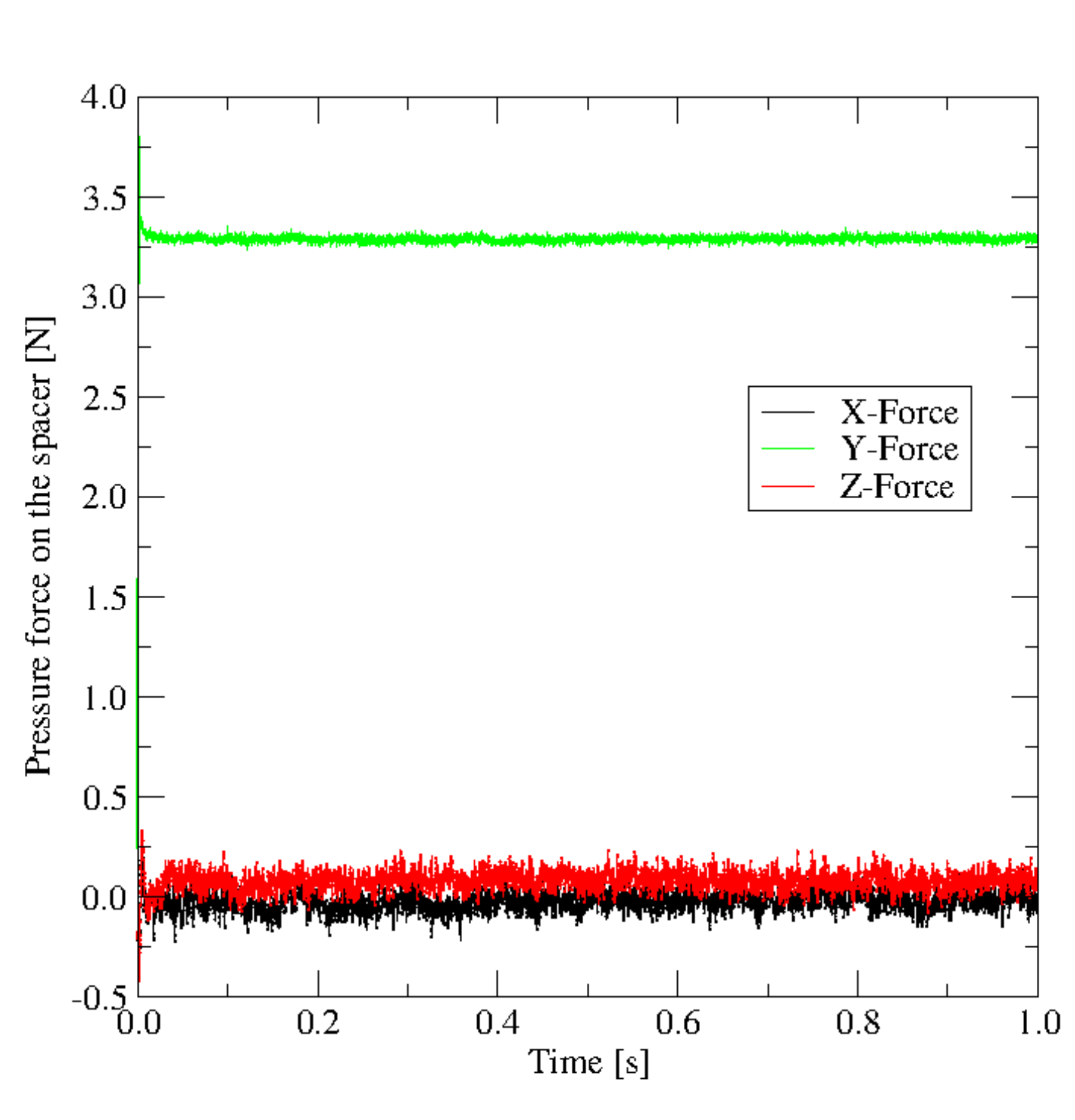}} \\
\subfloat[Shear force on the central rod.]{
\includegraphics[height=2.4in]{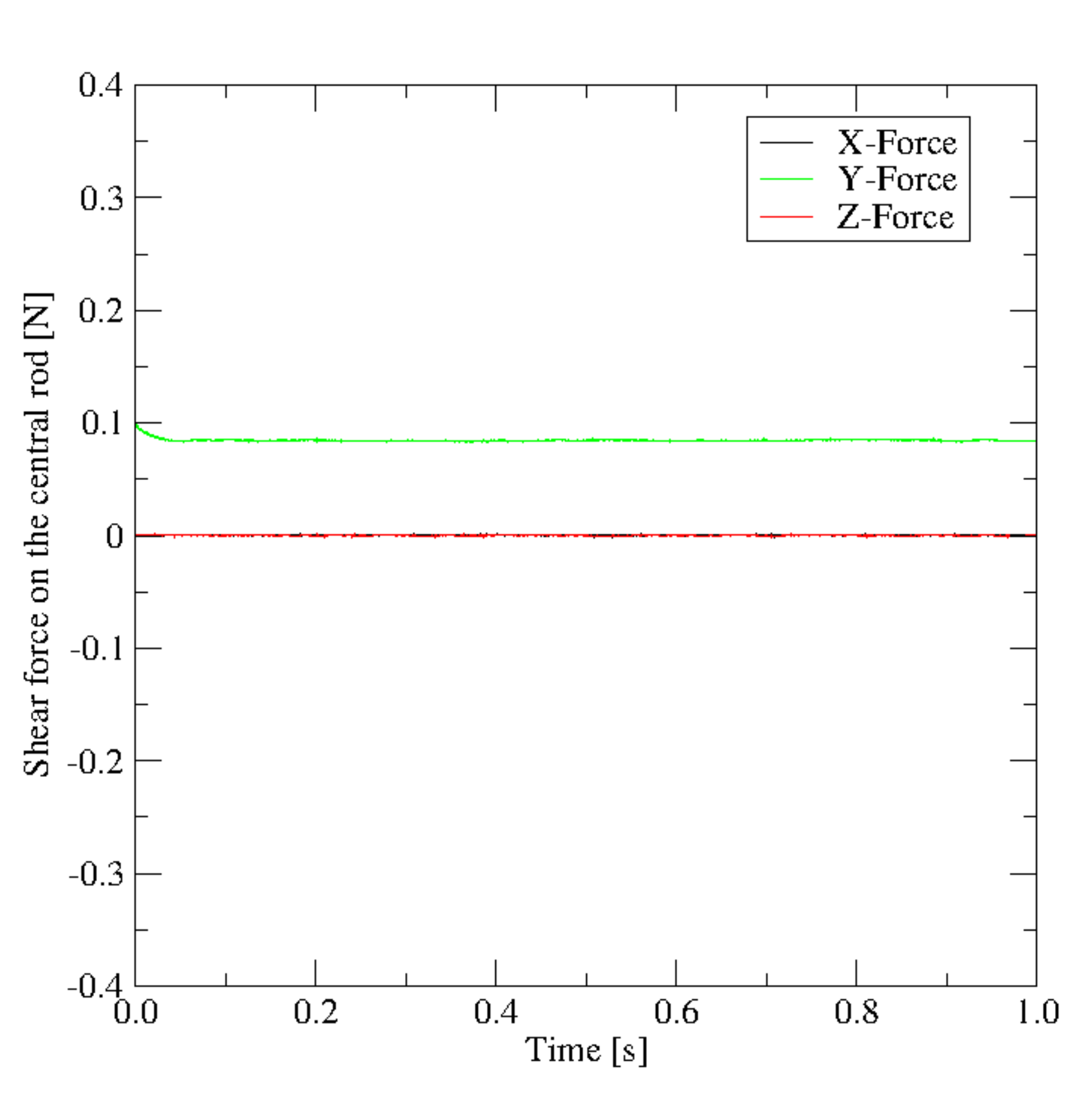}}
\subfloat[Shear force on the spacer.]{
\includegraphics[height=2.4in]{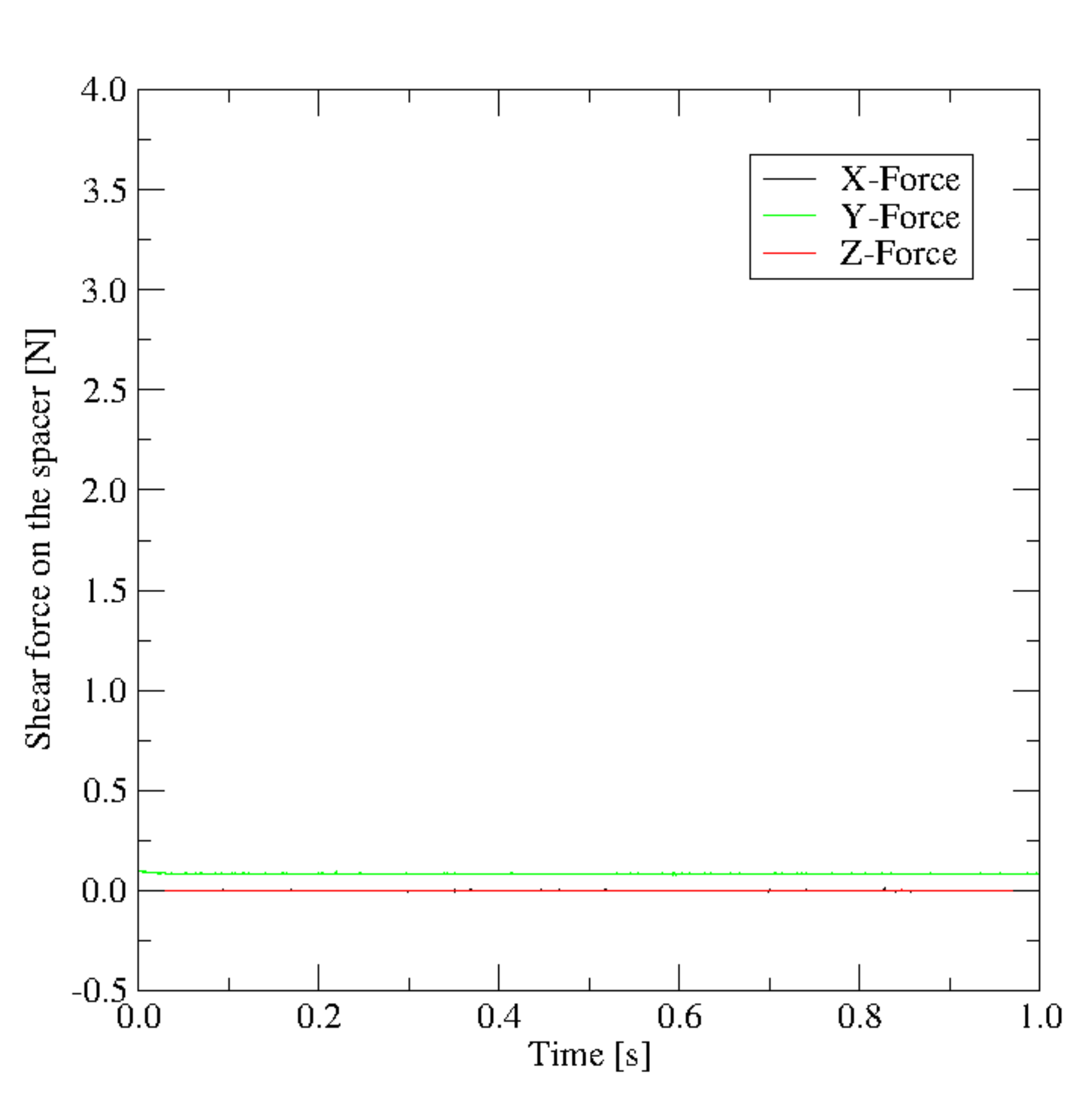}} 
\end{center}
\caption{Total, pressure, and shear force time histories on the
  central rod and spacer for the 7M Spider mesh.
\label{fig:spider_total_forces}}
\end{figure*}

The RMS pressure along the rod is plotted in Figure
\ref{fig:spider_rmspress}(a) for three Spider meshes. The fluctuating pressure
force is probably the most important quantity to compute accurately for a
reasonable representation of the forces acting on the fuel rods. The RMS
pressure indicates the deviation in the pressure relative to the mean, and is
correlated in an incompressible flow to local eddying motion. The figure shows
that the RMS pressure peaks at the downstream end of the spacer for the 7M and
14M meshes. This is expected, since this is where the flow is separated with
complex eddy structures, and where the level of turbulent kinetic energy is the
largest. While the downstream locations of the peaks are somewhat aligned for
the varying meshes, their amplitudes and downstream evolution are quite
different. The 2M mesh is too coarse to adequately capture the second pressure
moment.  At this point, we are not in a position to draw any conclusions
regarding the grid-convergence of the RMS pressure. Regardless, the turbulent
kinetic energy and the RMS pressure must decay downstream as no energy
production occurs downstream of the mixing vanes.

The total force and its two components, the pressure and viscous forces, have
been extracted in time on the central rod and the spacer. Surface forces are
computed by integrating pressure and shear stress over the given surface:
\begin{equation}
F_i(t) = -\int p(t)n_i\mathrm{d}A + 2\int\mu S_{ij}(t)n_j\mathrm{d}A,
\label{eq:force}
\end{equation}
where $\bv{F}$, $p$, $\bv{n}$, $A$, and $S_{ij}=(v_{i,j}+v_{j,i})/2$
denote the total force, pressure, outward surface normal, surface
area, and the strain rate of the instantaneous velocity, $\bv{v}$,
respectively. This gives the force time history that can be used to
compute power spectral distributions or fed directly into structural
dynamics codes to compute the rod dynamics response and ultimately
-- wear. The total, pressure, and viscous force
time-histories for the 7M case are presented in Figure
\ref{fig:spider_total_forces}, which shows that the mean forces are
similar to those computed using the Cubit meshes presented in
\cite{GTRF_2011}. On the other hand the pressure force acting on the
central rod, probably the most important quantity for the GTRF
problem, shows much larger fluctuations about the mean for the Spider
mesh relative to the Cubit results.

The total forces have also been integrated in 12 one-inch segments downstream of
the mixing vanes. This gives details on the spatial distribution of the forces
loading the central rod and allows for a more direct comparison with the
Star-CCM+ LES results in \cite{elmahdi:2011}. In Figure
\ref{fig:spider_rmspress}(b) the RMS total force is given in segments for the
7M, 14M, and 27M Spider meshes, compared to that of the Star-CCM+ LES results of
Elmahdi, et al.\ \cite{elmahdi:2011} using a 47M-cell mesh. The 2M Spider mesh
(not plotted) is inadequate to provide meaningful second moments of the force
loading the road. The RMS forces computed by Hydra-TH using the 7M, 14M, and 27M
meshes are quite close to those for Star-CCM+, but obtained with significantly
coarser meshes.

Additional insight into the fluctuating velocity field is found by
examining the turbulent kinetic energy and Reynolds stresses, shown in Figure
\ref{fig:spider_reynoldsstress}. In Figure
\ref{fig:spider_reynoldsstress}(a), the downstream spatial evolution of the
turbulent kinetic energy (TKE) is plotted for the 2M, 7M, and 14M meshes.
Similar to the pressure fluctuations in Figure \ref{fig:spider_rmspress}, the
TKE, $k=\irmean{\bv{v}\cdot\bv{v}}/2$, peaks in the vicinity of the mixing vanes
and stays at a relatively high value until approximately $0.2 \mathrm{m}$
downstream. This reinforces the earlier observation that the highest level of
TKE occurs close to the downstream edge of mixing vanes. Figure
\ref{fig:spider_reynoldsstress}(a) also indicates that the 2M-cell mesh is too
coarse to produce a qualitatively correct TKE evolution; similar to the RMS
pressure, the TKE should also decay downstream.

Figure \ref{fig:spider_reynoldsstress}(b) depicts the downstream evolution of
the different components of the Reynolds stress tensor, $\irmean{\bv{v}\bv{v}}$
for the 14M mesh. The figure shows that the flow downstream of the mixing vanes
remains highly anisotropic until the end of the computational domain: almost all
kinetic energy is in the streamwise component, $\irmean{vv}$, of the velocity,
$\bv{v}=(u,v,w)$, i.e., the streamwise fluctuations are large compared that of
both cross-stream components, $\irmean{uu}$, $\irmean{ww}$, in $x$ and $z$
directions, respectively.

\subsection{LES on the $5\times5$ Spider meshes}
\label{sec:5x5_spider}

\begin{figure}
\begin{center}
\setlength{\abovecaptionskip}{0pt}
\setlength{\belowcaptionskip}{-0.7cm}
\includegraphics[width=1.0\columnwidth]{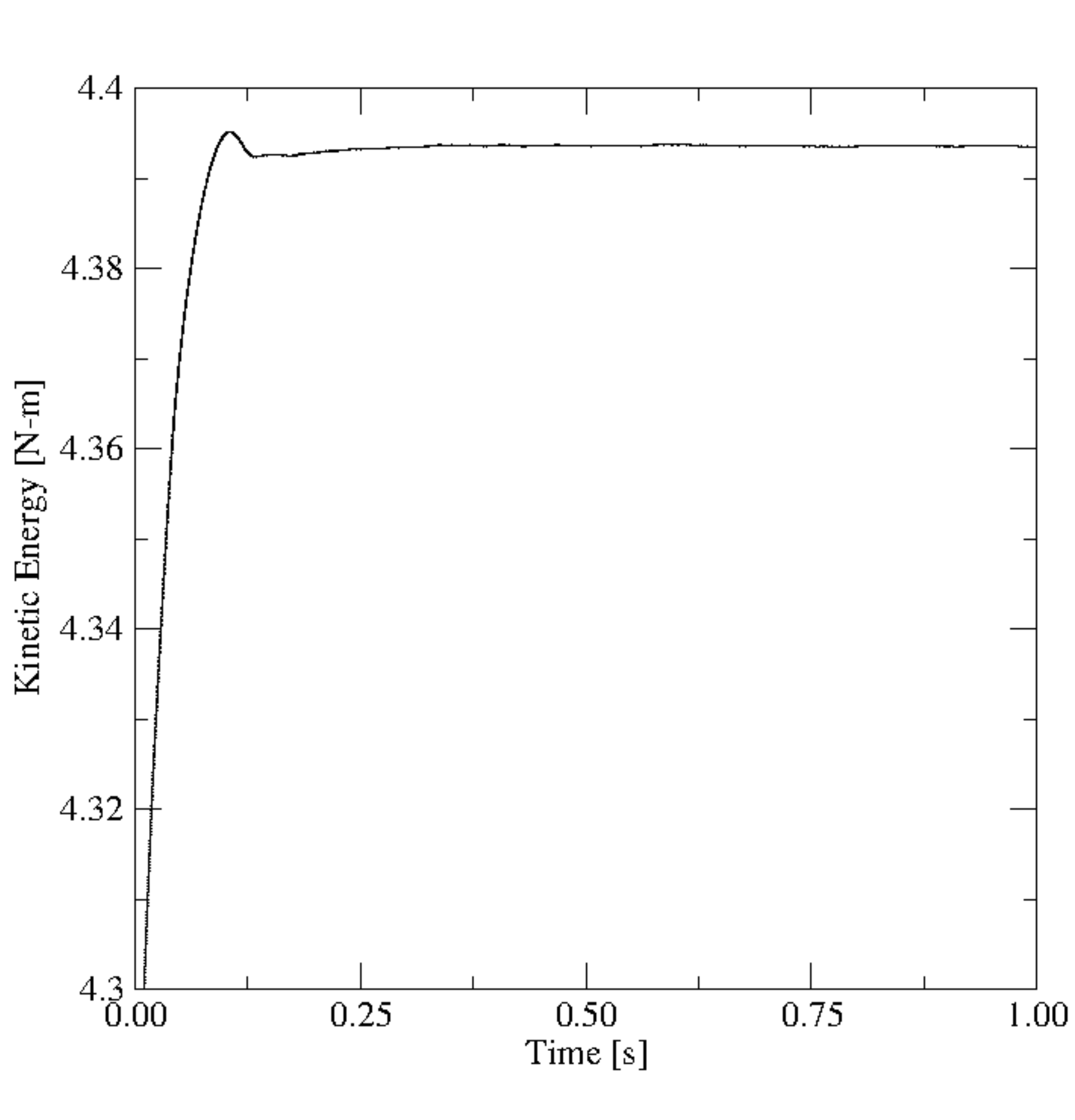}
\caption{Domain-integrated kinetic energy,
         $\int\rho\bv{v}\!\cdot\!\bv{v}/2 \ \mathrm{d} \Omega$,
         vs. time for the 14M $5 \times 5$ rod bundle.}
\label{fig:5x5_14M_KE}
\end{center}
\end{figure}

\begin{figure*}[t]
\begin{center}
\setlength{\abovecaptionskip}{0pt}
\setlength{\belowcaptionskip}{-0.7cm}
\includegraphics[width=0.85\textwidth]{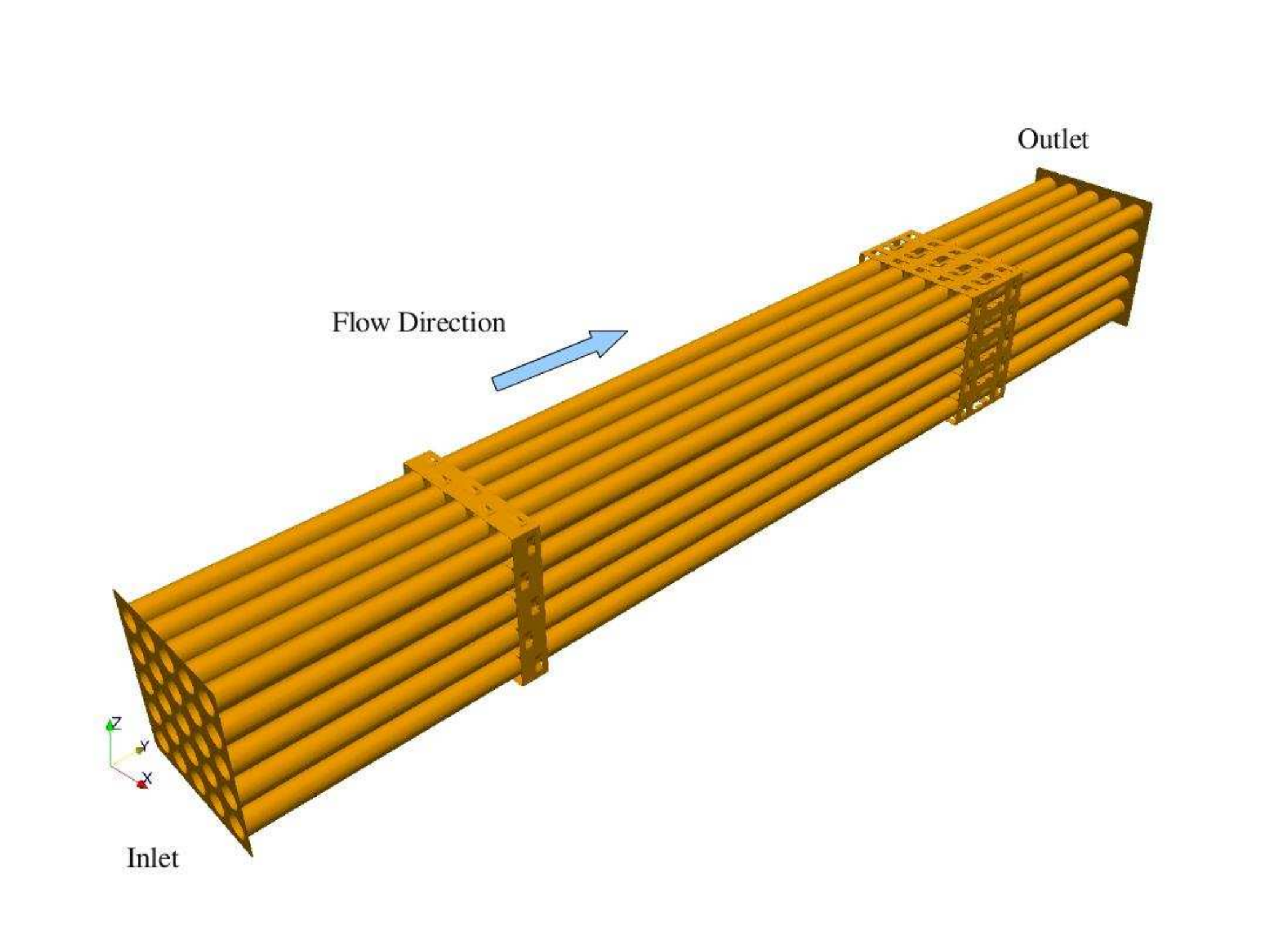}
\caption{Flow domain for the $5\times5$ rod bundle showing the rods, the
  inlet/outlet planes, the support, and the spacer grid.}
\label{fig:5x5_domain}
\end{center}
\end{figure*}

This section discusses the calculations for the $5 \times 5$ fuel-rod bundle.
The geometry was provided in CAD format by Westinghouse, and corresponds to the
experimental configuration used at Texas A\&M, where PIV measurements were
carried out. The flow domain is shown in Figure \ref{fig:5x5_domain}. Not shown
here are the exterior walls of the flow housing used in the experimental
facility. Additional details on the experimental configuration and results may
be found in Conner, et al.\ \cite{conner:2011} and Yan, et al.\ \cite{yan:2012}.

At the inlet of the flow domain, a constant prescribed velocity $(0.0, 2.48,
0.0)\mathrm{m/s}$ is applied with the fluid properties for water at
$24^\circ\mathrm{C}$ and atmospheric pressure. This corresponds to a Reynolds
number of approximately $28,000$ based on the hydraulic diameter for the rod
bundle. At the surfaces of the flow housing, rods, support and spacer grids,
no-slip/no-penetration velocity conditions were prescribed. Homogeneous Neumann
conditions for velocity along with a zero-pressure condition were prescribed at
the outflow plane. A fixed (maximum) $CFL=4$
condition was used with automatic time-step control for all computations. The
flow domain, meshed with Spider is illustrated in Figure \ref{fig:mesh}(b).

\begin{figure*}
\begin{center}
\setlength{\belowcaptionskip}{-0.7cm}
\subfloat[14M Mesh.]{
\includegraphics[width=0.90\textwidth]{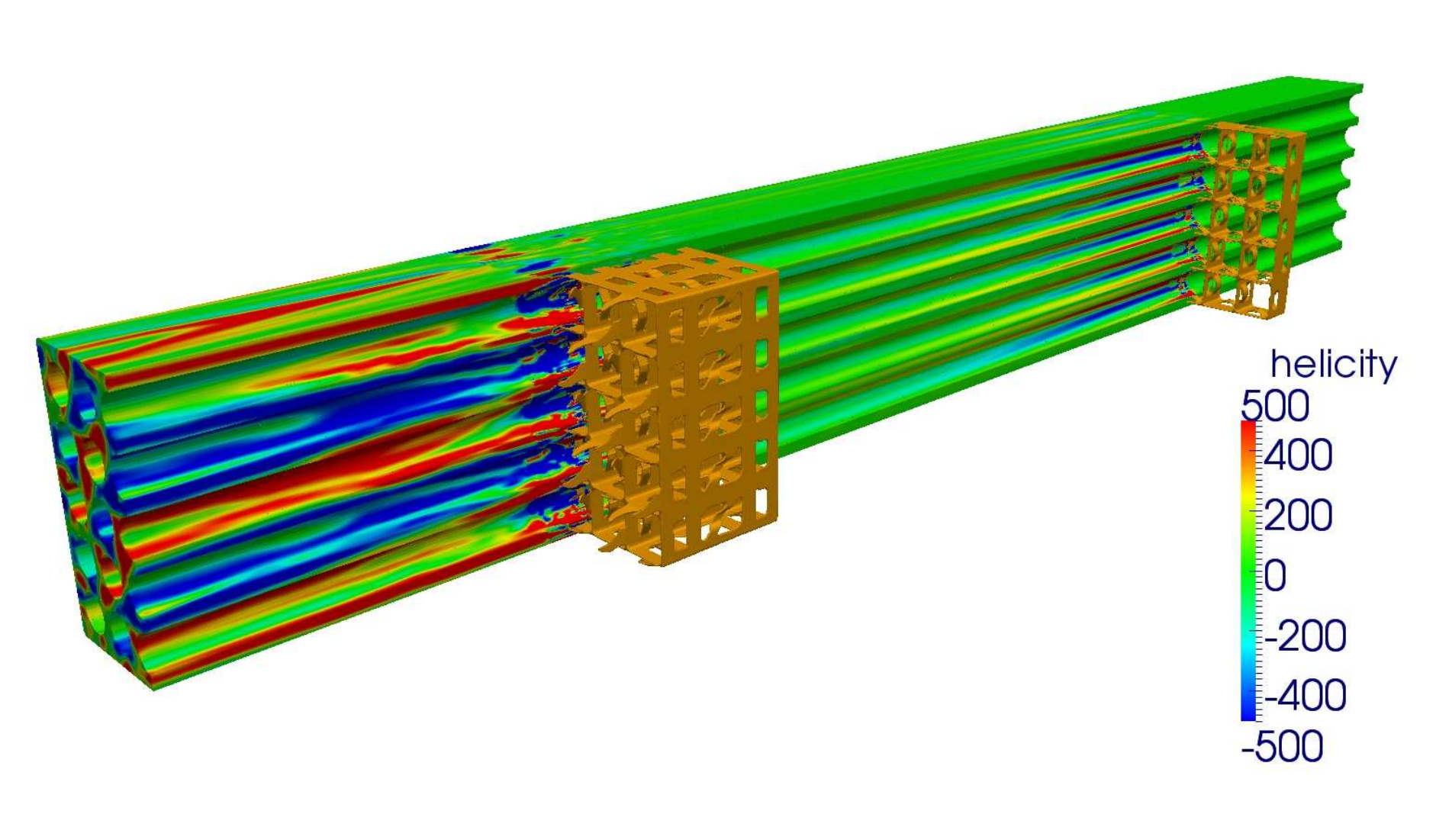}} \\
\subfloat[96M Mesh.]{
\includegraphics[width=0.90\textwidth]{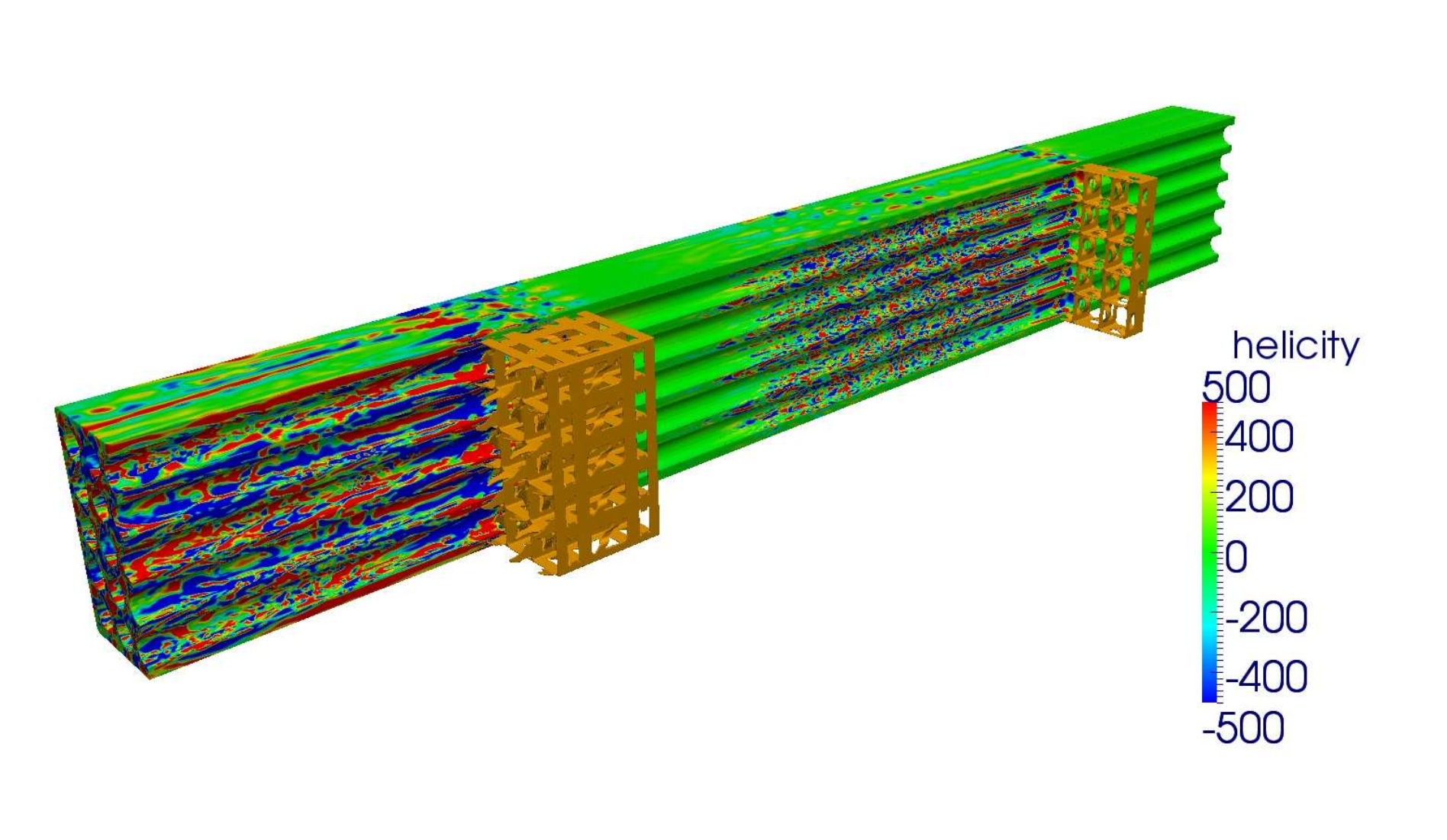}} \\
\caption{Snapshots of the instantaneous helicity field for the 14M and
  96M element meshes.}
\label{fig:5x5_helicity}
\end{center}
\end{figure*}

Following the procedures to perform LES calculations on the $3\times3$
rod-bundle, outlined in \cite{GTRF_2011}, a series of preliminary coarse-mesh
calculations were conducted to determine when a stationary turbulent state would
be achieved and to test the sensitivity to mesh resolution and the time-step
size. Figure \ref{fig:5x5_14M_KE} shows the domain-average kinetic energy,
$\int\rho\bv{v}\!\cdot\!\bv{v}/2 \mathrm{d}\Omega$, as a function of time. Here
$\Omega$ denotes the volume of the flow domain. Based on these preliminary
calculations, we chose a time of approximately $0.2\mathrm{s}$
as the starting point for collecting time-averaged flow statistics
until the end of the simulation at $t=1.0\mathrm{s}$. The initial
$0.2\mathrm{s}$ corresponds to approximately one flow transit after which the
domain-integrated kinetic energy has reached a statistically stationary state.

In order to illustrate the impact of increasing mesh resolution on the flow,
Figure \ref{fig:5x5_helicity} shows snapshots of the instantaneous helicity
field for the $5\times5$ rod bundle. For the 14M mesh, there are relatively
large coherent structures downstream of the support and spacer grid. In contrast,
the flow structures captured by the 96M mesh are significantly smaller
and appear more randomly distributed spatially. In both cases, the influence of
the mixing vanes on the spacer grid is apparent.

\begin{figure*}
\begin{center}
\setlength{\belowcaptionskip}{-0.7cm}
\subfloat[Weld nugget location.]{
\includegraphics[width=0.45\textwidth]{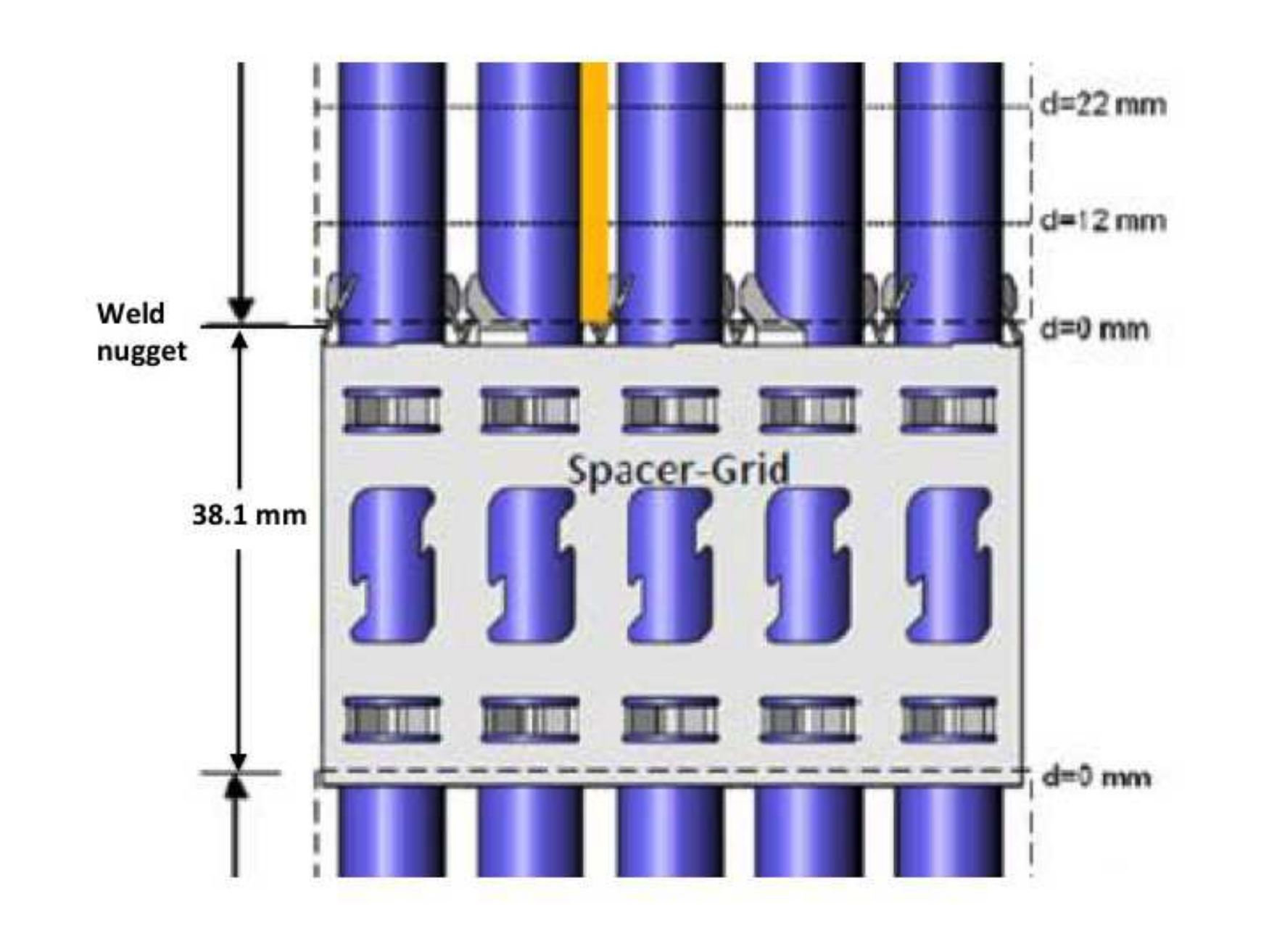}}
\subfloat[Sample points.]{
\includegraphics[width=0.45\textwidth]{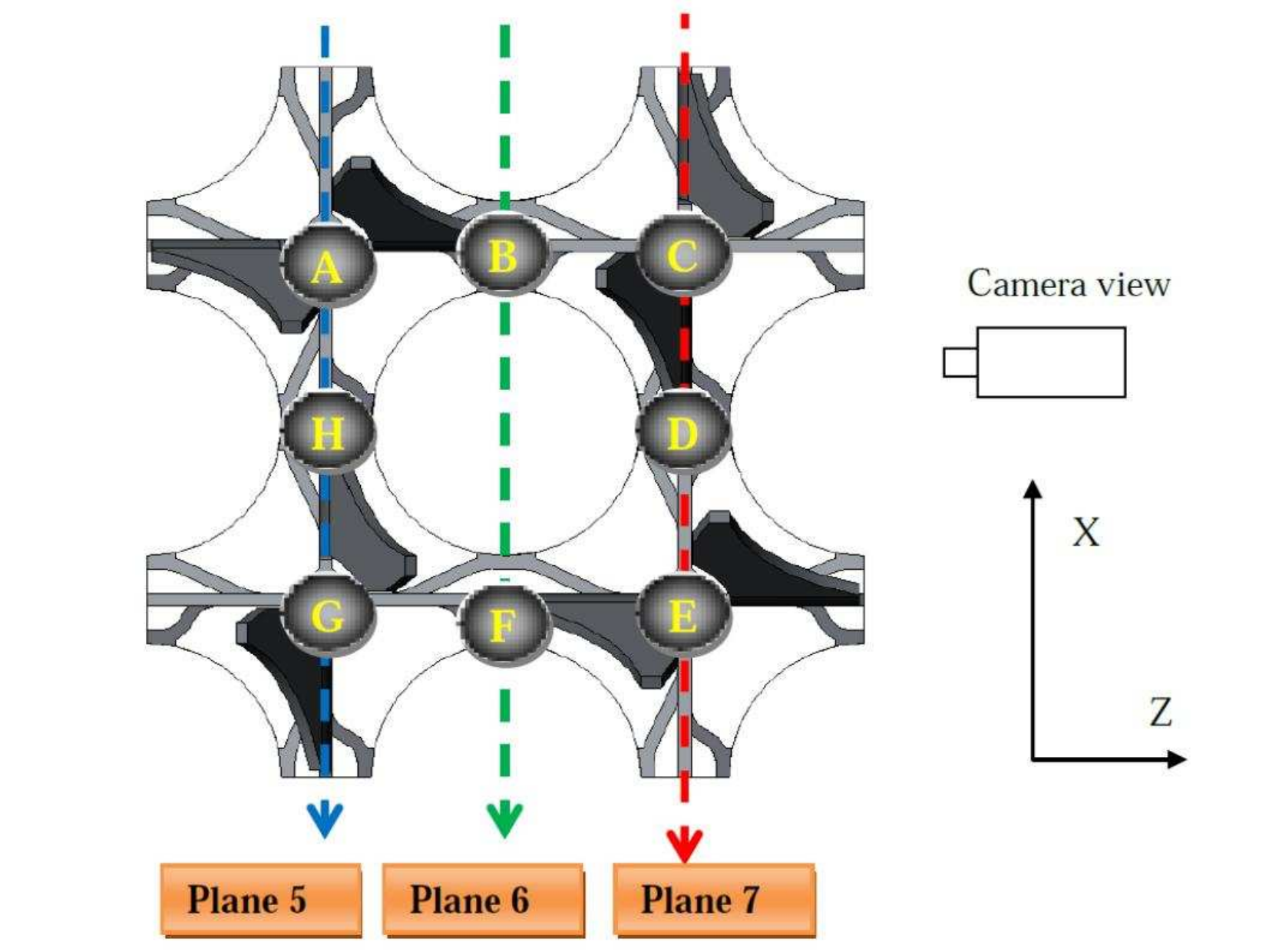}}
\caption{Locations relative to the ``weld nugget'' used for extracting
data along planes 5, 6 and 7. (Reproduced from \cite{yan:2012}) without
permission.)}
\label{fig:5x5_sample_locations}
\end{center}
\end{figure*}

In order to compare to the experimental data, discussed in \cite{yan:2012}, a
series of line plots were extracted from the mean velocity field for the
14M-mesh $5\times5$ run at locations that fall in the planes of the PIV
measurements.  All line data were measured relative to the so-called
``weld-nugget'' located on the spacer grid. The ``weld nugget'' is located at
$38.1$ mm from the bottom of the spacer grid \cite{yan:2012b}, as shown in Figure
\ref{fig:5x5_sample_locations}(a). The line-data extracted from the computation
was located at the positions indicated in Figure \ref{fig:5x5_sample_locations}.
The coordinates of the sample points A -- H are shown in Table
\ref{tbl:5x5_sample_locations} and are relative to the center of rod 13 in
Figure 3 of \cite{yan:2012}. In the flow direction, the line-data is extracted
for $0.05 \leq y \leq 0.09 \ m$ corresponding to the region where PIV data is
available in the region downstream of the spacer grid.

\begin{figure*}[t]
\setlength{\abovecaptionskip}{0pt}
\setlength{\belowcaptionskip}{0pt}
\begin{center}
\begin{tabular}{cc}
\includegraphics[width=0.4\textwidth]{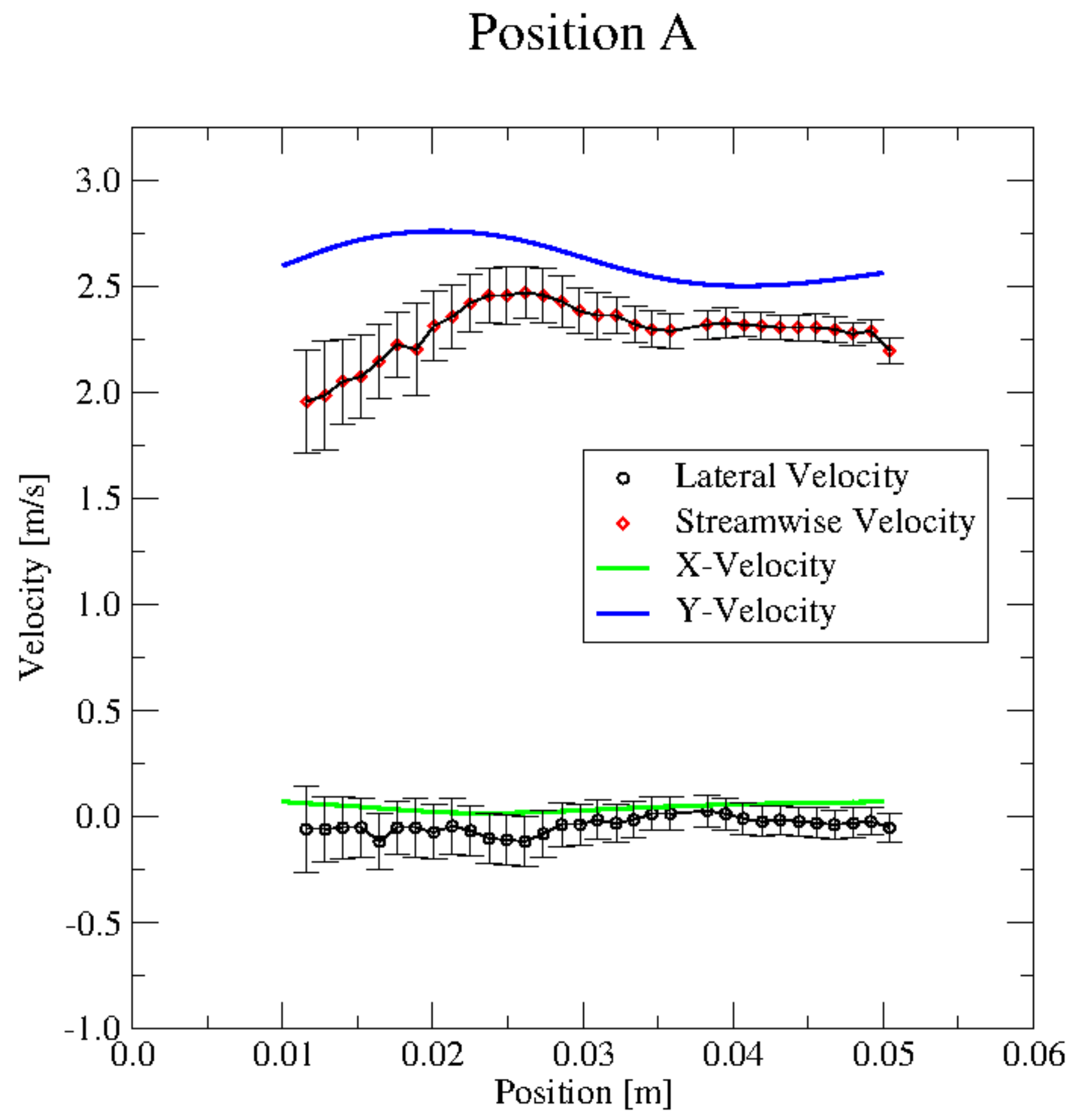} &
\includegraphics[width=0.4\textwidth]{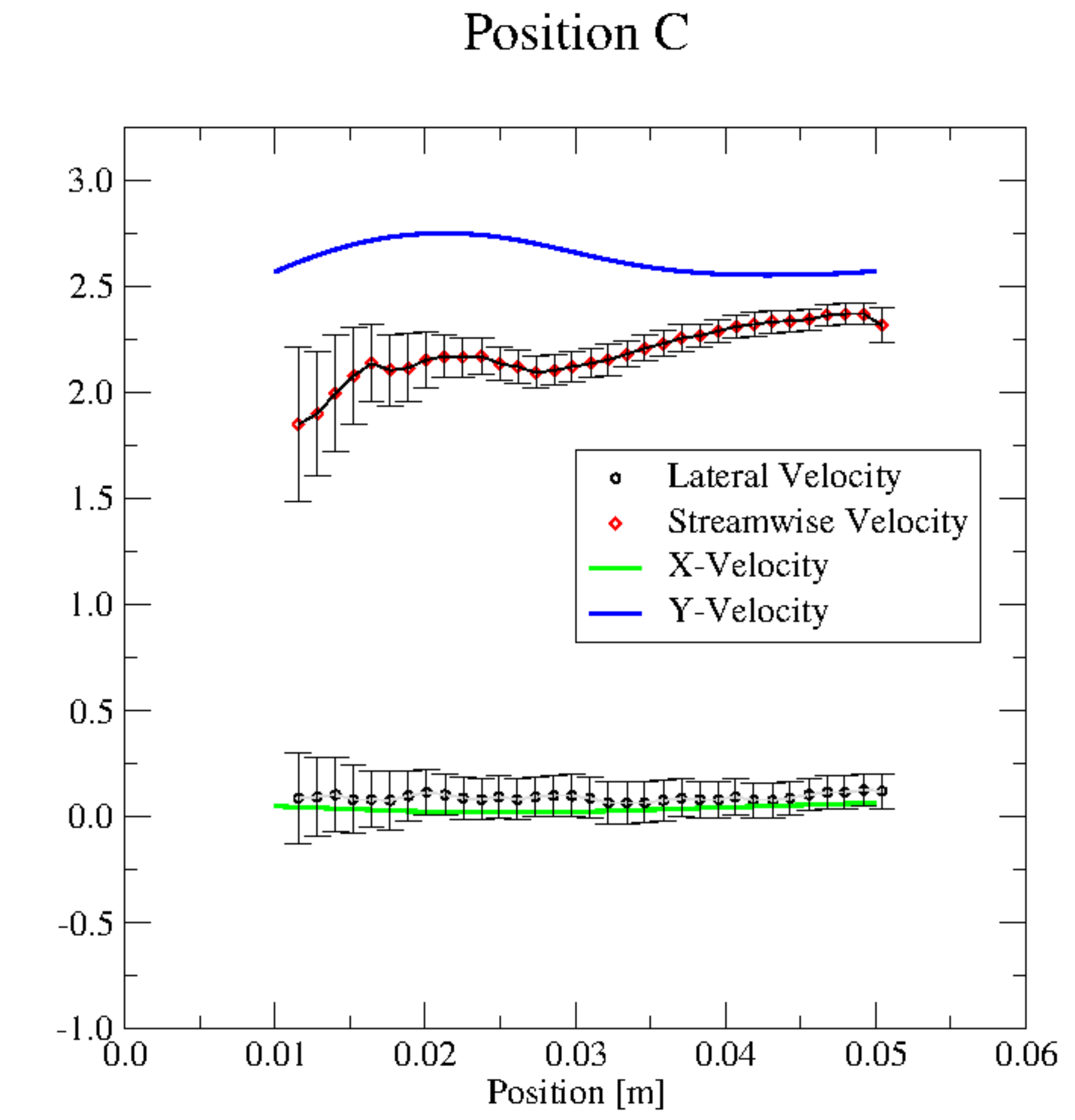} \\
\includegraphics[width=0.4\textwidth]{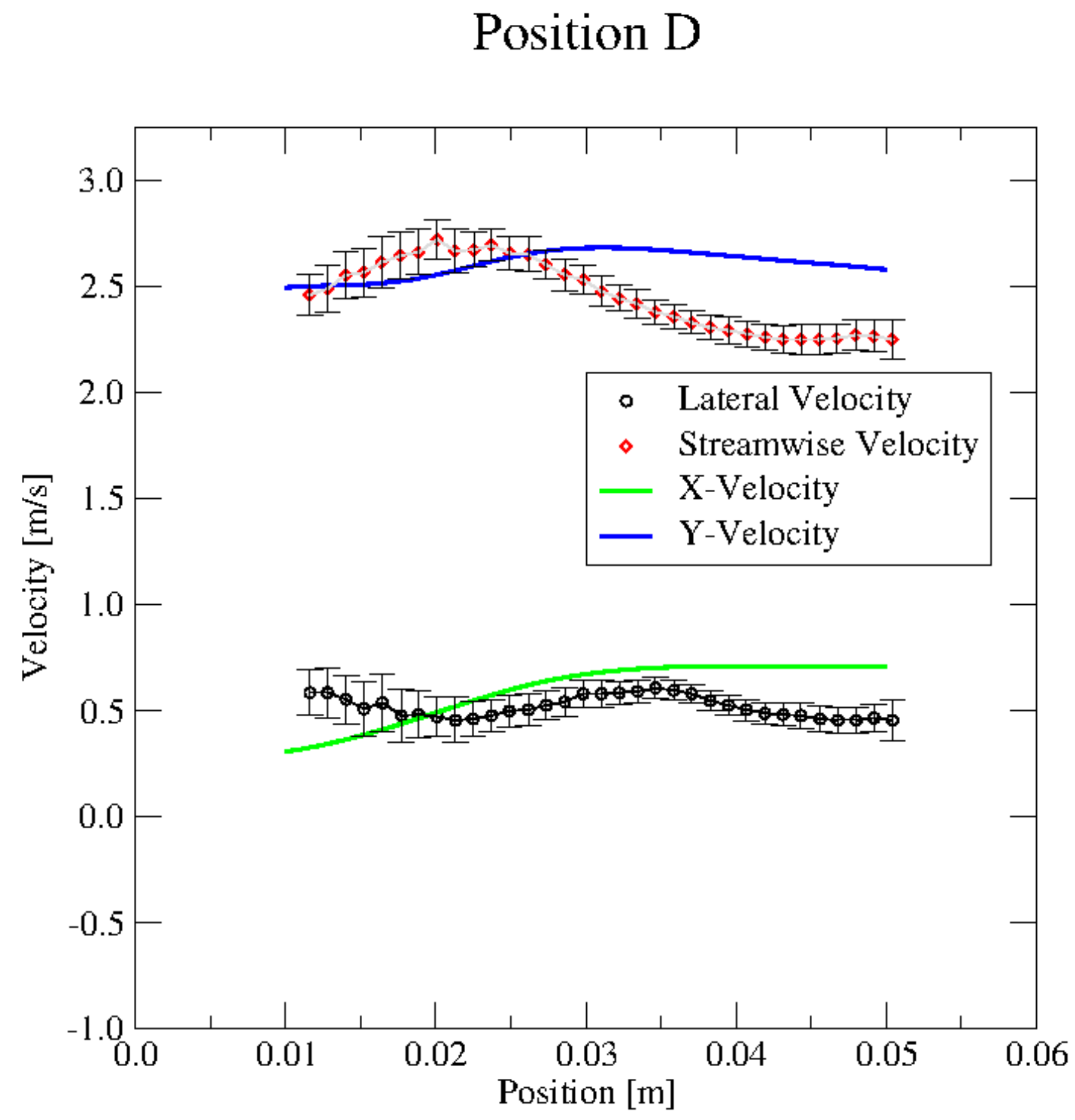} &
\includegraphics[width=0.4\textwidth]{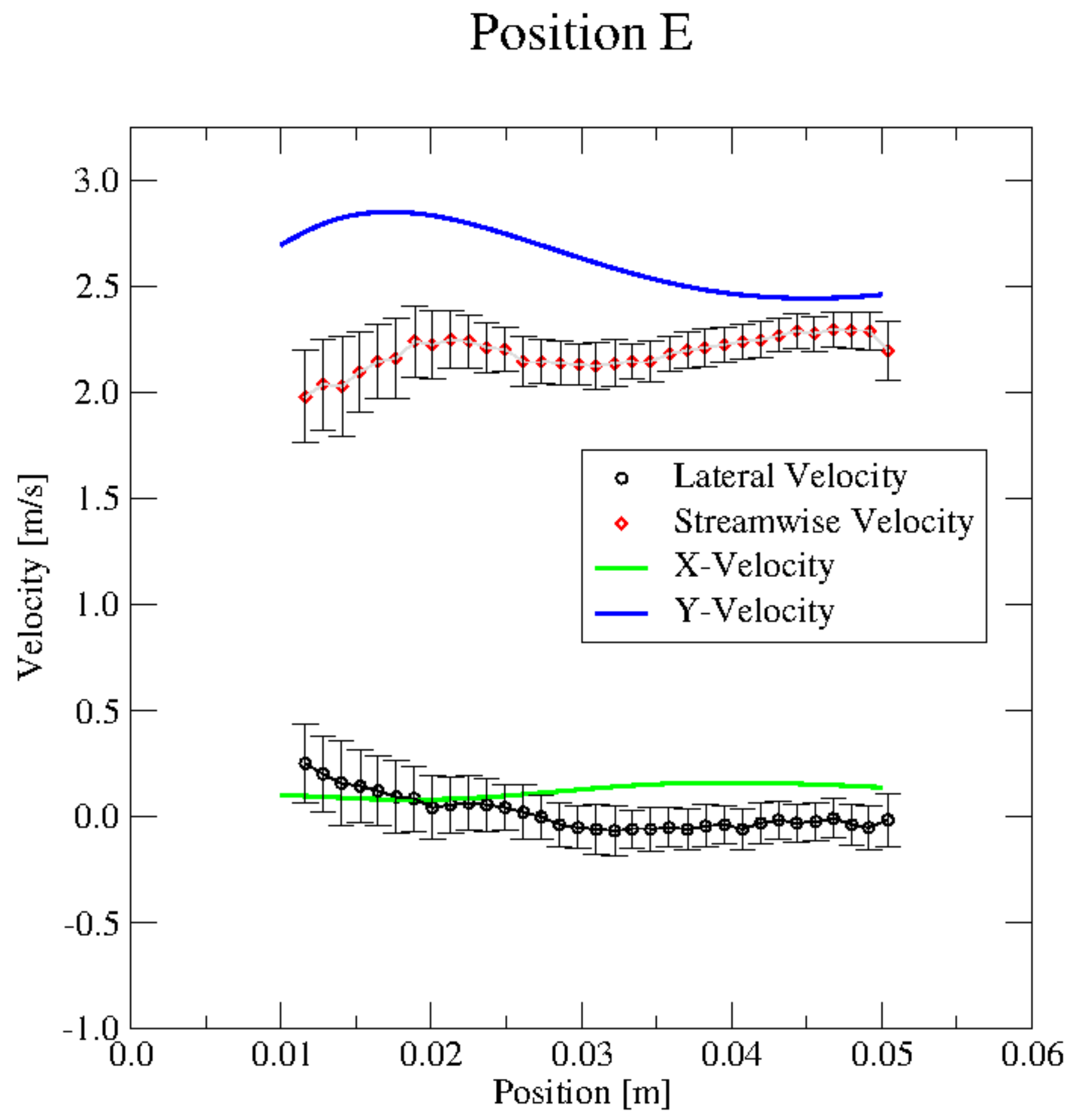} \\
\includegraphics[width=0.4\textwidth]{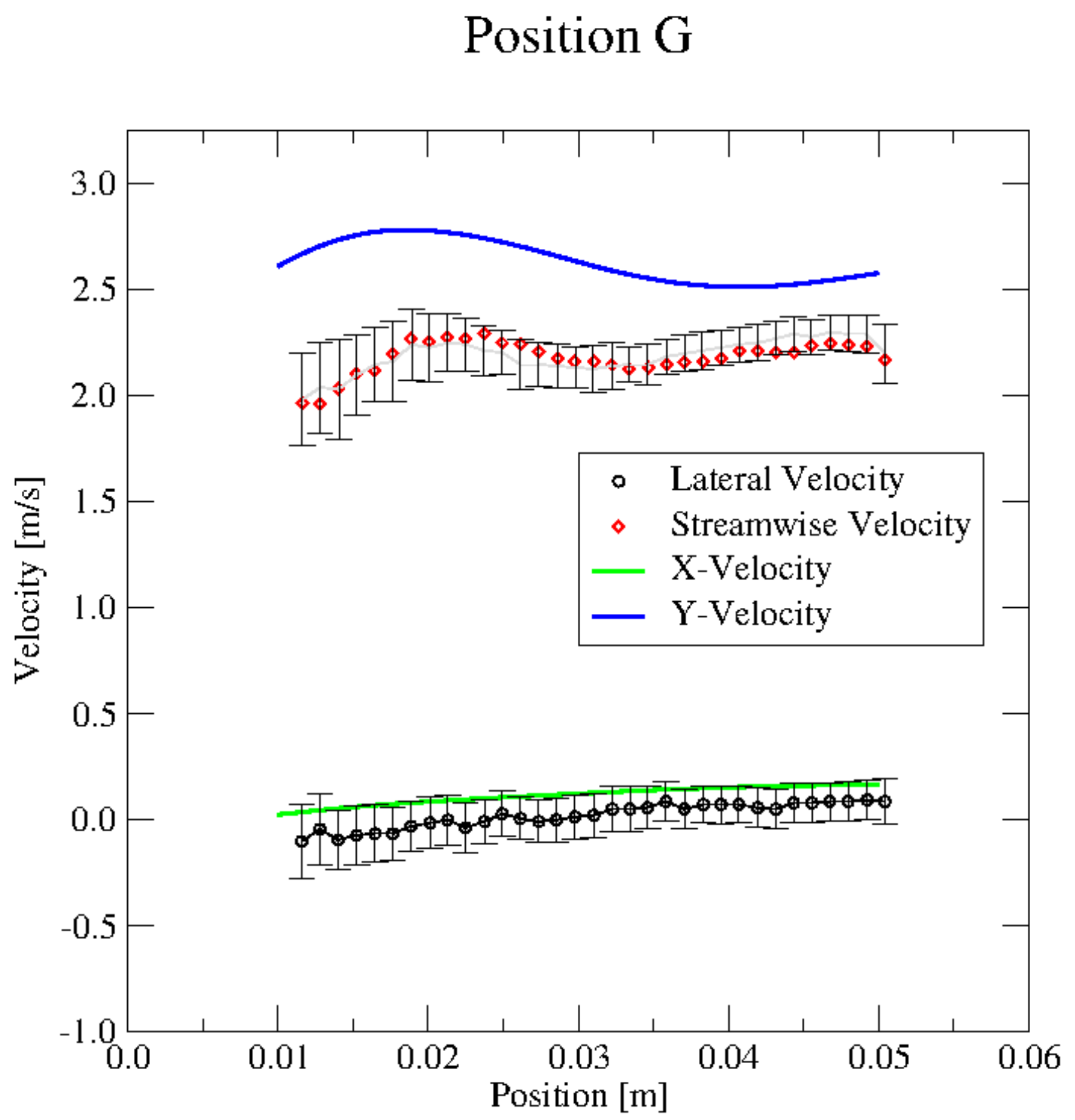} &
\includegraphics[width=0.4\textwidth]{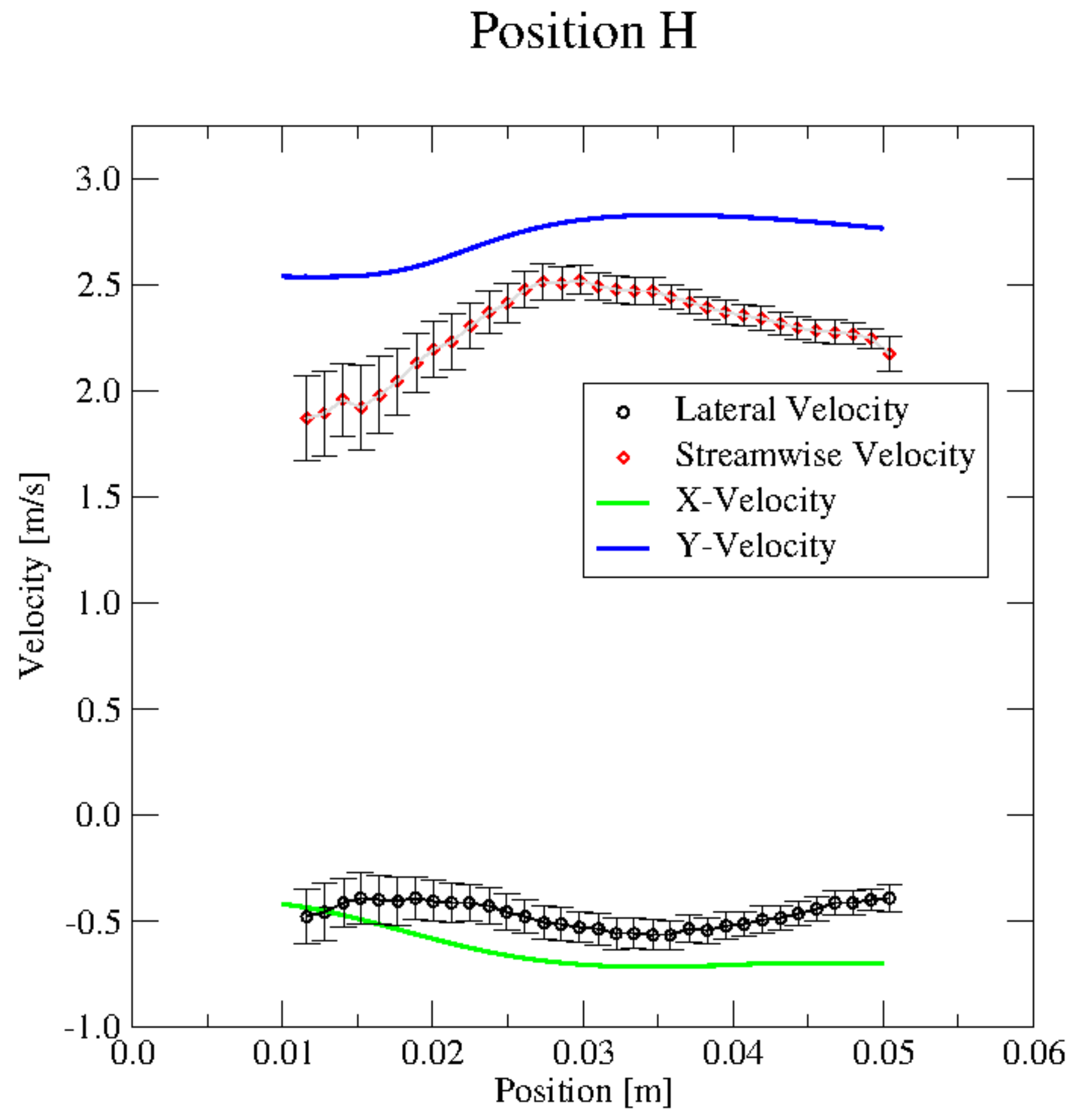} \\
\end{tabular}
\caption{Mean axial and lateral velocity profiles at positions A, C,
  D, E, G and H for the 14M mesh.}
\label{fig:5x5_vel_lines}
\end{center}
\end{figure*}

\begin{table}
\setlength{\abovecaptionskip}{0pt}
\setlength{\belowcaptionskip}{-0.5cm}
\begin{center}
\begin{tabular}{c|c}
Point & $(x,z)$ Position [$10^{-3}m$] \\
\hline
A & (-6.3,  6.3) \\
B & (-6.3,  0.0) \\
C & (-6.3, -6.3) \\
D & ( 0.0, -6.3) \\
E & ( 6.3, -6.3) \\
F & ( 6.3,  0.0) \\
G & ( 6.3,  6.3) \\
H & ( 0.0,  6.3) \\
\end{tabular}
\end{center}
\caption{Sample points A -- H used to extract line-data for comparison
  with experimental data.\label{tbl:5x5_sample_locations}}
\end{table}

\begin{figure*}
\setlength{\abovecaptionskip}{0pt}
\setlength{\belowcaptionskip}{0pt}
\begin{center}
\begin{tabular}{cc}
\includegraphics[width=0.4\textwidth]{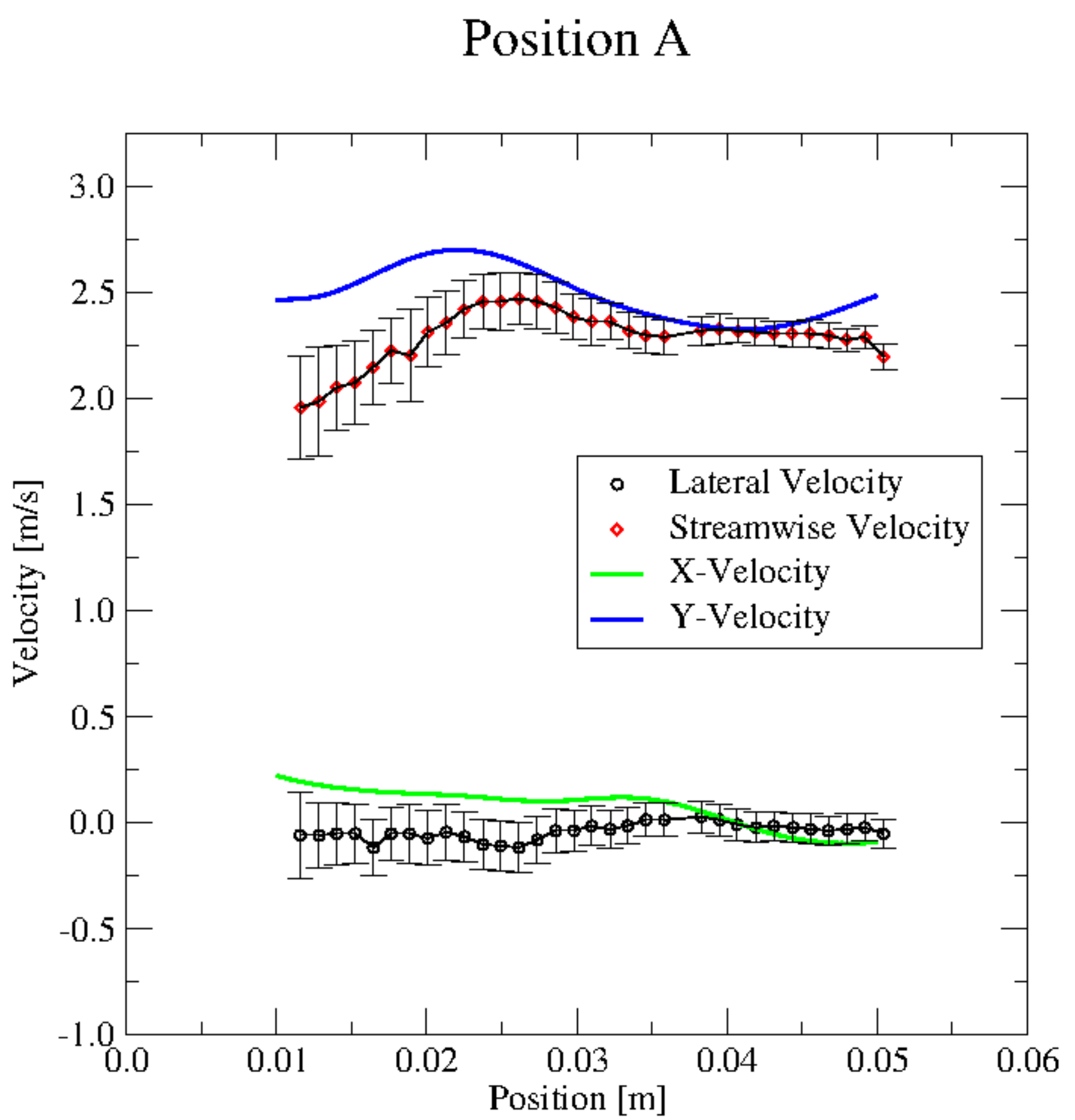} &
\includegraphics[width=0.4\textwidth]{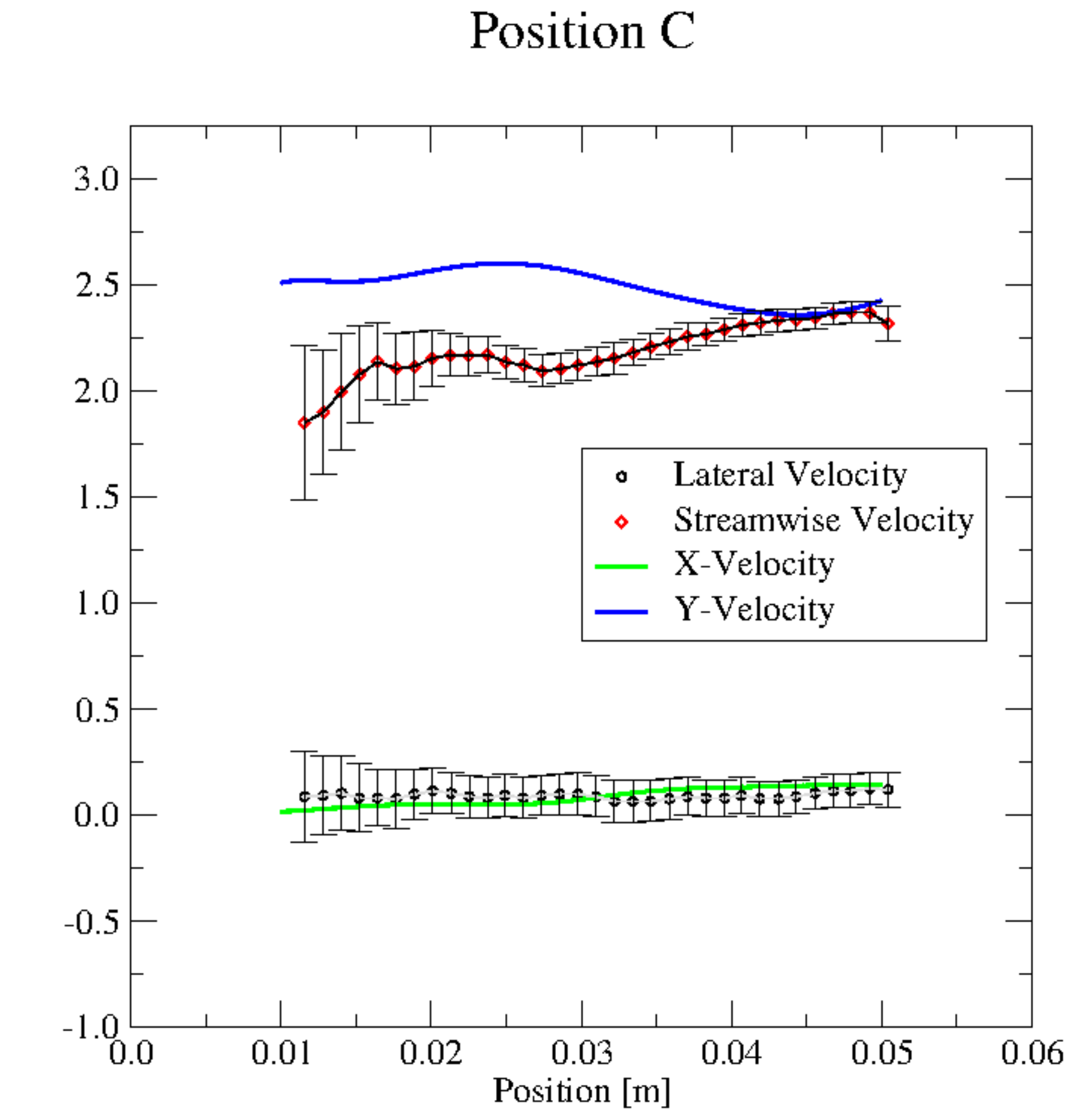} \\
\includegraphics[width=0.4\textwidth]{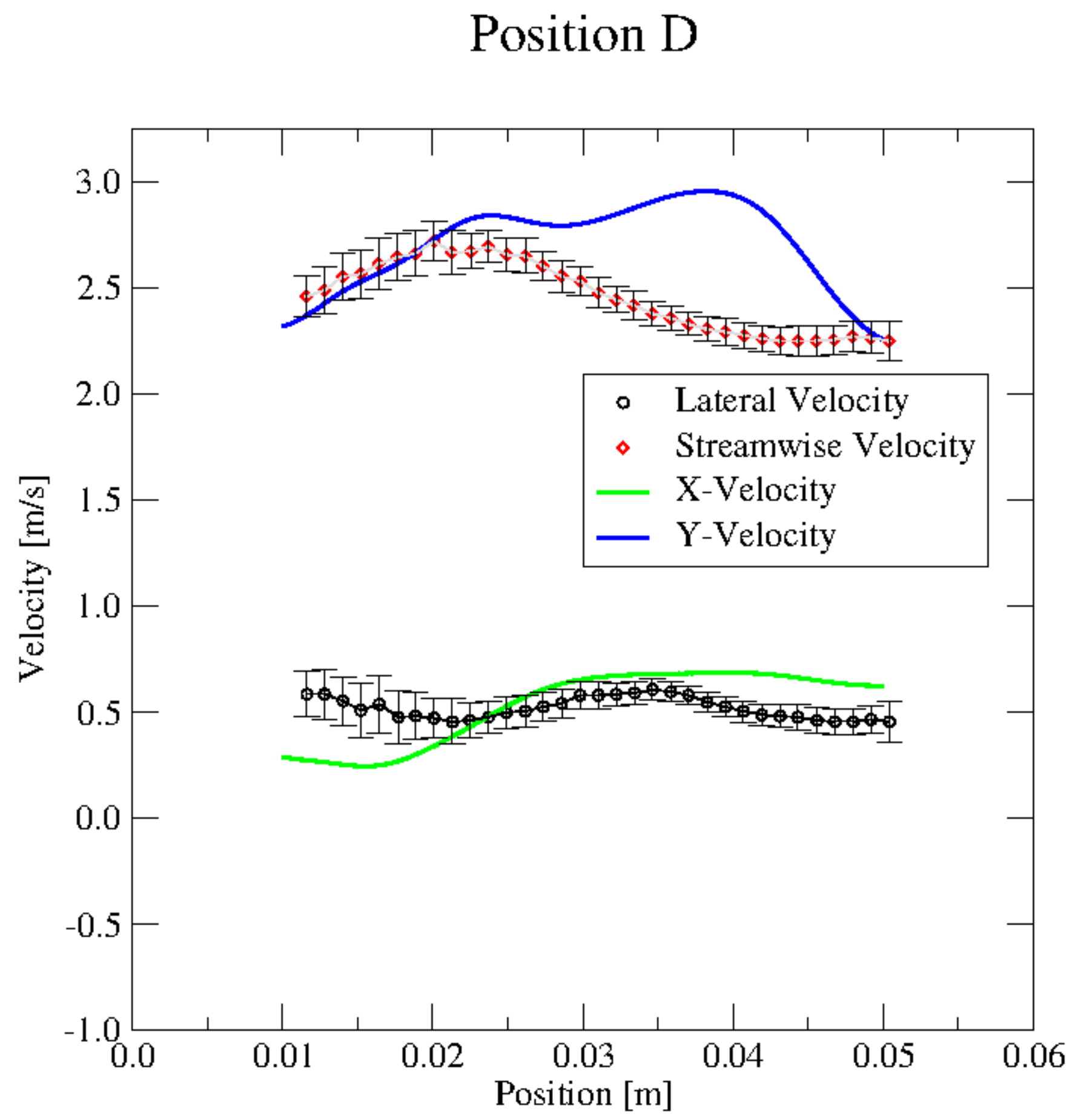} &
\includegraphics[width=0.4\textwidth]{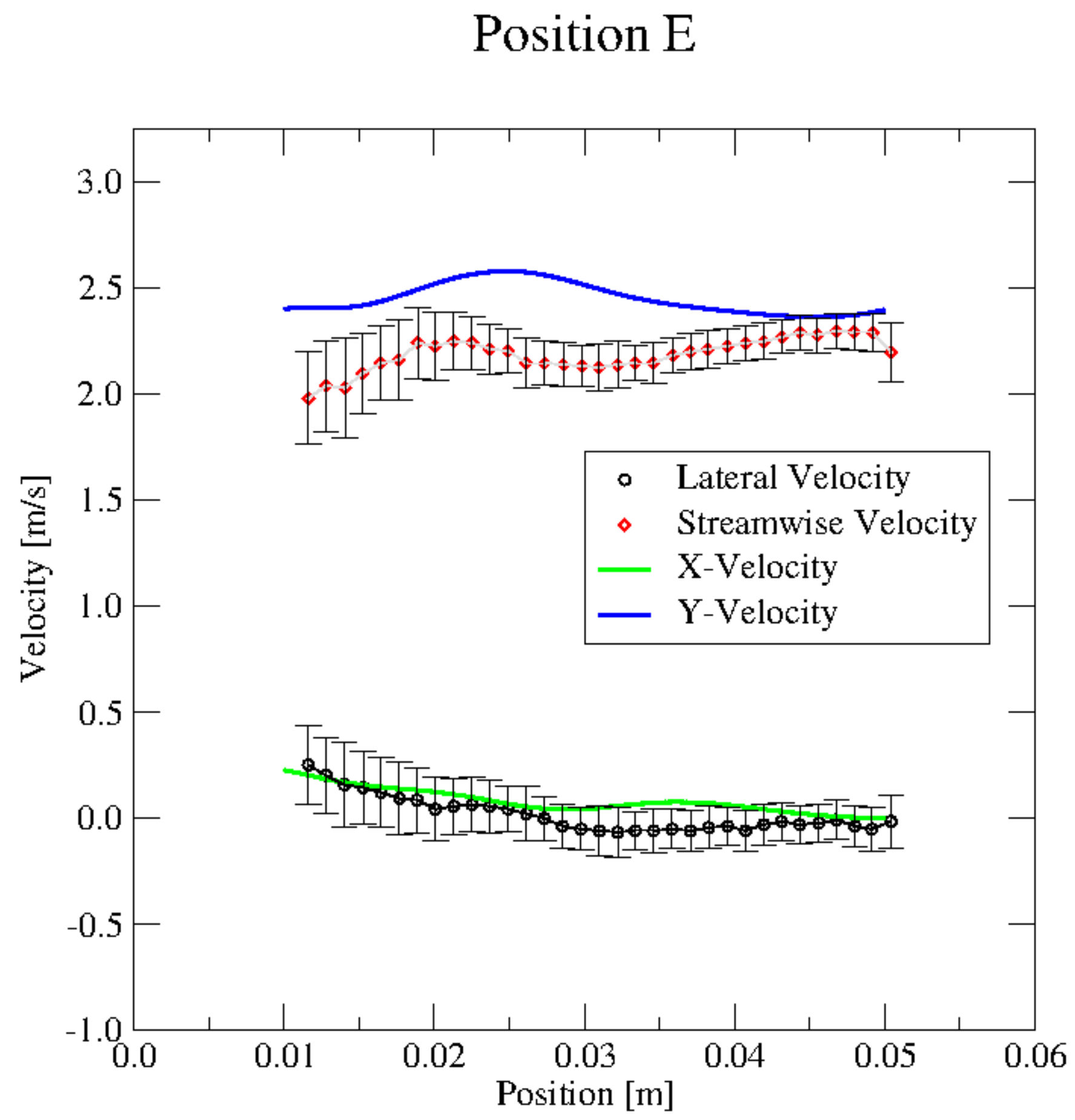} \\
\includegraphics[width=0.4\textwidth]{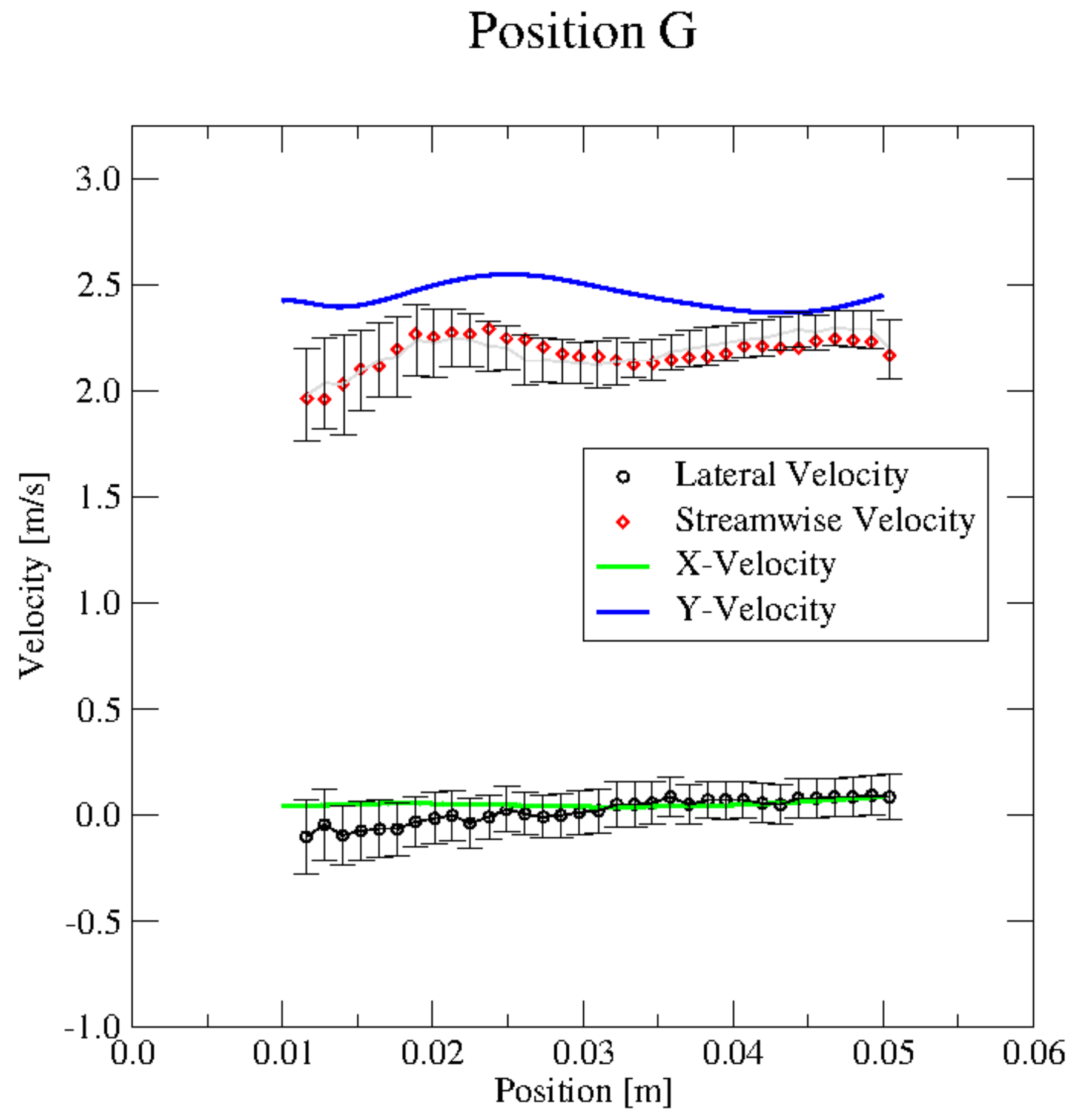} &
\includegraphics[width=0.4\textwidth]{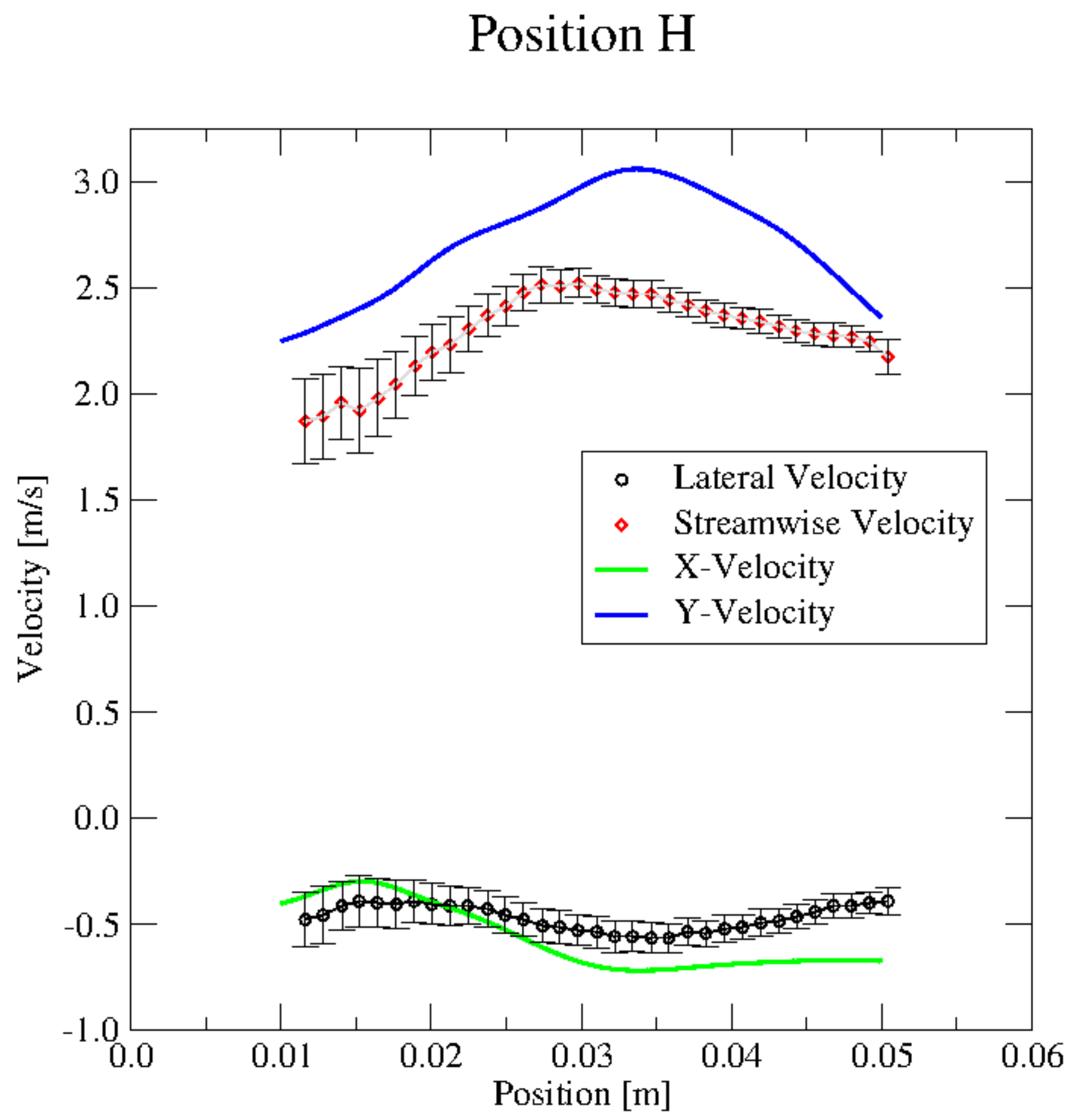} \\
\end{tabular}
\caption{Mean axial and lateral velocity profiles at positions A, C,
  D, E, G and H for the 96M mesh.}
\label{fig:5x5_96M_vel_lines}
\end{center}
\end{figure*}

\begin{figure*}
\setlength{\belowcaptionskip}{-0.7cm}
\begin{center}
\subfloat[Experimental Axial Velocity]{
\includegraphics[width=0.39525\textwidth]{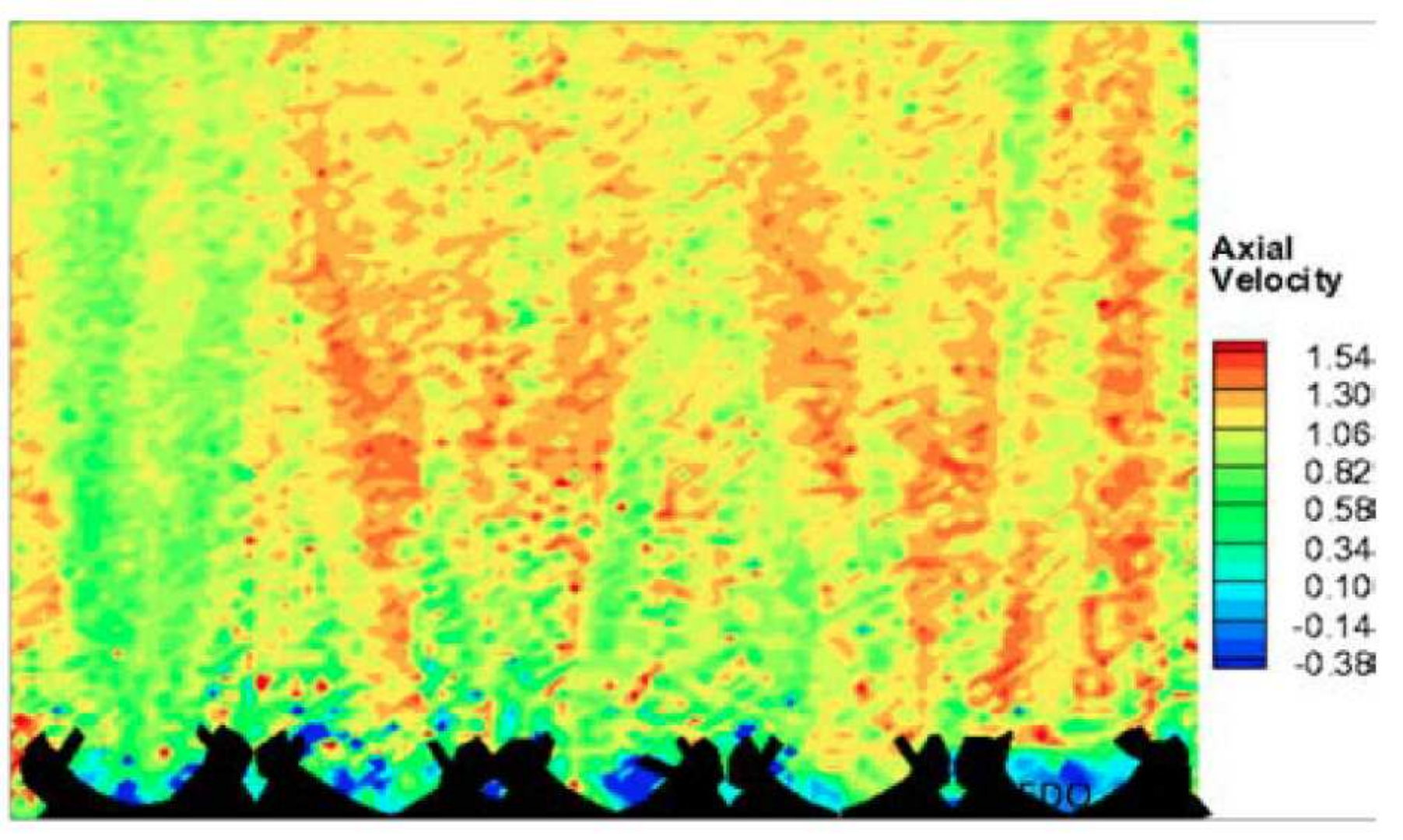}}
\subfloat[Hydra-TH Y-Velocity]{
\includegraphics[width=0.39525\textwidth]{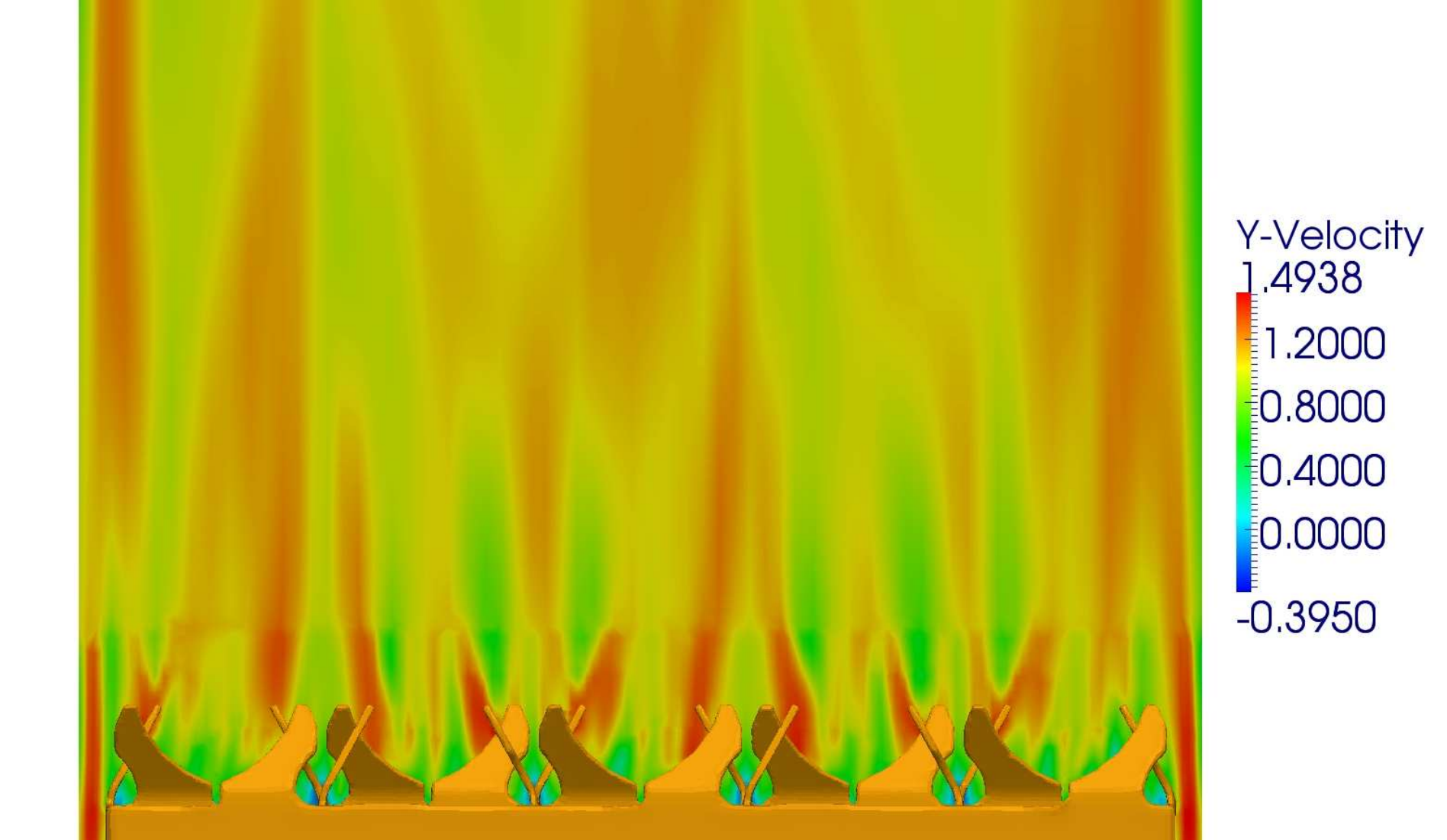}} \\
\subfloat[Experimental Lateral Velocity]{
\includegraphics[width=0.39525\textwidth]{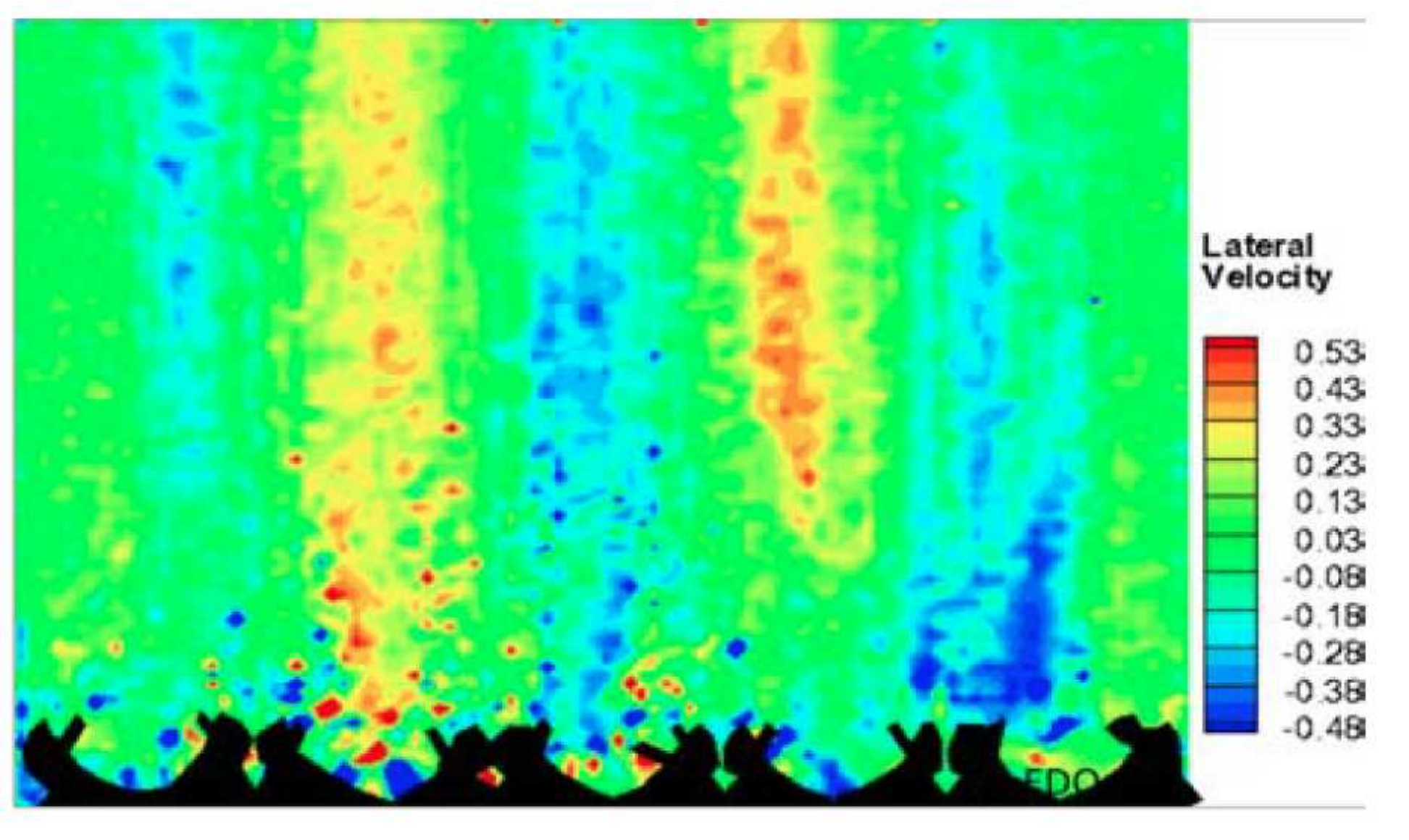}}
\subfloat[Hydra-TH X-Velocity]{
\includegraphics[width=0.39525\textwidth]{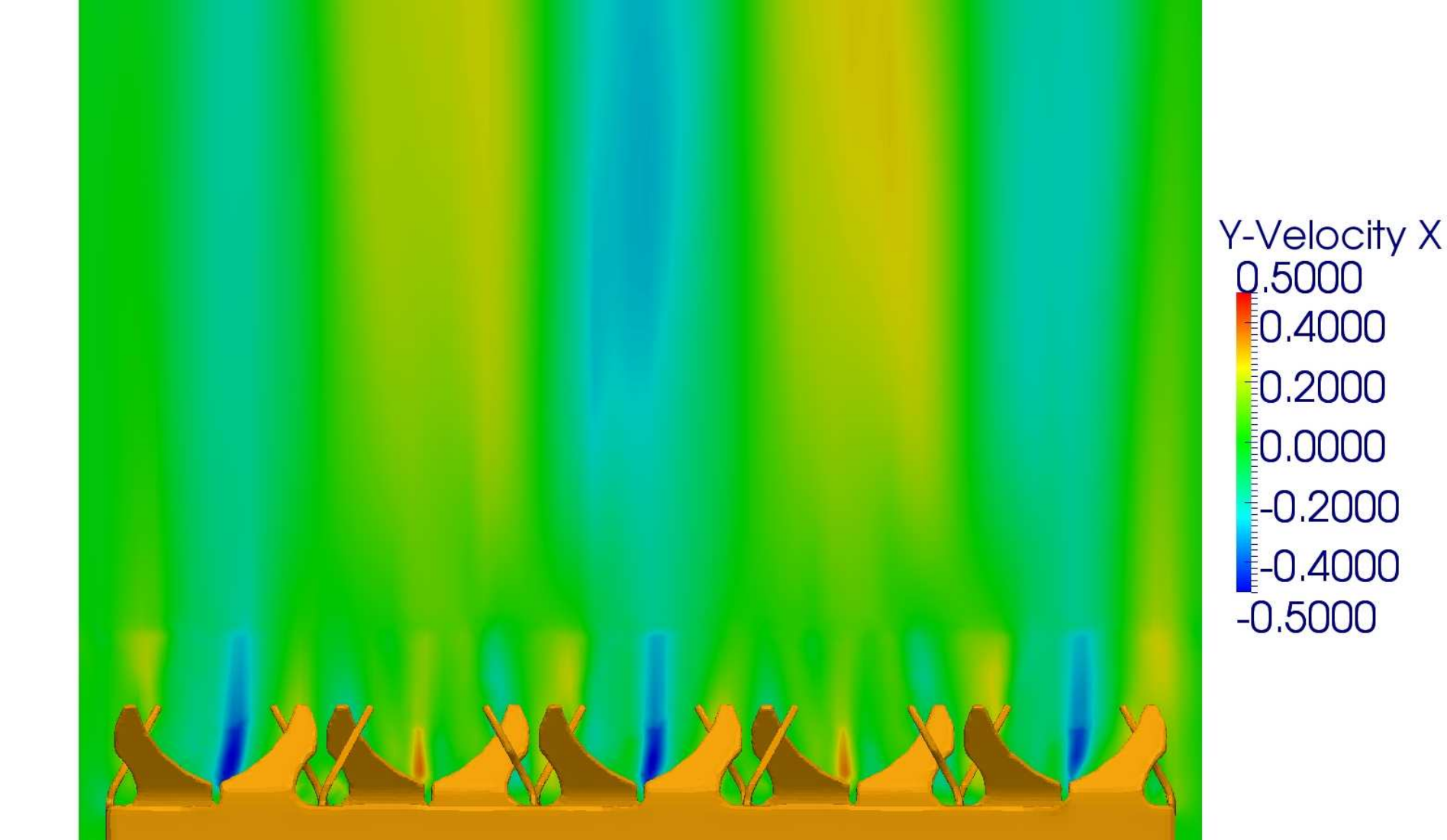}}\\
\caption{Experimental and computed axial (y-direciton) time-averaged velocities
on plane 5. Velocity magnitude has been scaled relative to the $2.48 \
m/s$ inlet velocity.} \addvspace{0.3cm}
\label{fig:5x5_velocity_plane5}
\end{center}
\end{figure*}

\begin{figure*}
\setlength{\belowcaptionskip}{-0.7cm}
\begin{center}
\subfloat[Experimental Axial Velocity]{
\includegraphics[width=0.39525\textwidth]{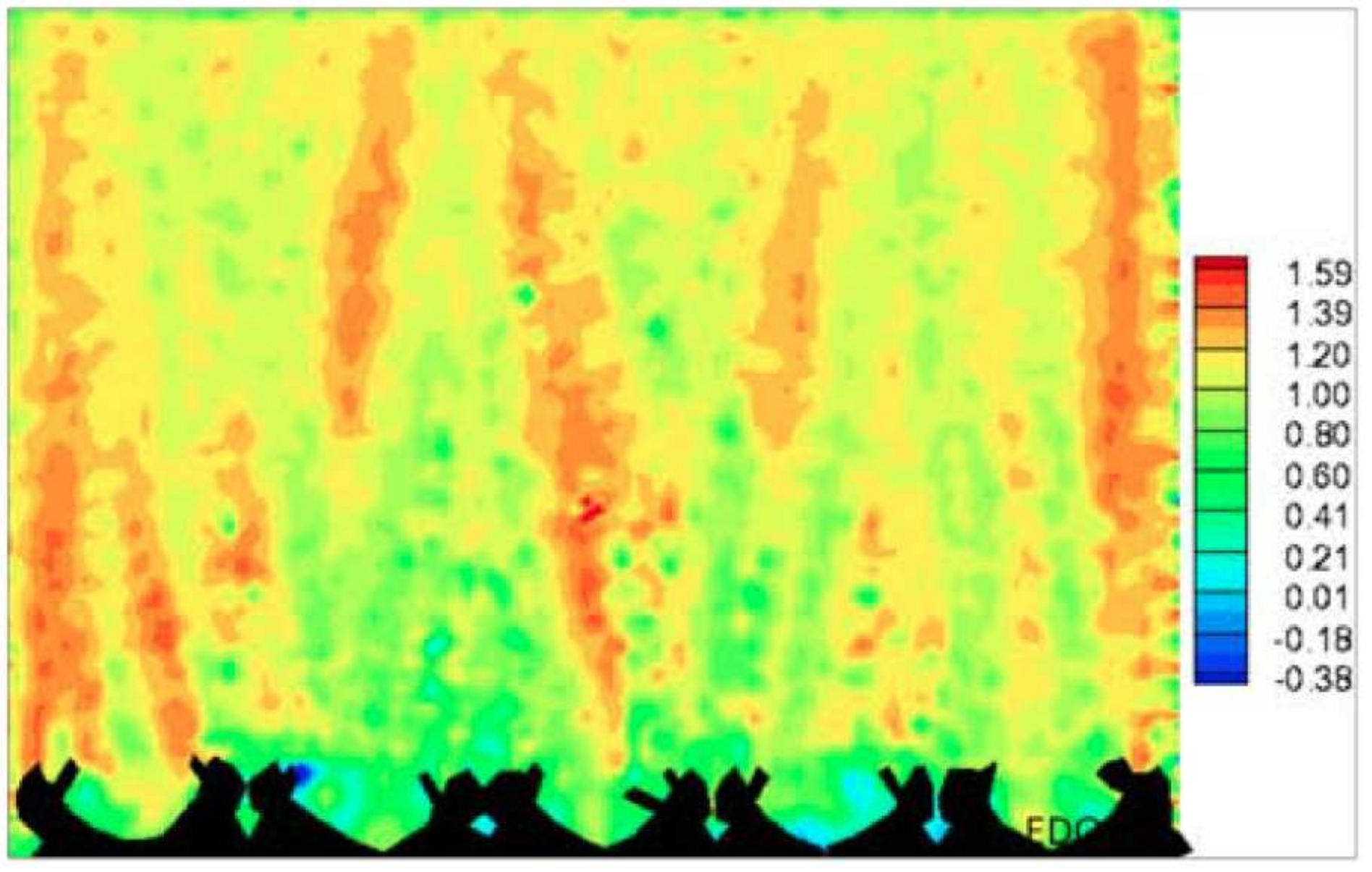}}
\subfloat[Hydra-TH Y-Velocity]{
\includegraphics[width=0.39525\textwidth]{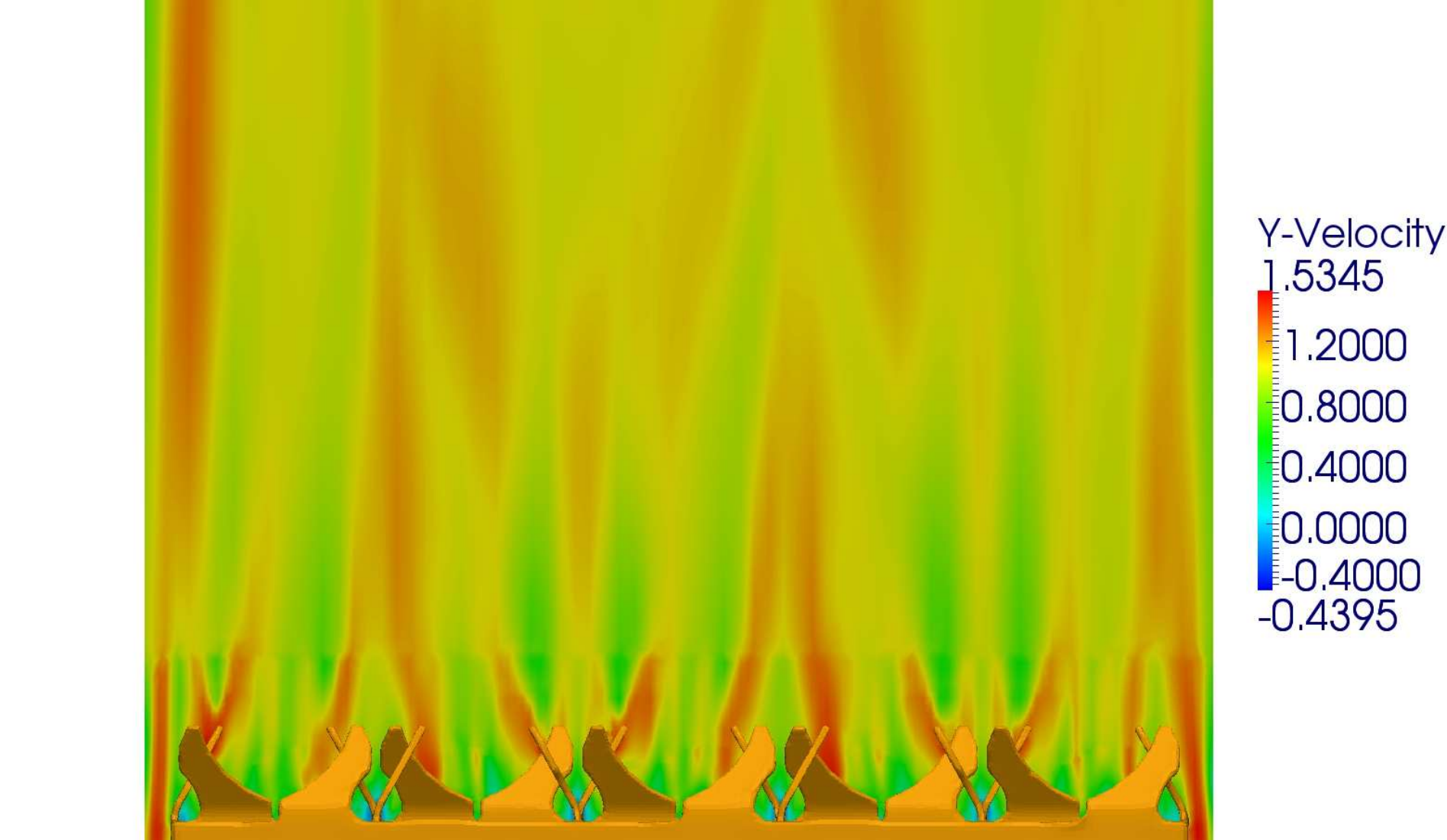}} \\
\subfloat[Experimental Lateral Velocity]{
\includegraphics[width=0.39525\textwidth]{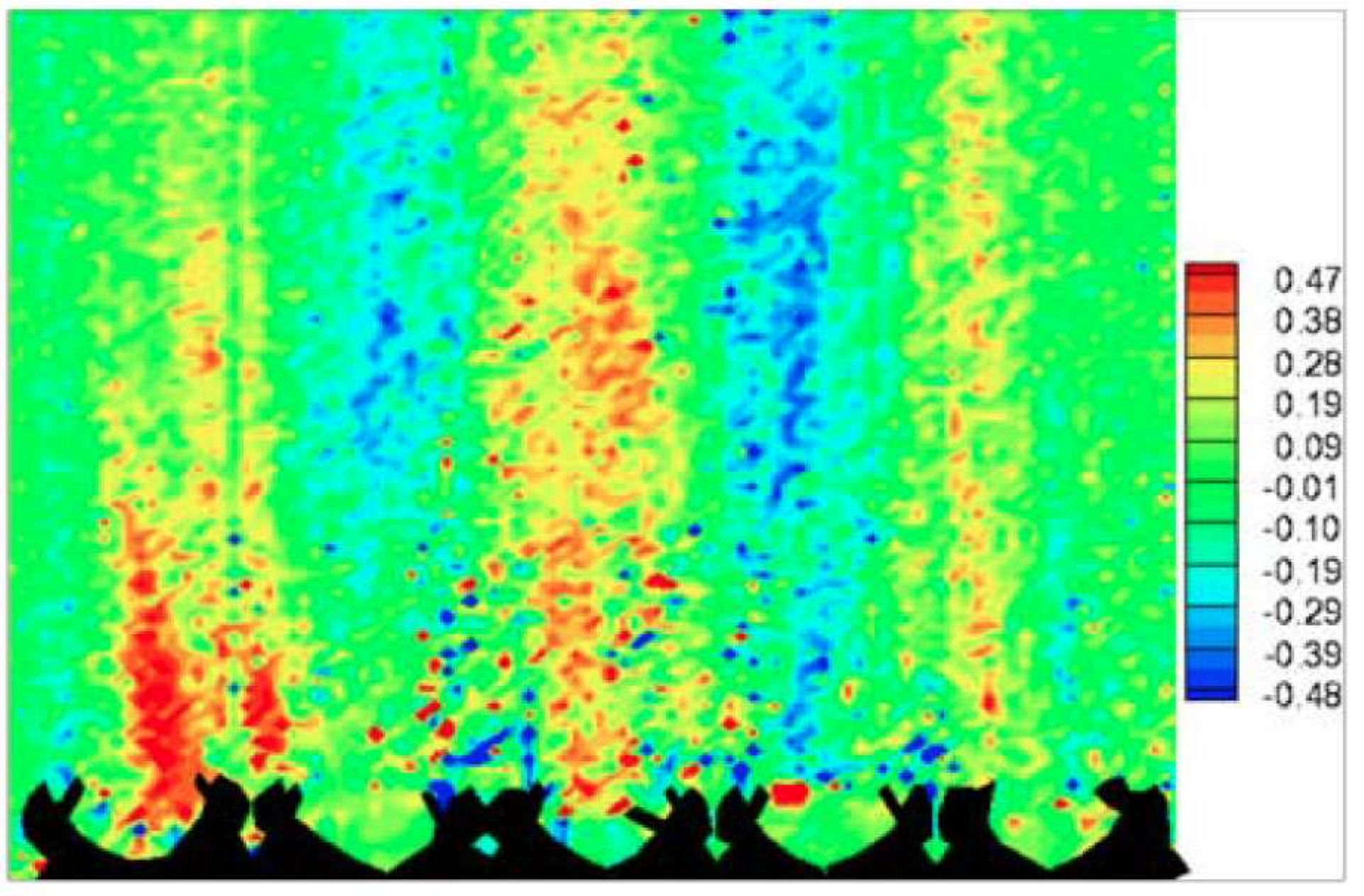}}
\subfloat[Hydra-TH X-Velocity]{
\includegraphics[width=0.39525\textwidth]{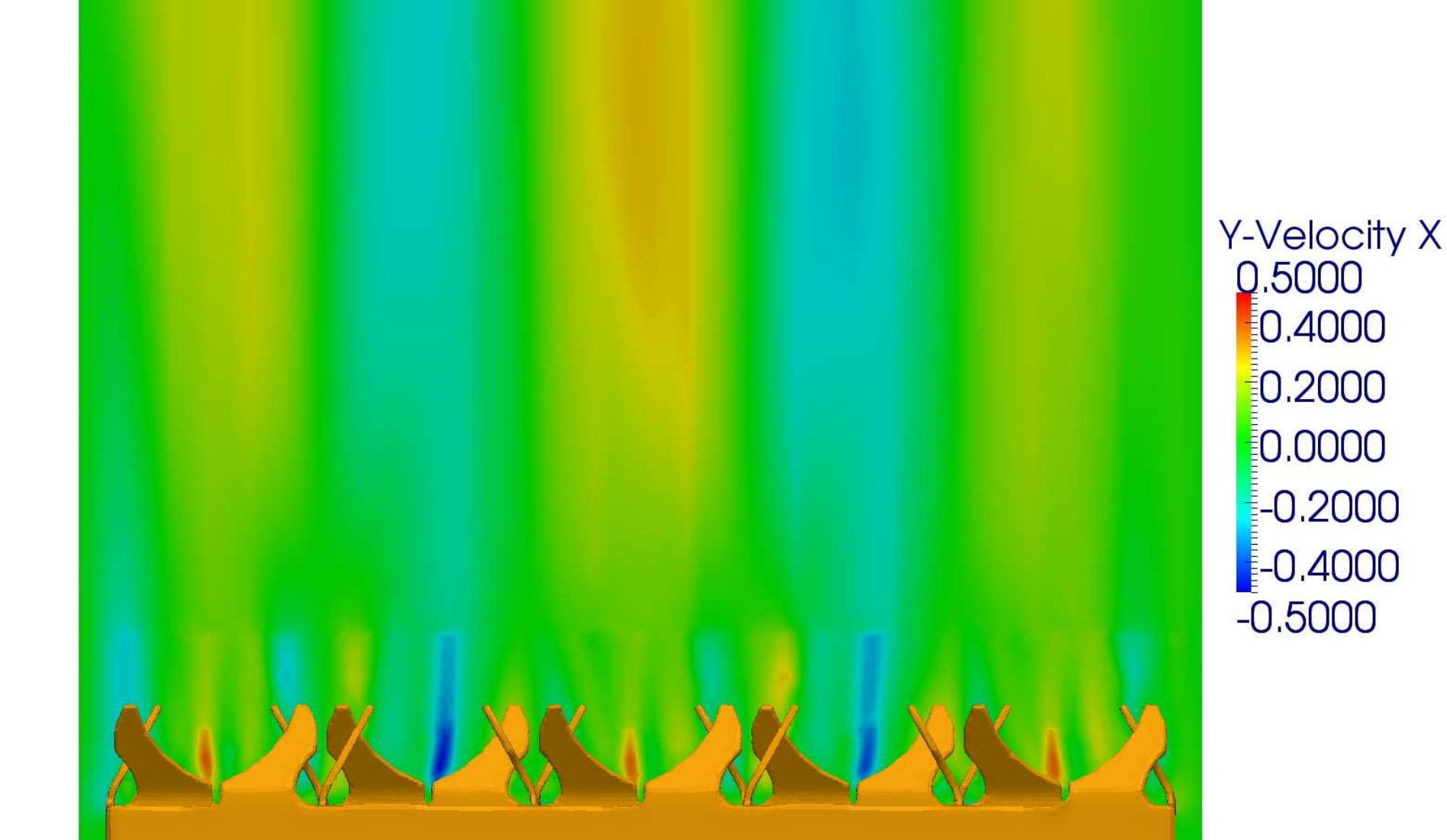}}\\
\caption{Experimental and computed axial (y-direciton) time-averaged velocities
on plane 7. Velocity magnitude has been scaled relative to the $2.48 \
m/s$ inlet velocity.}
\label{fig:5x5_velocity_plane7}
\end{center}
\end{figure*}

Following Yan, et al.\ \cite{yan:2012}, mean velocities are compared at
points A, C, D, E, G, and H, as shown in Figure
\ref{fig:5x5_vel_lines}. Here, the streamwise velocity in the
experiments corresponds to the $y$-velocity in the computation, while
the lateral velocity corresponds to the $x$-velocity. Yan, et al.\
\cite{yan:2012}, estimated the systematic uncertainty in the
velocities due to the PIV measurments, software acquisition, etc., to
be a maximum of $0.199\mathrm{m/s}$.  The statistical uncertainty,
which is a function of the number of snapshots of the velocity, is
estimated to be $\pm0.167V_\mathrm{inlet}$ in the lateral direction,
and $\pm0.15V_\mathrm{inlet}$ in the axial direction, where
$V_\mathrm{inlet}=2.48\mathrm{m/s}$. All experimental data has been
plotted with the uncertainty bounds provided by Dominguez-Ontiveros
and Hassan, see also \cite{conner:2011}.

Line plots of the velocity are presented in Figure \ref{fig:5x5_vel_lines} for
stations A -- H. Inspection of Figure \ref{fig:5x5_vel_lines} indicates that
experimental and computed $x$-velocities correlate relatively well, although for
points A, C, E, and G, the $x$-velocities are near zero. For this relatively
coarse mesh, the streamwise velocities do not compare as well, however, the
general trends appear to be similar. Note that typically, the $y$-velocity is
overpredicted in the streamwise directions, which is not surprising for this
coarse mesh.  In comparison, the mesh used by Yan, et al.\ \cite{yan:2012},
contained approximately 76M hexahedral elements.

The line plots of velocity for the 96M mesh are presented in Figure
\ref{fig:5x5_96M_vel_lines} for stations A -- H. In comparison to the velocity
profiles in Figure \ref{fig:5x5_vel_lines}, the 96M results match the
experimental data more closely at all points A -- H.
However, the stream-wise velocity still appears to be slightly overpredicted.
In contrast, the $x$-velocities fall within the uncertainty bounds for points A,
C, E, and G, while the $x$-velocities at points D and H have similar profiles, but
are not quite within the uncertainty bounds.  Overall, the 96M results compare
very well to the experimental data.

Time-averaged velocities in plane-5, see Figure \ref{fig:5x5_sample_locations}
from \cite{conner:2011}, are shown in Figure \ref{fig:5x5_velocity_plane5} with
the computed time-averaged mean velocity fields. Similarly, the experimental and
computed mean velocity fields on plane-7 are shown in Figure
\ref{fig:5x5_velocity_plane7}. The data in the figures have been scaled relative
to the $2.48\mathrm{m/s}$ inlet velocity. The peak velocities in the axial
direction are slightly under-predicted in the Hydra-TH computations, while the
lateral velocities are slighly over-predicted.  This is likely due to the very
coarse mesh used in this LES calculation.  While the peak velocities appear to
be relatively close to those found experimentally, inspection of Figures
\ref{fig:5x5_velocity_plane5} and \ref{fig:5x5_velocity_plane7} indicates that
the deflection in the velocity vectors due to the mixing vanes and the flow
housing is well-captured by the simulation.

\section{Summary\label{sec:summary}}
A series of isothermal turbulent flow calculations have been carried out using
Hydra-TH, a thermal hydraulics code developed at Los Alamos National Laboratory.
Our main goal is to understand the fluid dynamics of the flow-induced vibration
problem leading to grid-to-rod-fretting (GTRF), a major cause of nuclear plant
shutdowns.

We found that the mesh quality is extremely important for the accurate
computation of turbulent fluctuations and the resulting dynamic load on the fuel
rods.

Progress is reported here on several fronts towards a simulation capability for
advanced thermal-hydraulics methods in the nuclear engineering industry:
\begin{enumerate}
  \item \emph{Mesh generation.} Numeca's Hexpress/Hybrid mesh generator, a.k.a.\
        ``Spider'', has been used for the first time to generate computational
        meshes for the GTRF problem. Spider is easy to use, fast, and
        automatically generates high-quality meshes with optional
        power-law-graded boundary layers. Output is saved in the latest
        HDF5/ExodusII format, capable of storing meshes in the billion-cell
        range. Spider can also generate meshes, in one run, for both fluid and
        solid parts of a domain, which allows mesh generation for
        fluid-structure interaction and conjugate heat transfer problems
        \cite{numeca:2012}.
  \item \emph{Quantitative a priori mesh assessment.} A method for quantitative
        assessment of complex unstructured meshes with no-slip walls has been
        developed and used to assess a series of meshes generated for the GTRF
        problem by two mesh generators.
  \item \emph{New GTRF flow calculations.} A series of turbulent flow
        simulations have been carried out on both $3\times3$ and $5\times5$ rod
        bundle geometries. Various statistics of the fluctuating flow field have
        been analyzed and compared to data from computations carried out by
        Westinghouse using Star-CCM+ and from experiments at Texas A\&M
        University.
  \item \emph{RMS forces on rod order-of-magnitude different between Cubit and
        Spider meshes.} Arguably the most important quantity for coupling the
        current results to a structural code is the fluctuations of the pressure
        force loading the rods. Using the same algorithm and code, we found the
        predicted RMS forces, integrated for the whole rod, an order of
        magnitude larger using the higher-quality Spider meshes compared to
        the Cubit meshes, see \cite{GTRF_2011}.
\end{enumerate}

\section{Future work\label{sec:future}}
Future work on GTRF will focus on coupling the structural response of the fuel
rods at different dynamic levels of approximation (e.g.\ one-way, two-way),
along with coupling different wear models developed by collaborators in the CASL
project. The largest mesh we have run to date for the $3\times3$ and $5\times5$
problems have approximately 47 and 96 million computational cells, respectively.
In order to adequately resolve the turbulent flow features and the heat transfer
along the turbulent boundary layers, we believe meshes of 100 million to 1
billion elements may be required, depending on plant operating conditions. To
incorporate the effects of boiling, multiphase flow models are also being
developed and implemented in the Hydra software toolkit.

\section{Acknowledgments\label{sec:acknowledgments}}
This research is supported by the Consortium for Advanced Simulation of Light
Water Reactors (CASL), a U.S.\ Department of Energy Innovation Hub. The authors
gratefully acknowledge the help in visualization and high-performance computing
issues from Ross Toedte and Ramanan Sankaran, respectively, at Oak Ridge
National Laboratory; Elvis Dominguez-Ontiveros and Yassin Hassan at Texas A\&M
University for providing the experimental data; and the help in meshing from
Roger Pawlowski and Tom Smith at Sandia National Laboratories.

\section*{References}
\bibliographystyle{elsarticle-num}
\bibliography{paper}

\end{document}